\documentclass[12pt,a4paper]{article}
\usepackage[margin=0.6in]{geometry}

\renewcommand{\title}[1]{\vbox{\center\LARGE{#1}}\vspace{5mm}}
\renewcommand{\author}[1]{\vbox{\center#1}\vspace{5mm}}
\newcommand{\address}[1]{\vbox{\center\em#1}}
\newcommand{\email}[1]{\vbox{\center\tt#1}\vspace{5mm}}

\renewcommand{\date}[1]{\vbox{\center#1}}

\usepackage{amssymb}
\usepackage{amsmath,bm}
\usepackage{amssymb}
\usepackage{graphicx}
\usepackage{amsfonts}         
\usepackage{fancybox}   

\usepackage{enumitem}

\usepackage{slashed}

\usepackage{epstopdf}
\usepackage[usenames,dvipsnames]{xcolor}
\usepackage[pagebackref, bookmarks={false}, pdfauthor={Diptarka Das}, pdftitle={IslandJAM}]{hyperref}
\hypersetup{colorlinks=true, linkcolor=darkblue, citecolor=red, filecolor=OliveGreen, urlcolor=blue, filebordercolor={.8 .8 1}, urlbordercolor={.8 .8 0}}
\usepackage{soul}
\setstcolor{Red}

\usepackage{wrapfig}

\definecolor{jazzberryjam}{rgb}{0.65, 0.04, 0.37}
\definecolor{lust}{rgb}{0.9, 0.13, 0.13}
\definecolor{sandybrown}{rgb}{0.96, 0.64, 0.38}
\definecolor{mountainmeadow}{rgb}{0.19, 0.73, 0.56}
\definecolor{glaucous}{rgb}{0.38, 0.51, 0.71}
\definecolor{chromeyellow}{rgb}{1.0, 0.65, 0.0}
\definecolor{emerald}{rgb}{0.31, 0.78, 0.47}
\definecolor{deepsaffron}{rgb}{1.0, 0.6, 0.2}
\definecolor{darkgreen}{rgb}{0,0.4,0}
\definecolor{darkred}{rgb}{0.4,0,0}
\definecolor{darkblue}{rgb}{0,0,0.4}
\definecolor{lightblue}{rgb}{.6,.6,0.9}

\definecolor{uglybrown}{rgb}{0.8,  0.7,  0.5}

\definecolor{palatinatepurple}{rgb}{0.41, 0.16, 0.38}
\definecolor{celebrationcolor}{rgb}{0.75,  0.0,  0.9}

 \usepackage{mdframed}

\usepackage{framed}
\definecolor{shadecolor}{rgb}{0.90,0.90,0.90}

\usepackage{tkz-euclide}

\usepackage{makeidx}
\usepackage{cite}
\usepackage{bm}
\usepackage{geometry}
\usepackage{braket}
\usepackage[margin=20pt,small]{caption}

\usepackage[toc]{appendix}

\usepackage{tikz}
\usetikzlibrary{calc,decorations.markings,arrows.meta,shapes.misc,decorations.pathmorphing,calc,bending}

\usetikzlibrary{arrows.meta,shapes.misc,decorations.pathmorphing,calc,bending}

\tikzset{
  branch point/.style={cross out,draw=black,fill=none,minimum size=2*(#1-\pgflinewidth),inner sep=0pt,outer sep=0pt}, 
  branch point/.default=5
}
\tikzset{
  branch cut/.style={
    decorate,decoration=snake,
    to path={
      (\tikztostart) -- (\tikztotarget) \tikztonodes
    },
    }
  }

\input epsf

\newlength{\extraspace}
\newlength{\extraspaces}
\def\be{\begin{equation}}
\def\ee{\end{equation}}

\newcommand{\bea}{\begin{eqnarray}}
\newcommand{\eea}{\end{eqnarray}}

\def\II{\relax{I\kern-.10em I}}

\def\IB{\relax{\rm I\kern-.18em B}}

\def\ID{\relax{\rm I\kern-.18em D}}
\def\IE{\relax{\rm I\kern-.18em E}}
\def\IF{\relax{\rm I\kern-.18em F}}
\def\IG{\relax\hbox{$\inbar\kern-.3em{\rm G}$}}
\def\IGa{\relax\hbox{${\rm I}\kern-.18em\Gamma$}}
\def\IH{\relax{\rm I\kern-.18em H}}
\def\II{\relax{\rm I\kern-.18em I}}
\def\IK{\relax{\rm I\kern-.18em K}}

\def\inbar{\,\vrule height1.5ex width.4pt depth0pt}

\def\lp10{\ell_p^{10}}
\def\lp11{\ell_p^{11}}
\def\R11{R_{11}}

\def\frac#1#2{{#1 \over #2}}

\newdimen\tableauside\tableauside=1.0ex
\newdimen\tableaurule\tableaurule=0.4pt
\newdimen\tableaustep
\def\phantomhrule#1{\hbox{\vbox to0pt{\hrule height\tableaurule width#1\vss}}}
\def\phantomvrule#1{\vbox{\hbox to0pt{\vrule width\tableaurule height#1\hss}}}
\def\sqr{\vbox{%
  \phantomhrule\tableaustep
  \hbox{\phantomvrule\tableaustep\kern\tableaustep\phantomvrule\tableaustep}%
  \hbox{\vbox{\phantomhrule\tableauside}\kern-\tableaurule}}}
\def\squares#1{\hbox{\count0=#1\noindent\loop\sqr
  \advance\count0 by-1 \ifnum\count0>0\repeat}}
\def\tableau#1{\vcenter{\offinterlineskip
  \tableaustep=\tableauside\advance\tableaustep by-\tableaurule
  \kern\normallineskip\hbox
    {\kern\normallineskip\vbox
      {\gettableau#1 0 }%
     \kern\normallineskip\kern\tableaurule}%
  \kern\normallineskip\kern\tableaurule}}
\def\gettableau#1 {\ifnum#1=0\let\next=\null\else
  \squares{#1}\let\next=\gettableau\fi\next}

 \def\eqnn#1{\xdef #1{(\secsym\the\meqno)}\writedef{#1\leftbracket#1}%
 \global\advance\meqno by1\wrlabeL#1}
 \def\eqna#1{\xdef #1##1{\hbox{$(\secsym\the\meqno##1)$}}
 \writedef{#1\numbersign1\leftbracket#1{\numbersign1}}%
 \global\advance\meqno by1\wrlabeL{#1$\{\}$}}
 \def\eqn#1#2{\xdef #1{(\secsym\the\meqno)}\writedef{#1\leftbracket#1}%
 \global\advance\meqno by1$$#2\eqno#1\eqlabeL#1$$}

\global\newcount\itemno \global\itemno=0

\def\itemaut#1{\global\advance\itemno by1\noindent\item{\the\itemno.}#1}

\def\({\left(}
\def\){\right)}

\def\lsim{\mathrel{\mathstrut\smash{\ooalign{\raise2.5pt\hbox{$<$}\cr\lower2.5pt\hbox{$\sim$}}}}}
\def\gsim{\mathrel{\mathstrut\smash{\ooalign{\raise2.5pt\hbox{$>$}\cr\lower2.5pt\hbox{$\sim$}}}}}

\def\overleftrightarrow#1{\vbox{\ialign{##\crcr
     $\leftrightarrow$\crcr\noalign{\kern-0pt\nointerlineskip}
     $\hfil\displaystyle{#1}\hfil$\crcr}}}
     
     \def\overleftarrow#1{\vbox{\ialign{##\crcr
     $\leftarrow$\crcr\noalign{\kern-0pt\nointerlineskip}
     $\hfil\displaystyle{#1}\hfil$\crcr}}}

\usepackage[yyyymmdd,hhmmss]{datetime}
\newif{\ifeq}           
\eqtrue                 
\newcounter{lecturecounter}

\usepackage[utf8]{inputenc}
\usepackage{comment}
\usepackage{float, xcolor}
\usepackage{physics}
\usepackage{ragged2e}
\usepackage{booktabs}
\usepackage{latexsym}
\usepackage{mathtools}

\usepackage{subcaption}
\usepackage{graphicx}
\usepackage{cleveref}
\usepackage{amsmath}

\begin{document}
\title{  Reflected Entropy for Communicating Black Holes I: Karch-Randall Braneworlds }

\author{Mir Afrasiar${}^{\dagger}$, Jaydeep Kumar Basak${}^{\ddagger,\star,\dagger}$, Ashish Chandra${}^{\dagger}$ and Gautam Sengupta${}^{\dagger}$}

\address{ \vspace{0.4cm}
	{\it $\ddagger$
		Department of Physics,\\ National Sun Yat-Sen University, \\ Kaohsiung 80424, Taiwan\\}
	{\it $\star$
		Center for Theoretical and Computational Physics, \\ Kaohsiung 80424, Taiwan\\}
	{\it $\dagger$
		Department of Physics,\\Indian Institute of Technology Kanpur, \\208016, India\\
}}

\date{}			

\email{\href{mailto:afrasiar@iitk.ac.in}{afrasiar@iitk.ac.in}, \href{mailto:jkb.hep@gmail.com}{jkb.hep@gmail.com}, \href{mailto:achandra@iitk.ac.in}{achandra@iitk.ac.in},\\ \href{mailto:sengupta@iitk.ac.in}{sengupta@iitk.ac.in}}

\abstract{ 
	
	We obtain the reflected entropy for bipartite mixed state configurations of two adjacent and disjoint intervals at a finite temperature in $BCFT_2$s with two distinct boundaries through a replica technique in the large central charge limit. Subsequently these field theory results are reproduced from bulk computations involving the entanglement wedge cross section in the dual BTZ black hole geometry truncated by two Karch-Randall branes. Our result confirms the holographic duality between the reflected entropy and the bulk entanglement wedge cross section in the context of the $AdS_3/BCFT_2$ scenario. We further investigate the critical issue of the holographic Markov gap between the reflected entropy and the mutual information for these configurations from the bulk braneworld geometry and study its variation with subsystem sizes and time.

}

\newpage
\tableofcontents

\vfill\eject

\section{Introduction}
Black hole information paradox \cite{Hawking:1975vcx,Hawking:1976ra} has been one of the central problems in understanding various aspects of quantum gravity. Recently there has been significant progress towards a resolution of this issue in the context of lower dimensional effective theories of quantum fields coupled to semi classical gravity. An evaporating black hole together with its radiation flux as a closed quantum system is expected to respect
unitarity under time evolution. This suggests that for an evaporating black hole, the time evolution of the entanglement entropy of Hawking radiation should follow the Page curve \cite{Page:1993wv,Page:2013dx,Page:1993df}. In the last few years a possible resolution of this puzzle was proposed utilizing ``entanglement islands" mechanism which restores the unitarity 
of black hole evaporation for certain simple models. The quantum corrected version of the Ryu-Takayanagi proposal \cite{Ryu:2006bv,Ryu:2006ef,Hubeny:2007xt,Faulkner:2013ana} has been the main motivation behind this ``island'' formalism. This involves the generalized fine-grained entropy of a subregion in quantum field theories coupled to semi-classical gravity followed by an extremization over the location of the ``islands" which appear in the entanglement wedge of the subregion after the Page time \cite{Penington:2019npb,Almheiri:2019psf,Almheiri:2019hni,Almheiri:2019yqk,Almheiri:2020cfm}. The corresponding formula for the fine-grained entropy of a subregion $\mathcal{R}$ in the radiation flux of an evaporating black hole is given by,
\begin{align}\label{IsformEE}
	S[\mathcal{R}]=\min \left\{\operatorname{ext}_{Is(\mathcal{R})}\left[\frac{\operatorname{Area}[\partial Is(\mathcal{R})]}{4 G_{N}}+S_{\textit{eff}}[\operatorname{\mathcal{R}} \cup Is(\mathcal{R})]\right]\right\},
\end{align}
where $Is(\mathcal{R})$ is called the entanglement entropy island region which is included in entanglement wedge of the subregion $\mathcal{R}$. This formula has also been derived utilizing the gravitational path integral approach which describes certain non trivial replica wormhole saddle contributions in \cite{Almheiri:2020cfm,Penington:2019kki,Almheiri:2019qdq,Kawabata:2021vyo}.

Recently, it has been communicated in the literature that the entanglement island formula naturally emerges in the context of the $AdS/BCFT$ correspondence where the dual $CFT$ is defined on a manifold with a boundary ($BCFT$s) \cite{Takayanagi:2011zk, Anous:2022wqh}. The holographic dual geometry of a $BCFT$ with a single boundary involves a codimension-one surface called End-of-the-World (EOW) brane in the bulk $AdS$ spacetime \cite{Takayanagi:2011zk,Suzuki:2022xwv}. In a Karch-Randall braneworld geometry this EOW brane is termed a Karch-Randall (KR) brane which supports
a lower dimensional black hole induced from the bulk $AdS$ geometry \cite{Karch:2000ct,Karch:2000gx}. In this context, the dual $BCFT$ acts as a radiation bath for the evaporating black hole on the KR brane. It is then expected that the holographic entanglement entropy computed through the RT prescription should be consistent with the one obtained by utilizing the island formula in the framework of the $AdS/BCFT$ scenario and accordingly, the Page curve for the radiation region should be reproduced.

In this article, we consider one such model in a KR braneworld geometry described in \cite{Geng:2021iyq}. The authors considered a $BCFT_2$ defined on a manifold with two distinct boundaries. The bulk dual geometry is then described by an $AdS_3$ space time truncated by two KR branes containing matter $CFT_2$s with a constant Lagrangian. These matter fields on the KR branes are further coupled to the $BCFT_2$ through transparent boundary conditions at the junctions \cite{Almheiri:2019hni,Almheiri:2019yqk}. At a finite temperature, two copies of such $BCFT_2$s with two boundaries constitute a thermo-field double (TFD) state whose dual geometry corresponds to a wedge region of a bulk eternal BTZ black hole bounded by the two KR branes. Effectively this corresponds to two-dimensional black holes on the KR branes induced from the higher dimensional eternal BTZ black hole. In the effective two-dimensional theory, the two copies of the $BCFT_2$s serve as radiation baths for the induced black holes on the KR branes. Remarkably both the bath $BCFT_2$s with the KR branes appear to be gravitating from the mutual perspectives of each other \footnote{In this direction, application of the island formalism involving gravitating baths coupled to semiclassical gravity has been explored thoroughly in the article \cite{Geng:2020fxl}.}. In this framework, the authors of \cite{Geng:2021iyq} considered an interval in the bath $BCFT_2$s and obtained the corresponding entanglement entropy from both the field theory and the gravitational perspectives through corresponding replica techniques and wedge holography respectively \cite{Akal:2020wfl,Miao:2020oey}. 

On a separate note, from quantum information theory, the entanglement entropy is known to be a unique measure for pure states only and is invalid for mixed state entanglement. In this context, several valid mixed state entanglement and correlation measures such as entanglement negativity, reflected entropy, entanglement of purification have been proposed in quantum information theory \cite{Vidal:2002zz,Terhal:2002,Plenio:2007zz,Horodecki:2009zz,Dutta:2019gen} as well as for conformal field theories \cite{Dutta:2019gen,Calabrese:2012ew,Calabrese:2012nk,Calabrese:2014yza,Hirai:2018jwy,Caputa:2018xuf} and in holography \cite{Rangamani:2014ywa,Chaturvedi:2016rcn,Chaturvedi:2016rft,Jain:2017aqk,Takayanagi:2017knl,Malvimat:2018txq,Malvimat:2018ood,Kudler-Flam:2018qjo,Kusuki:2019zsp,Dutta:2019gen,KumarBasak:2020eia,KumarBasak:2021lwm,Dong:2021clv,Basu:2021axf,Basu:2021awn,Basu:2022nds}. 
In \cite{Afrasiar:2022ebi}, the authors have explored the model \cite{Geng:2021iyq} and computed the holographic entanglement negativity for various bipartite mixed state configurations in the bath $BCFT_2$s where the analogue of Page curves for the entanglement negativity was reproduced for communicating black holes, revealing some intriguing properties of the entanglement structure. 

Recently an island formulation for the reflected entropy ($S_R$)  a mixed state correlation measure introduced in \cite{Dutta:2019gen} was proposed and further explored in the literature \cite{Chandrasekaran:2020qtn,Li:2020ceg,Ling:2021vxe,Akers:2022zxr} which leads to the analogue of the Page curve for black hole evaporation \cite{Akers:2022max,Vardhan:2021mdy}. In the present article, we compute the reflected entropy for various mixed state configurations in the bath $BCFT_2$s for the communicating black hole model described earlier. Subsequently, we substantiate these field theory results from a bulk computation of the minimal entanglement wedge cross-section (EWCS) for the wedge region in the eternal BTZ black hole geometry subtended by the mixed states under consideration. We further obtain the holographic mutual information ($I$) for the mixed states in the $BCFT$s and compare the results with the reflected entropy which corresponds to an important feature that most multipartite entanglement implies, known as the ``\textit{Markov gap}" and it is defined as the difference $S_R-I$ \cite{Hayden:2021gno,Bueno:2020fle,Camargo:2021aiq,Bueno:2020vnx}.

This article is organized as follows. In \cref{review}, we review some earlier works relevant to our article. We begin with the review of a braneworld model described in \cite{Geng:2021iyq} in \cref{model} followed by a brief recapitulation of reflected entropy described in \cref{Reflected}. Subsequently we review the construction of the EWCS from \cref{EWCS} and the issue of the Markov gap described in \cref{Markov}. The next \cref{SrEwInGeng} describes
the detailed computations of the reflected entropy for various mixed state configurations involving two adjacent and disjoint intervals in the communicating black hole braneworld model using the field theory replica technique described in \cite{Dutta:2019gen,Jeong:2019xdr}. Furthermore the corresponding EWCS for the intervals in the 
bulk braneworld geometry are also computed to obtain the holographic reflected entropy and compared with the field theory results. In \cref{MarkovInGeng} we compare our results for the holographic reflected entropy with that of the holographic mutual information for different scenarios involving the adjacent and disjoint intervals and comment on the corresponding Markov gap observed. Finally we summarize our results in \cref{discussion} and discuss some future open issues.


\section{Review of earlier results}\label{review}
We begin with a brief review of a model involving two communicating black holes introduced in the article \cite{Geng:2021iyq}. The authors considered a TFD state which consists of two copies of $BCFT_2$s, each with two distinct boundaries at a finite temperature. The bulk dual geometry for this configuration is defined by an eternal $AdS_3$ BTZ black hole truncated by two Karch-Randall (KR) branes. Subsequently, we review the mixed state correlation measure termed reflected entropy and the replica technique for its computation
in $CFT_2$s described in \cite{Dutta:2019gen}. Following this we briefly review the entanglement wedge cross section (EWCS) \cite{Takayanagi:2017knl}
which describes the bulk holographic dual of the reflected entropy and also
review the Markov gap between the reflected entropy and the mutual information in the context of the $AdS_3/CFT_2$ correspondence.

\subsection{Braneworld model}\label{model}
The authors of \cite{Geng:2021iyq} considered a $BCFT_2$ at a zero temperature on a manifold containing two boundaries with distinct boundary conditions imposed on both. The holographic dual for this configuration involves two Karch-Randall (KR) branes truncating the bulk $AdS_3$ geometry. These KR branes are further coupled to $CFT_2$ matter fields with constant Lagrangians describing the brane tensions. From the two dimensional perspective, the gravitating KR branes are connected to a non-gravitating bath described by the $BCFT_2$. Transparent boundary conditions \cite{Almheiri:2019hni,Almheiri:2019yqk} are imposed at the junctions of the two branes with the bath $BCFT_2$. However, the $BCFT_2$ along with one of the two KR branes appears to be a gravitating bath from the perspective of the other brane. 

\begin{figure}[H]
	\centering
	\includegraphics[scale=.8]{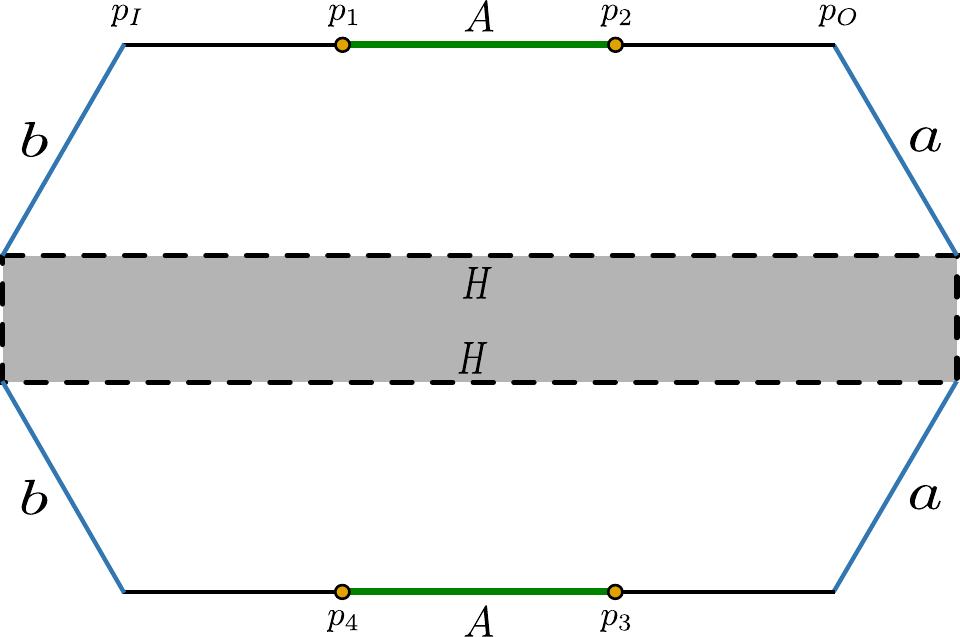}
	\caption{Diagram depicts a constant time slice of an eternal $AdS_3$ BTZ black hole truncated by two Karch-Randall (KR) branes namely $a$ and $b$ branes. At the asymptotic boundaries of the bulk BTZ black hole, two copies of $BCFT_2$s are located which constitute a TFD state. Each of the $BCFT_2$s lives on a manifold with two boundaries where the boundary points $p_I=r_I=1$ and $p_O=r_O=2$ contain different boundary conditions $b$ and $a$ respectively. An interval $A=[p_1,p_2]\cup [p_3,p_4]$ in the two copies of the $BCFT_2$s is also shown in the diagram. (Figure modified from \cite{Geng:2021hlu,Geng:2021iyq})}\label{geng}
\end{figure}

It is now possible to construct a TFD state at a finite temperature described by two copies of the above $BCFT_2$s each with two distinct boundaries. The holographic dual for this TFD state is described by an eternal BTZ black hole in the bulk $AdS_3$ braneworld geometry as depicted in \cref{geng}. The bulk eternal BTZ black hole induces two-dimensional black holes on the KR branes which are at different temperatures arising from the distinct boundary conditions imposed at the two boundaries of the $BCFT_2$s \cite{Rozali:2019day}. Once again from the two dimensional perspective, the KR branes are connected to the two copies of the $BCFT_2$s which serves as bath regions for the radiation flux from the induced black holes on the KR branes.

In the above framework, the authors of \cite{Geng:2021iyq} considered an interval in the $BCFT_2$s at the conformal boundary and computed the entanglement entropy of the corresponding bipartite state at zero and finite temperatures through a suitable replica technique. Subsequently these field theory results were substantiated through a bulk computation involving wedge holography. The entanglement entropy of the interval at a finite temperature in the bath $BCFT_2$s then characterized the communication of information between the two induced black holes on the KR branes.





\subsection{Reflected entropy}\label{Reflected} 
In this subsection, we provide a brief review of the reflected entropy which is a mixed state correlation measure in quantum information theory. For this purpose it is required to consider a bipartite system $A\cup B$ in a mixed state defined in a Hilbert space $\mathcal{H}_A \otimes \mathcal{H}_B$ with the density matrix $\rho_{AB}$. The canonical purification of the mixed state $\rho_{AB}$ is described by the density matrix $\sqrt{\rho_{AB}}$ in a Hilbert space $\mathcal{H}_A \otimes \mathcal{H}_B \otimes \mathcal{H}_A^* \otimes \mathcal{H}_B^*$  where $A^*$ and $B^*$ are the CPT conjugate copies of the intervals $A$ and $B$. The reflected entropy between the intervals $A$ and $B$ may then be defined as the von Neumann entropy between $A$ and $A^*$ in the pure state $\sqrt{\rho_{AB}}$ as follows\cite{Dutta:2019gen,Jeong:2019xdr}
\begin{align}
	S_R(A: B) = S(AA^*)_{\sqrt{\rho_{AB}}}.
\end{align}

In two-dimensional $CFT$s, the reflected entropy may be obtained by a replica technique developed in \cite{Dutta:2019gen}. In this context, it is required to consider two disjoint intervals $A \equiv [z_1,z_2]$ and $B \equiv [z_3,z_4]$ in a $CFT_{2}$. To obtain the reflected entropy for this configuration we require to compute the R\'enyi reflected entropy in terms of the partition function on an $mn$-sheeted replica manifold,
\begin{equation}
	S_{n}(AA^{*})_{\psi_{m}}=\frac{1}{1-n}\log \frac{Z_{n,m}}{(Z_{1,m})^n} \, .
\end{equation}
This may further be expressed in term of a four point correlation function of the twist operators located at the end points of the intervals described above as follows,
\begin{equation}\label{SR}
	S_{n}(AA^{*})_{\psi_{m}}=\frac{1}{1-n} \log \frac{\left<\sigma_{g_{A}}(z_{1})\sigma_{g_{A}^{-1}}(z_{2})\sigma_{g_{B}}(z_{3})\sigma_{g_{B}^{-1}}(z_{4})\right>_{CFT^{\otimes mn}}}{\left<\sigma_{g_{m}}(z_{1})\sigma_{g_{m}^{-1}}(z_{2})\sigma_{g_{m}}(z_{3})\sigma_{g_{m}^{-1}}(z_{4})\right>^n_{CFT^{\otimes m}}}
	=S_R^{(n,m)}(A:B) \, .
\end{equation}
Here $\sigma_{g_{A}}$s and $\sigma_{g_{m}}$s are the twist operators inserted at the end points of the intervals in the $nm$-sheeted and the $m$-sheeted replica manifolds respectively with conformal dimensions given by,
\begin{equation}\label{reflected-twist-field}
	h_{g_A^{-1}}=h_{g^{}_B}=\frac{n \,c}{24}\left(m-\frac{1}{m}\right),
	\quad h_{{g^{}_B} g_A^{-1}}=\frac{2 \,c}{24}\left(n-\frac{1}{n}\right), \quad h_{g_m}=\frac{c}{24}\left(m-\frac{1}{m}\right).
\end{equation}

Now, following the analytic continuations in the $n$ and $m$ replica indices, the corresponding reflected entropy may be obtained as\footnote{The two replica limits $n \to 1$ and $m \to 1$ do not commute with each other as discussed in \cite{Kusuki:2019evw, Akers:2021pvd, Dutta:2019gen, Akers:2022max}. In this work, we compute the reflected entropy by first considering $n \to 1$ and subsequently $m \to 1$ limit as suggested in \cite{Kusuki:2019evw, Akers:2021pvd, Dutta:2019gen, Akers:2022max}.}
\begin{equation}\label{ReplicaLimit}
	S_R(A:B)=\lim_{ n \to 1 }\lim_{m \to 1 }\,S_R^{(n,m)}(A:B) \, .
\end{equation}

In the next subsection we review the the entanglement wedge cross section (EWCS) which describes the bulk holographic dual of the reflected entropy  in the context of the $AdS/CFT$ correspondence. Note that besides the reflected entropy other entanglement/correlation measures such as  entanglement of purification (EoP) \cite{terhal2002entanglement}, odd entropy \cite{Tamaoka:2018ned} and the entanglement negativity \cite{Vidal:2002zz,Plenio:2005cwa} in dual $CFT$s have also been conjectured to be holographically related to the bulk EWCS in the $AdS/CFT$ scenario.

\subsection{Entanglement wedge cross section}\label{EWCS} 
\begin{figure}[H]
	\centering
	\includegraphics[scale=.3]{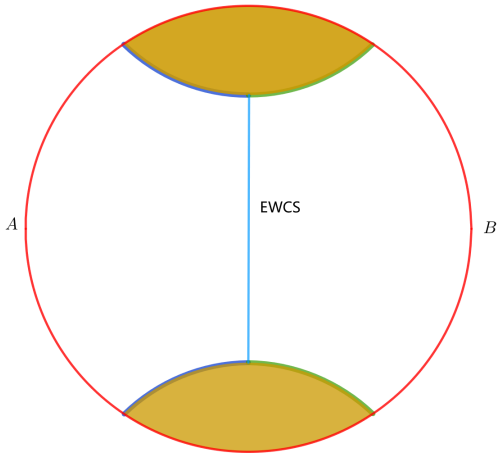}
	\caption{The unshaded region represents the entanglement wedge of the interval $A\cup B$. (Adapted from \cite{Chen:2022fte})}\label{ewcs}
\end{figure}

The entanglement wedge is described as the bulk dual of the density matrix $\rho_{AB}$ corresponding to a interval $A\cup B$ in the dual $CFT$ \cite{Czech:2012bh}. If $\gamma_{AB}$ is the codimension-two bulk minimal surface homologous to the interval $A\cup B$, then the entanglement wedge is defined by the bulk region enclosed by $A\cup B\cup \gamma_{AB}$. Subsequently, the entanglement wedge cross section ($E_W$) is described by the minimum cross sectional area of the entanglement wedge of the interval $A\cup B$,
\begin{equation}
	E_W(A:B)= \frac{\text{Area}(\Sigma^{\text{min}}_{AB})}{4 G_N}.
\end{equation}
Here $\Sigma^{\text{min}}_{AB}$ is the codimension-two minimal surface that divides the corresponding entanglement wedges of the intervals $A$ and $B$ as depicted in the \cref{ewcs}.

Consider the disjoint intervals $A$ and $B$ specified as $A\equiv [a_1,a_2]$ and $B\equiv [b_1,b_2]$ such that $a_1<a_2<b_1<b_2$. The authors of \cite{Takayanagi:2017knl} showed that for such disjoint intervals, the EWCS is given as
\begin{equation}\label{EW}
	E_W(A:B)=\frac{c}{6}  \log \left(1+2 z+2 \sqrt{z (z+1)}\right).
\end{equation}
Here $z$ is the finite temperature cross ratio defined as
\begin{equation}\label{CrRatio}
	z= \frac{\text{sinh}[\frac{\pi}{\beta}(a_2-a_1)]\,\text{sinh}[\frac{\pi}{\beta}(b_2-b_1)]}{\text{sinh}[\frac{\pi}{\beta}(b_1-a_2)]\,\text{sinh}[\frac{\pi}{\beta}(b_2-a_1)]}.
\end{equation}

In the context of the $AdS/CFT$ scenario the EWCS was proposed as the holographic dual of the reflected entropy $S_R(A:B)$ in \cite{Dutta:2019gen}. The authors further proved this duality from the gravitational path integral techniques developed in \cite{Lewkowycz:2013nqa} and is described as
\begin{align}\label{Sr_Ew}
	S_{R}(A: B)=2 E_{W}(A: B)\,,
\end{align}
where $E_{W}(A: B)$ is the EWCS for the bipartite quantum state $A\cup B$.

In the following subsection we describe a crucial issue of the holographic Markov gap \cite{Hayden:2021gno} between the reflected entropy and the mutual information in the context of the $AdS/CFT$ correspondence.

\subsection{Markov gap}\label{Markov}
Consider a bipartite mixed state $\rho_{A B}$ in the Hilbert space $H_A\otimes H_B$. As discussed in sec\cref{Reflected}, there exists a canonical purification $\sqrt{\rho_{A B}}$ in the doubled Hilbert space $H_A\otimes H_B\otimes H_{A^*}\otimes H_{B^*}$ such that the von Neumann entropy of $\rho_{A A^*}$ leads to the reflected entropy $S_R(A:B)$.

Now in the context of quantum information theory, a \textit{Markov recovery map} may be thought of as a quantum channel from a one-party interval into a two-party interval, provided with a three-party quantum state $\rho_{ABC}$. For example, consider the quantum channel defined by the map $\mathcal{R}_{B\rightarrow BC}$ whose action takes the system $B$ into the system $BC$. We may produce a tripartite state by the action of this quantum channel 
$\mathcal{R}$ on the bipartite reduced density matrix $\rho_{A B}$ as
\begin{align}\label{MRP1}
	\tilde{\rho}_{A B C}=\mathcal{R}_{B \rightarrow B C}\left(\rho_{A B}\right)\,.
\end{align}
In this connection, the \textit{Markov recovery process} corresponds to 
the production of the state $\rho_{A B C}$ by the action of a Markov map ($\mathcal{R}_{B \rightarrow B C}$ in the above example) on one of its bipartite reduced density matrix ($\rho_{A B}$ in the above example). A perfect \textit{Markov recovery process} happens when $\tilde{\rho}_{A B C}$ in \cref{MRP1} becomes equal to the tripartite state ${\rho}_{A B C}$. In this scenario, ${\rho}_{A B C}$ is termed as a quantum Markov chain with the ordering $A\rightarrow B\rightarrow C$ satisfying the eq.,
\begin{align}\label{MRP2}
	\rho_{A B C}=\mathcal{R}_{B \rightarrow B C}\left(\rho_{A B}\right)\,.
\end{align}
As discussed in \cite{Petz:1986tvy}, the above statement is true only when the conditional mutual information $I(A:C|B)$ vanishes. Furthermore, as demonstrated in \cite{Fawzi_2015}, there exists an upper bound to the mutual information given as
\begin{align}\label{fidelity}
	I(A: C \mid B) \geq-\max _{\mathcal{R}_{B \rightarrow B C}} \log F\left(\rho_{A B C}, \mathcal{R}_{B \rightarrow B C}\left(\rho_{A B}\right)\right)\,.
\end{align}
Here $F\left(\rho_{A B C}, \mathcal{R}_{B \rightarrow B C}\left(\rho_{A B}\right)\right)$, the quantum Fidelity of the optimal \textit{Markov recovery process}, lies between zero and one. This Fidelity is one for a perfect \textit{Markov recovery process} and zero when the density matrices have support on orthogonal subspaces. The inequality in \cref{fidelity} further implies some constraint conditions on the conditional mutual information given by,
\begin{align}
	I(A: B \mid B^*) &\geq-\max _{\mathcal{R}_{B \rightarrow B B^*}} \log F\left(\rho_{A B B^*}, \mathcal{R}_{B \rightarrow B B^*}\left(\rho_{A B}\right)\right)\,\label{MGap1},\\
	S_R(A:B)-I(A:B)&\geq-\max _{\mathcal{R}_{B \rightarrow B B^*}} \log F\left(\rho_{A B B^*}, \mathcal{R}_{B \rightarrow B B^*}\left(\rho_{A B}\right)\right)\,\label{MGap2},
\end{align}
where $\rho_{A B B^*}=Tr_{A^*}(|\sqrt{\rho_{AB}}\rangle\langle\sqrt{\rho_{AB}}| )$ is the reduced density matrix appearing in the canonical purification. The left hand side in \cref{MGap1} may be expressed in terms of the reflected entropy and the mutual information in \cref{MGap2} which is interpreted as the \textit{Markov Gap} in \cite{Hayden:2021gno}. In the framework of the $AdS_3/CFT_2$ scenario, the authors also demonstrated that the bound described in \cref{MGap2} may be expressed geometrically as
\begin{align}\label{MG}
	S_{R}(A: B)-I(A: B) \geq \frac{\log (2) \ell_{\text {AdS }}}{2 G_{N}} \times(\# \text { of boundaries of  the EWCS })+\mathcal{O}\left(\frac{1}{G_{N}}\right)\,,
\end{align}
where $l_{\text {AdS }}$ is the $AdS$ radius. The number of non-trivial endpoints of the EWCS ending in the bulk $AdS_3$ geometry defines the ``$\# \text { of boundaries of the EWCS}$'' in the above expression. Note that the endpoints at the asymptotic boundary are not to be considered here since they are located at spatial infinity.

\section{Reflected entropy and the EWCS in the braneworld model}\label{SrEwInGeng}
In this section, we first compute the reflected entropy for bipartite states involving adjacent and disjoint intervals in $BCFT_2$s for the communicating black hole configuration described in \cref{model}. In this context, we obtain the reflected entropy for various finite temperature mixed state configurations specific to channels for the corresponding twist field correlation functions in \cref{SR}. Subsequently, we demonstrate that the field theory results are exactly reproduced from a computation of the entanglement wedge cross section (EWCS) for the bulk eternal BTZ black hole geometries verifying the holographic duality described in \cref{Sr_Ew}.
We then evaluate the holographic mutual information for the corresponding mixed states in the bath $BCFT_2$s and describe the behaviour of the Markov gap for these mixed states through the comparison of the reflected entropy and the mutual information with respect to different interval sizes and time.

In the following, we first itemize the various contributions to the reflected entropy for different channels of the twist field correlator and the corresponding bulk dual EWCS computations for two adjacent intervals in the bath $BCFT_2$s.

\subsection{Adjacent intervals} 
We first consider two adjacent intervals $A\equiv[p_1,p_2]\cup[p_5,p_6]$ and $B\equiv[p_2+\epsilon,p_3]\cup[p_4,p_5+\epsilon]$ in the bath $BCFT_2$s for the communicating black holes described earlier. In this context, we compute 
the contributions from the various dominant channels of the corresponding twist field correlator for the reflected entropy of the above bipartite mixed state in the $BCFT_2$s depending on the interval sizes. Subsequently we compute the corresponding EWCSs in the dual bulk eternal BTZ black hole geometry and compare these with the field theory results.

\subsubsection*{Configuration (a):}
\begin{figure}[H]
	\centering
	\includegraphics[scale=.9]{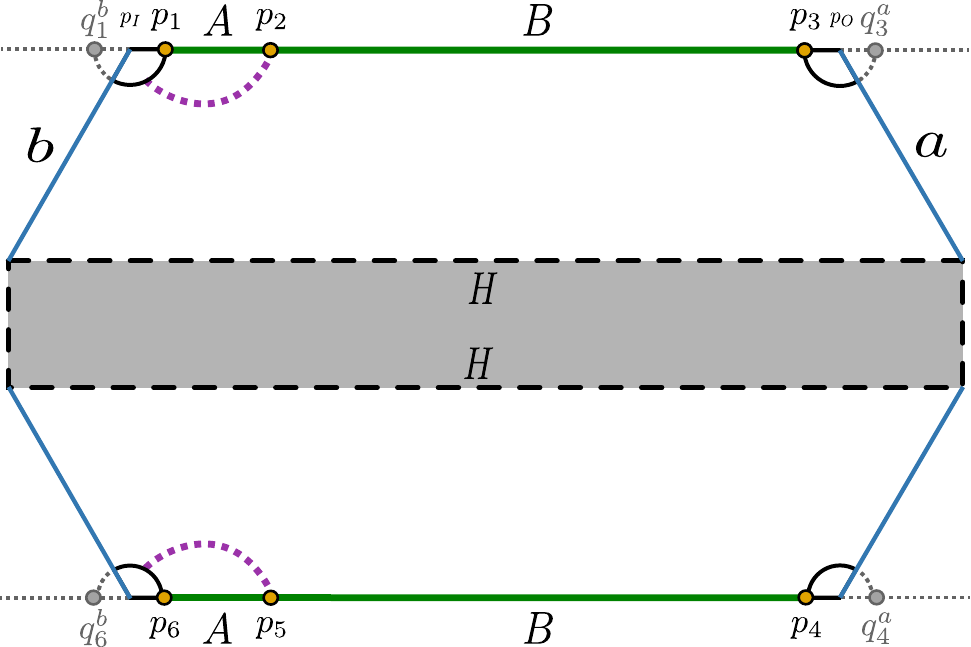}
	\caption{Dominant channel for the $S_R$. EWCS is denoted by the violet dotted curve. Here $q^b_{1}=r_I-\epsilon$ is the mirror image of the point $p_1=r_I+\epsilon$ with respect to the $b$-boundary at point $p_I=r_I$. Similarly, $q^a_{3}=r_O+\epsilon$ is the mirror image of the point $p_3=r_O-\epsilon$ with respect to the $a$-boundary at point $p_O=r_O$.}\label{fig_adjcase1a}
\end{figure}

We begin with the configuration as depicted in \cref{fig_adjcase1a} and obtain the reflected entropy for the adjacent intervals $A$ and $B$ in the boundary $BCFT_2$s with the twist operators located at their endpoints. In this case, we encounter a six-point twist correlator in the corresponding expression for the reflected entropy described in \cref{SR}. From the symmetry of the \cref{fig_adjcase1a}, it is observed that
the six-point function may be expressed in terms of certain three-point twist correlator in the large central charge limit. Following this the R\'enyi reflected entropy for this configuration may be obtained as,
\begin{align}
	S_R^{(n,m)}(A:B) &= 2\frac{1}{1-n}\log \frac{\left< \sigma_{g_A}(p_1)\sigma_{g_Bg_A^{-1}}(p_2)\sigma_{g_B^{-1}}(p_3) \right>_{\mathrm{BCFT}^{\bigotimes mn}}}{\left<\sigma_{g_m}(p_1)\sigma_{g_m^{-1}}(p_3) \right>^n_{\mathrm{BCFT}^{\bigotimes m}}}\notag\\
	&= 2\frac{1}{1-n}\log\frac{\left< \sigma_{g_A}(p_1)\sigma_{g_A^{-1}}(q^b_{1})\sigma_{g_Bg_A^{-1}}(p_2)\right>_{\mathrm{CFT}^{\bigotimes mn}}}{\left<\sigma_{g_m}(p_1)\sigma_{g_m^{-1}}(q^b_{1})\right>_{\mathrm{CFT}^{\bigotimes m}}^n}. \label{adjcase1a}
\end{align}
The factor two in the above equation arises due to the symmetry of this configuration as shown in the \cref{fig_adjcase1a}. Here we have utilized a factorization of the three-point function into a two-point and a one-point function in the $BCFT_2$ for the dominant channel as described in \cref{adjcasea} of appendix \ref{app_adj}.  The final expression in \cref{adjcase1a} was obtained by using the doubling trick which maps the $BCFT_2$ correlators to the entire complex plane. In the replica limit as described in \cref{ReplicaLimit}, the reflected entropy for the intervals $A$ and $B$ may then be obtained as
\begin{align}\label{SRcasea}
		S_R(A:B)= \frac{2c}{3} \log \left( 4\frac{\sinh \left(\frac{\pi  (r-r_I-\epsilon)}{\beta }\right) \sinh \left(\frac{\pi  (r_I-r-2\epsilon)}{\beta }\right)}{\sinh \left(\frac{\pi \epsilon}{\beta }\right) \sinh \left(\frac{\pi (-2\epsilon) }{\beta }\right)}\right) ,
\end{align}
where the factor four arises from the OPE coefficient of the three-point function in \cref{adjcase1a}.

The EWCS for this configuration is described by a co-dimension two surface which starts from the common point of the two adjacent intervals $A$ and $B$ and lands on the RT surface for the interval $[p_I,p_1]\equiv[r_I,r_I+\epsilon]$ as shown in \cref{fig_adjcase1a}. We follow the procedure discussed in \cref{EWCS} to compute the corresponding EWCS from \cref {EW} which in the adjacent limit, for the above intervals reduces to,
\begin{align}\label{EWcase1a}
	E_W(A:B)=\frac{c}{3}  \log \left(4z \right).
\end{align}
Here  $z$ is the cross-ratio at a finite temperature $\beta$ may be given by,
\begin{equation}\label{CRadjj1c}
	z=\frac{\sinh \left(\frac{\pi  (p_2-p_1)}{\beta }\right) \sinh \left(\frac{\pi  (q_1^b-p_2-\epsilon)}{\beta }\right)}{\sinh \left(\frac{\pi  (p_2+\epsilon-p_2)}{\beta }\right) \sinh \left(\frac{\pi (q_1^b-p_1) }{\beta }\right)}
	=\frac{\sinh \left(\frac{\pi  (r-r_I-\epsilon)}{\beta }\right) \sinh \left(\frac{\pi  (r_I-r-2\epsilon)}{\beta }\right)}{\sinh \left(\frac{\pi \epsilon}{\beta }\right) \sinh \left(\frac{\pi (-2\epsilon) }{\beta }\right)}.
\end{equation}

\subsubsection*{Configuration (b):}
\begin{figure}[H]
	\centering
	\includegraphics[scale=.9]{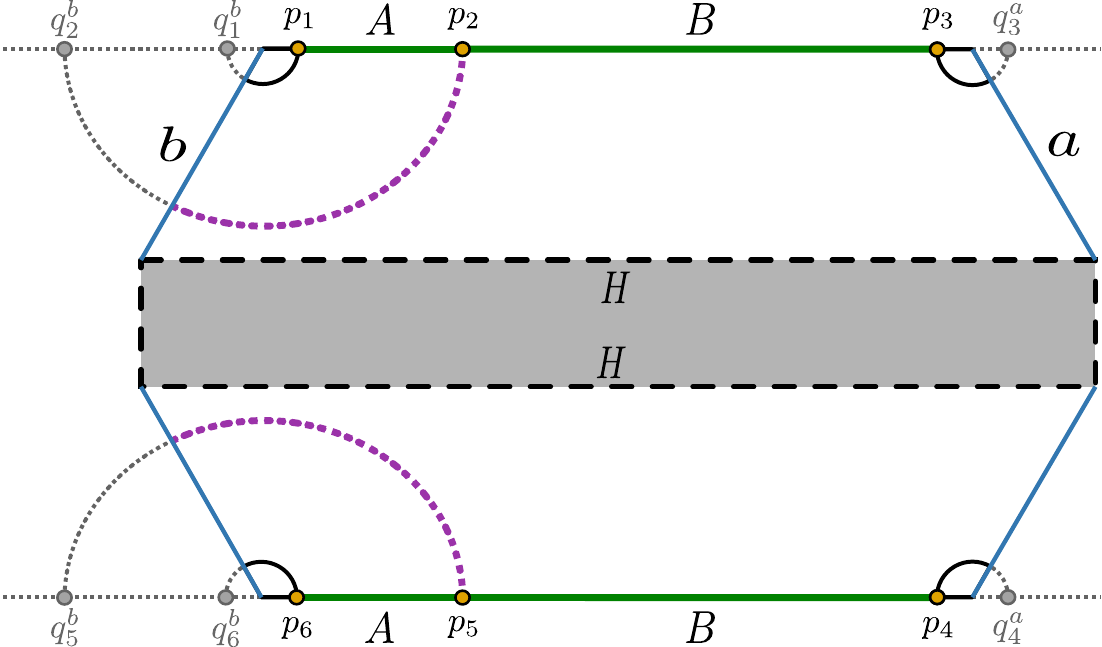}
	\caption{Dominant channel for the $S_R$. EWCS is denoted by the violet dotted curve. Here $q^b_{1}=r_I-\epsilon$ is the mirror image of the point $p_1=r_I+\epsilon$ with respect to the $b$-boundary at $p_I=r_I$. Similarly, $q^a_{3}=r_O+\epsilon$ is the mirror image of the point $p_3=r_O-\epsilon$ with respect to the $a$-boundary at $p_O=r_O$. Also, $q^b_{2}=2r_I-r$ is the mirror image of the point $p_2=r$ with respect to the $b$-boundary.}\label{fig_adjcase1b}
\end{figure}

This configuration describes another possible contribution to the reflected entropy $S_R(A:B)$ from a dominant channel of the corresponding twist field correlators. As earlier, from the symmetry of the configuration described in \cref{fig_adjcase1b}, the six-point correlation function in the $BCFT_2$s reduces to three-point function in the large central charge limit. The expression for the R\'enyi reflected entropy may then be given by
\begin{align}
	S_R^{(n,m)}(A:B) &= 2\frac{1}{1-n}\log \frac{\left< \sigma_{g_A}(p_1)\sigma_{g_Bg_A^{-1}}(p_2)\sigma_{g_B^{-1}}(p_3) \right>_{\mathrm{BCFT}^{\bigotimes mn}}}{\left<\sigma_{g_m}(p_1)\sigma_{g_m^{-1}}(p_3) \right>^n_{\mathrm{BCFT}^{\bigotimes m}}}\notag\\
	& = 2\frac{1}{1-n} \log\left(\left<\sigma_{g_Bg_A^{-1}}(p_2)\sigma_{g_Ag_B^{-1}}(q^b_{2})\right>_{\mathrm{CFT}^{\bigotimes mn}}\right). \label{adjcase1b}
\end{align}
The factor two in the above expression corresponds to the symmetry of the \cref{fig_adjcase1b} and we have utilized a factorization of  the three-point function in $BCFT_2$s  at large central charge limit to obtain a two-point and a one-point function as shown in \cref{adjcaseb} of appendix \ref{app_adj}. Finally using the doubling trick, we obtain the expression for the $S_R^{(n,m)}(A:B)$ in \cref{adjcase1b}. The key point to be noted here is that we utilize the doubling trick for the one-point function $\left<\sigma_{g_Bg_A^{-1}}(p_2)\right>$ with respect to the $b$-boundary.
We may then consider the replica limit $(n,m\to 1)$ of \cref{adjcase1b} to obtain the reflected entropy for the adjacent intervals $A$ and $B$ as follows
\begin{align}\label{SRcaseb}
	S_R(A:B) &= \frac{2c}{3}  \log \left( \frac{r^2-r_I^2}{r_I \epsilon } \right)+4 S_{\text{bdyb}}.
\end{align}

The EWCS in this configuration corresponds to a co-dimension two surface which starts from the common point of the two adjacent intervals and intersects the $b$-brane as depicted in \cref{fig_adjcase1b}. The expression for the EWCS may now be obtained using the area of the RT surface described in \cite{Geng:2021iyq} as follows,
\begin{equation}\label{EWcaseb}
		E_W(A:B)=\frac{ c}{3}  \log \left( \frac{r^2-r_I^2}{r_I \epsilon } \right)+2 S_{\text{bdyb}}.
\end{equation}

\subsubsection*{Configuration (c):}
We now discuss another configuration where a non-trivial contribution to the reflected entropy $S_R(A:B)$ arises from a different dominant channel of the corresponding twist field correlator. The R\'enyi reflected entropy in this configuration involves a six-point function with the twist operators located at the end points of the intervals in both the $BCFT_2$ copies. However, in the large central charge limit, this six-point function factorizes to four one-point and one two-point function in the $BCFT_2$s as shown in \cref{discasec} of appendix \ref{app_adj}. Finally, we utilize the doubling trick as earlier to map the $BCFT_2$ correlators to the entire complex plane. The R\'enyi reflected entropy for this configuration may then be expressed as follows
\begin{align}
	S_R^{(n,m)}(A:B) &= \frac{1}{1-n}\log \frac{\left< \sigma_{g_A}(p_1)\sigma_{g_Bg_A^{-1}}(p_2)\sigma_{g_B^{-1}}(p_3)\sigma_{g_B}(p_4)\sigma_{g_Ag_B^{-1}}(p_5)\sigma_{g_A^{-1}}(p_6) \right>_{\mathrm{BCFT}^{\bigotimes mn}}}{\left<\sigma_{g_m}(p_1)\sigma_{g_m^{-1}}(p_3)\sigma_{g_m}(p_4)\sigma_{g_m^{-1}}(p_6) \right>^n_{\mathrm{BCFT}^{\bigotimes m}}}\notag\\
	&= \frac{1}{1-n}\log\left( \left<\sigma_{g_Bg_A^{-1}}(p_2)\sigma_{g_Ag_B^{-1}}(p_5)\right>_{\mathrm{CFT}^{\bigotimes mn}}\right). \label{adjcase1c}
\end{align}
Implementing the replica limit in \cref{adjcase1c}, the corresponding reflected entropy for the intervals $A$ and $B$ may be computed as 
\begin{align}\label{SRcasec}
	S_R(A:B) &= \frac{2c}{3}  \log \left(\frac{2 r}{\epsilon } \cosh \frac{2 \pi  t}{\beta }\right).
\end{align}

\begin{figure}[H]
	\centering
	\includegraphics[scale=.9]{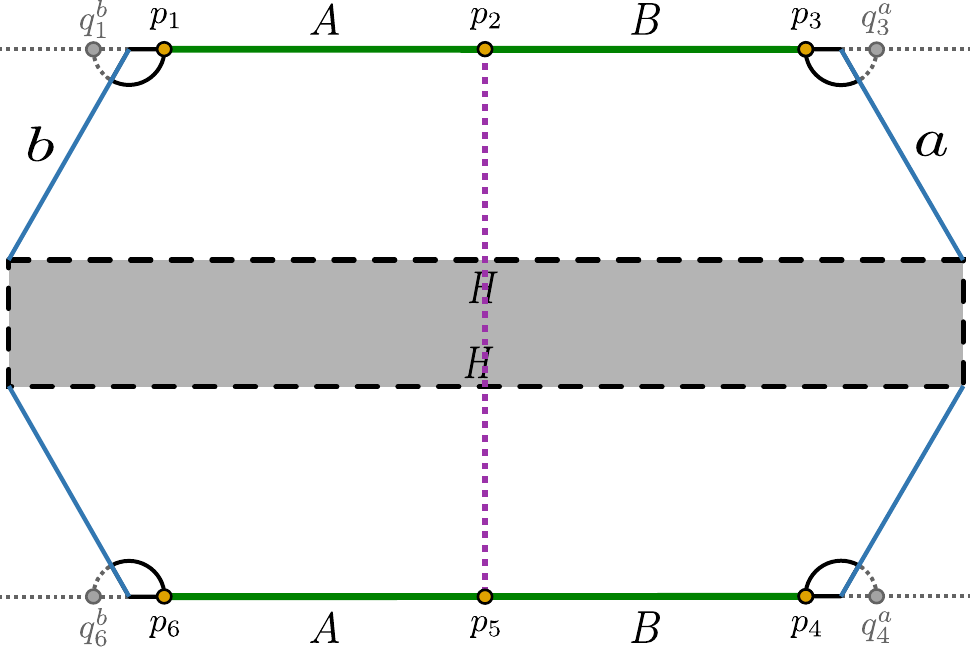}
	\caption{Possible channel for $S_R$. The violet dotted line defines the EWCS. Here $q^b_{1}=r_I-\epsilon$ and $q^b_{6}=r_I-\epsilon$ are the mirror images of the points $p_1=r_I+\epsilon$ and $p_6=r_I+\epsilon$ respectively with respect to the $b$-boundary. Similarly, $q^a_{3}=r_O+\epsilon$ and $q^a_{4}=r_O+\epsilon$ are the mirror images of the points $p_3=r_O-\epsilon$ and $p_4=r_O-\epsilon$ respectively with respect to the $a$-boundary. The point $p_2$ and $p_5$ are equal to $r$.}\label{fig_adjcase1c}
\end{figure}

For this configuration, the EWCS corresponds to the co-dimension two surface known as the Hartman-Maldacena (HM) surface which connects the common points of the two adjacent intervals in the asymptotic boundaries as depicted in \cref{fig_adjcase1c}. Once again the EWCS may be obtained from the area of the HM surface described in \cite{Geng:2021iyq} as
\begin{equation}\label{HM}
	E_W(A:B)=\frac{c}{3} \log \left(\frac{2 r}{\epsilon } \cosh \frac{2 \pi  t}{\beta }\right).
\end{equation}

\subsubsection*{Configuration (d):}
\begin{figure}[H]
	\centering
	\includegraphics[scale=.9]{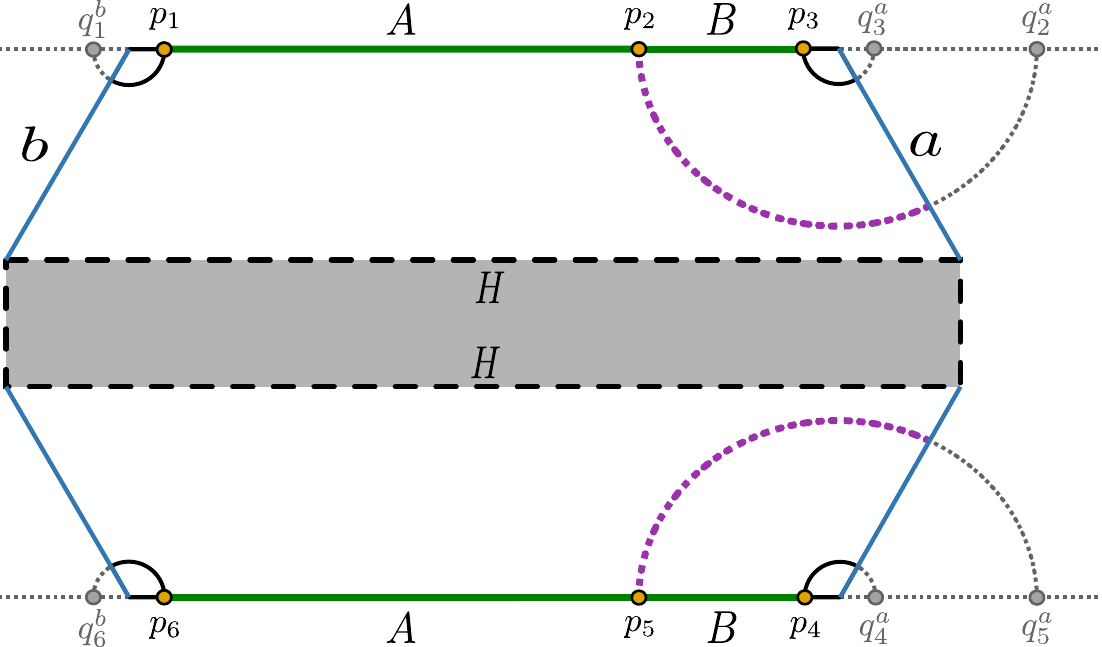}
	\caption{ Here the points and their corresponding mirror images are same as in the previous figure. Additionally, $q^a_{2}=2r_O-r$ and $q^a_{5}=2r_O-r$ are the mirror images of the points $p_2=r$ and $p_5=r$ with respect to the $a$-boundary at $p_O=r_O$.}\label{fig_adjcase1d}
\end{figure}

This scenario is similar to the configuration (b) where the one-point function $\langle\sigma_{g_Bg_A^{-1}}(p_2)\rangle$ in the $BCFT_2$ was written as a two-point function utilizing the doubling trick with respect to the $b$-boundary. In this case however, the doubling trick for the same one-point correlation function is performed with respect to the $a$-boundary of the $BCFT_2$ as depicted in \cref{fig_adjcase1d}. Once again, the factor two in the expression below arises from the symmetry of the figure.
\begin{align}
	S_R^{(n,m)}(A:B) &= 2\frac{1}{1-n}\log \frac{\left< \sigma_{g_A}(p_1)\sigma_{g_Bg_A^{-1}}(p_2)\sigma_{g_B^{-1}}(p_3) \right>_{\mathrm{BCFT}^{\bigotimes mn}}}{\left<\sigma_{g_m}(p_1)\sigma_{g_m^{-1}}(p_3) \right>^n_{\mathrm{BCFT}^{\bigotimes m}}}\notag\\
	& = 2\frac{1}{1-n}\log\left(\left<\sigma_{g_Bg_A^{-1}}(p_2)\sigma_{g_Ag_B^{-1}}(q^a_{2})\right>_{\mathrm{CFT}^{\bigotimes mn}}\right). \label{adjcase1d}
\end{align}
Now considering the replica limit as described in \cref{ReplicaLimit}, we obtain the reflected entropy for this configuration as
\begin{align}\label{SrCased}
	S_R(A:B) &= \frac{2c}{3}  \log \left( \frac{r_O^2-r^2}{r_O \epsilon } \right)+4 S_{\text{bdya}}.
\end{align}

The corresponding EWCS for this configuration is a co-dimension two surface which starts from the common point of the two adjacent intervals in the $BCFT_2$ and lands on the $a$ brane (\cref{fig_adjcase1d}). The computation for the EWCS follows from the results of \cite{Geng:2021iyq} as follows,
\begin{equation}\label{EwCased}
	E_W(A:B)=\frac{c}{3}\log \left( \frac{r_O^2-r^2}{r_O \epsilon } \right)+2 S_{\text{bdya}}.
\end{equation}

\subsubsection*{Configuration (e):}
\begin{figure}[H]
	\centering
	\includegraphics[scale=.9]{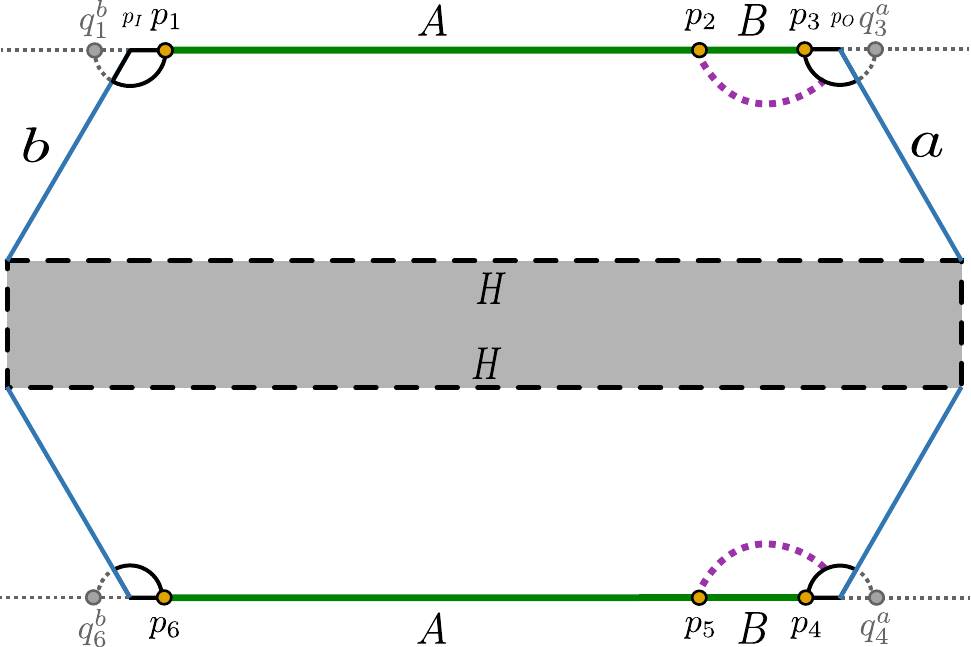}
	\caption{Possible channel for $S_R$. The violet dotted curve denotes the EWCS. All the points in the boundary theory are identified similarly as in \cref{fig_adjcase1a}.}\label{fig_adjcase1e}
\end{figure}

We proceed to another case which is analogous to the one discussed in configuration (a). However, in this configuration, the twist field correlators in the reflected entropy expression factorizes differently in the large central charge limit as shown in \cref{adjcasee} of appendix \ref{app_adj}.  This particular factorization  arises due to the locations of the twist field operators at the endpoints of the intervals in the $BCFT_2$s, depicted in \cref{fig_adjcase1e}. The corresponding R\'enyi reflected entropy for the bipartite mixed state configuration may be obtained as
\begin{align}
	S_R^{(n,m)}(A:B) &= 2\frac{1}{1-n}\log \frac{\left< \sigma_{g_A}(p_1)\sigma_{g_Bg_A^{-1}}(p_2)\sigma_{g_B^{-1}}(p_3) \right>_{\mathrm{BCFT}^{\bigotimes mn}}}{\left<\sigma_{g_m}(p_1)\sigma_{g_m^{-1}}(p_3) \right>^n_{\mathrm{BCFT}^{\bigotimes m}}}\notag\\
	&= 2\frac{1}{1-n}\log\frac{\left<\sigma_{g_Bg_A^{-1}}(p_2)\sigma_{g_B^{-1}}(p_3) \sigma_{g_B}(q^a_{3})\right>_{\mathrm{CFT}^{\bigotimes mn}}}{\left<\sigma_{g_m^{-1}}(p_3)\sigma_{g_m^{1}}(q^a_{3}) \right>_{\mathrm{CFT}^{\bigotimes m}}^n}. \label{adjcase1e}
\end{align}
As earlier, the pre-factor two in the above expression appears due to the symmetry of  \cref{fig_adjcase1e}. Once again we implement the replica limit in the \cref{adjcase1e} to compute the reflected entropy for the intervals $A$ and $B$ as
\begin{align}\label{SRcasee}
	S_R(A:B)=\frac{2c}{3}  \log \left(4 \frac{\sinh \left(\frac{\pi  (r-r_O-\epsilon)}{\beta }\right) \sinh \left(\frac{\pi  (r_O-r-2\epsilon )}{\beta }\right)}{\sinh \left(\frac{\pi  \epsilon }{\beta }\right) \sinh \left(\frac{\pi  (-2 \epsilon )}{\beta }\right) } \right),
\end{align}
where the factor four arises from the OPE coefficient of the three-point function in \cref{adjcase1e}.

We now discuss the bulk computation of the reflected entropy for this configuration where the violet dotted curve in \cref{fig_adjcase1e} describes the corresponding EWCS in the dual eternal BTZ black hole geometry. In this case, the corresponding EWCS starts from the common point of the two adjacent intervals and lands on the RT surface of the interval $[p_3,p_O]\equiv[r_O-\epsilon,r_O]$. Following the procedure as in configuration (a), the expression for the EWCS may then be obtained by considering the adjacent limit in \cref{EW} as,
\begin{align}\label{EwCasee}
	E_W(A:B)=\frac{c}{3}  \log \left(4 z  \right).
\end{align}
Here $z$ is the finite temperature cross ratio given by,
\begin{equation}\label{CrCasee}
	z= \frac{\sinh \left(\frac{\pi  (p_2-q_3^a)}{\beta }\right) \sinh \left(\frac{\pi  (p_3-p_2-\epsilon)}{\beta }\right)}{\sinh \left(\frac{\pi  (p_2+\epsilon-p_2)}{\beta }\right) \sinh \left(\frac{\pi  (p_3-q_3^a) }{\beta }\right)}
	=\frac{\sinh \left(\frac{\pi  (r-r_O-\epsilon)}{\beta }\right) \sinh \left(\frac{\pi  (r_O-r-2\epsilon )}{\beta }\right)}{\sinh \left(\frac{\pi  \epsilon }{\beta }\right) \sinh \left(\frac{\pi  (-2 \epsilon )}{\beta }\right) }.
\end{equation}

\subsubsection*{Configuration (f):} 
\begin{figure}[H]
	\centering
	\includegraphics[scale=.9]{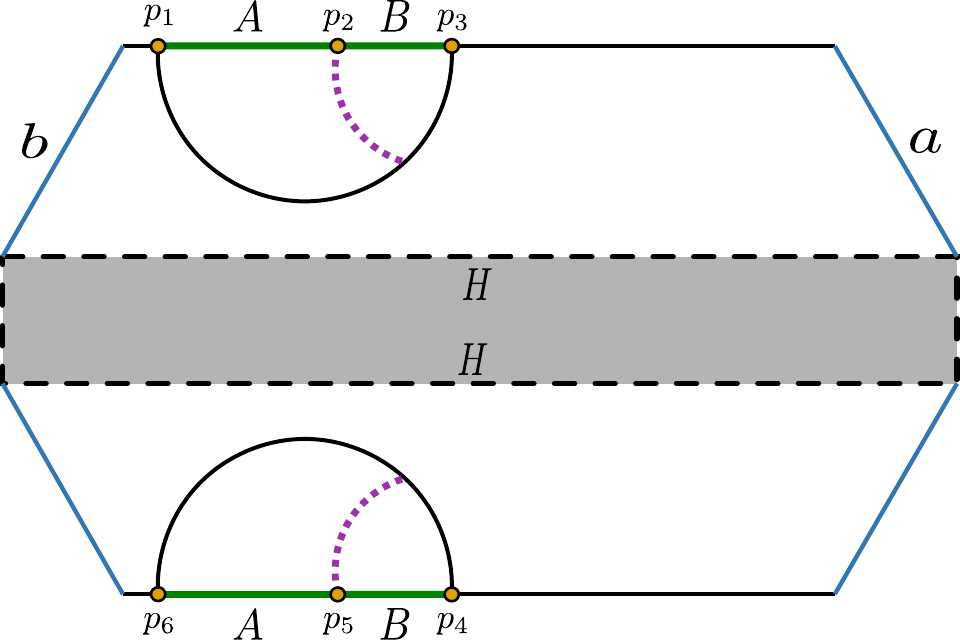}
	\caption{Possible channel for the $S_R$ when the interval sizes are small compared to the full bath $BCFT_2$s. The points in the diagram are specified as $p_1=r_I+\epsilon$, $p_2=r_1$ and $p_3=r_2$. Similarly in its TFD copy, the points are identified as $p_6=r_I+\epsilon$, $p_5=r_1$ and $p_4=r_2$}\label{fig_adjcase1f}
\end{figure}

The reflected entropy computation for this configuration is trivial since it reduces to the usual computation of $S_R$ for two adjacent intervals in a $CFT_2$. The R\'enyi reflected entropy then may be expressed as,
\begin{align}
	S_R^{(n,m)}(A:B) &= 2\frac{1}{1-n}\log \frac{\left< \sigma_{g_A}(p_1)\sigma_{g_Bg_A^{-1}}(p_2)\sigma_{g_B^{-1}}(p_3) \right>_{\mathrm{BCFT}^{\bigotimes mn}}}{\left<\sigma_{g_m}(p_1)\sigma_{g_m^{-1}}(p_3) \right>^n_{\mathrm{BCFT}^{\bigotimes m}}}\notag\\
	&= 2\frac{1}{1-n}\log \frac{\left< \sigma_{g_A}(p_1)\sigma_{g_Bg_A^{-1}}(p_2)\sigma_{g_B^{-1}}(p_3) \right>_{\mathrm{CFT}^{\bigotimes mn}}}{\left<\sigma_{g_m}(p_1)\sigma_{g_m^{-1}}(p_3) \right>^n_{\mathrm{CFT}^{\bigotimes m}}}, \label{adjcase2a}
\end{align}
with the symmetry factor two originating from the two copies of the $BCFT_2$s in a TFD state as depicted in \cref{fig_adjcase1f}. The reflected entropy for this configuration may then be obtained considering the replica limit as follows
\begin{align}
	S_R(A:B) &=\frac{2c}{3}  \log \left(4 \frac{\text{sinh}\left(\frac{\pi}{\beta}(r_1-r_I-\epsilon)\right)\,\text{sinh}\left(\frac{\pi}{\beta}(r_2-r_1-\epsilon)\right)}{\text{sinh}\left(\frac{\pi}{\beta}(\epsilon)\right)\,\text{sinh}\left(\frac{\pi}{\beta}(r_2-r_I-\epsilon)\right)} \right),
\end{align}
where the factor four arises from the OPE coefficient of the three-point function in \cref{adjcase2a}.

In the dual bulk geometry, the interval $A\cup B$ supports dome-type RT surface as depicted in \cref{fig_adjcase1f} and the EWCS for this configuration may be obtained from \cref{EW} which can be expressed in the adjacent limit as
\begin{align}
	E_W(A:B)=\frac{c}{3}  \log \left(4 z  \right),
\end{align}
with $z$ being the finite temperature cross ratio given by,
\begin{equation}\label{Cr}
	z=\frac{\text{sinh}\left(\frac{\pi}{\beta}(p_2-p_1)\right)\,\text{sinh}\left(\frac{\pi}{\beta}(p_3-p_2-\epsilon)\right)}{\text{sinh}\left(\frac{\pi}{\beta}(p_2+\epsilon-p_2)\right)\,\text{sinh}\left(\frac{\pi}{\beta}(p_3-p_1)\right)}
	=\frac{\text{sinh}\left(\frac{\pi}{\beta}(r_1-r_I-\epsilon)\right)\,\text{sinh}\left(\frac{\pi}{\beta}(r_2-r_1-\epsilon)\right)}{\text{sinh}\left(\frac{\pi}{\beta}(\epsilon)\right)\,\text{sinh}\left(\frac{\pi}{\beta}(r_2-r_I-\epsilon)\right)}.
\end{equation}

\subsubsection*{Configuration (g):}
We proceed to another possible contribution to the reflected entropy which is analogous to the one obtained in configuration (a). However in this case, the factorization of the multipoint correlator in the reflected entropy expression is quite different due to a different dominant channel as shown in \cref{adjcaseg} of appendix \ref{app_adj}. Particularly, the six-point function of the twist operators, located at endpoints of the intervals in the two copies of $BCFT_2$s, factorizes into three lower-point functions. Nevertheless, the final result of the reflected entropy turns out to be the same as that obtained in configuration (a) which is given as
\begin{align}
	S_R^{(n,m)}(A:B) &= \frac{1}{1-n}\log \frac{\left< \sigma_{g_A}(p_1)\sigma_{g_Bg_A^{-1}}(p_2)\sigma_{g_B^{-1}}(p_3)\sigma_{g_B}(p_4)\sigma_{g_Ag_B^{-1}}(p_5)\sigma_{g_A^{-1}}(p_6) \right>_{\mathrm{BCFT}^{\bigotimes mn}}}{\left<\sigma_{g_m}(p_1)\sigma_{g_m^{-1}}(p_3)\sigma_{g_m}(p_4)\sigma_{g_m^{-1}}(p_6) \right>^n_{\mathrm{BCFT}^{\bigotimes m}}}\notag\\
	&= 2\frac{1}{1-n}\log \frac{\Big< \sigma_{g_A}(p_1)\sigma_{g^{-1}_A}(q^b_1)\sigma_{g_Bg_A^{-1}}(p_2)\Big>_{\mathrm{CFT}^{\bigotimes mn}}}{\Big<\sigma_{g_m}(p_1)\sigma_{g^{-1}_m}(q^b_1)\Big>^n_{\mathrm{CFT}^{\bigotimes m}}}. \label{adjcase2b}
\end{align}
Finally considering the replica limit of the above equation the reflected entropy $S_R(A:B)$ is given by the expression described in \cref{SRcasea}.

\begin{figure}[H]
	\centering
	\includegraphics[scale=.9]{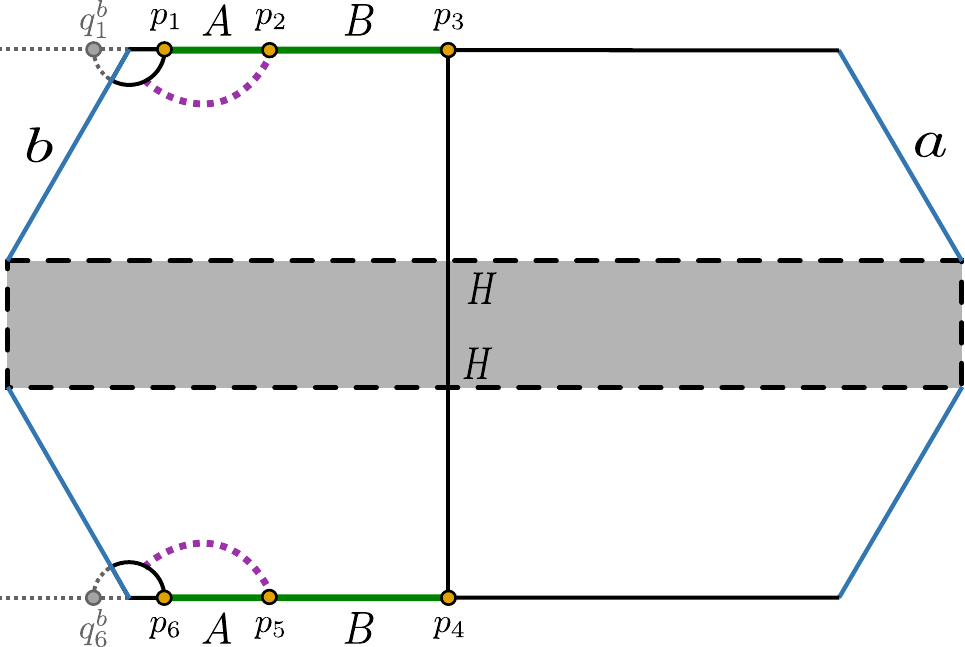}
	\caption{Here $q^b_{1}=r_I-\epsilon$ and $q^b_{6}=r_I-\epsilon$ are the mirror images of the points $p_1=r_I+\epsilon$ and $p_6=r_I+\epsilon$ respectively with respect to the $b$-boundary. Also the other points in the two $BCFT_2$s are identified as $p_2=r$, $p_3=r_1$, $p_5=r$ and $p_4=r_1$.}\label{fig_adjcase1g}
\end{figure}
As depicted in \cref{fig_adjcase1g} the EWCS for the entanglement wedge of $A\cup B$ is identical to that described in the configuration (a). Consequently, the computation for the same follows a similar procedure and the EWCS for this may be obtained from \cref{EWcase1a} as,
\begin{align}
	E_W(A:B)=\frac{c}{3}  \log \left(4 z  \right),
\end{align}
where $z$ is the cross ratio at a finite temperature $\beta$ given by the \cref{CRadjj1c}.

\subsubsection*{Configuration (h):} 
This configuration involves an analogous computation for the reflected entropy as in configuration (b) with a slightly different factorization of the correlator in the expression of the corresponding $S_R^{(n,m)}(A:B)$. We refer to the \cref{adjcaseh} of appendix \ref{app_adj} for the corresponding factorization employed here. However, the final expression for the R\'enyi reflected entropy is identical to that of configuration (b) and is described as
\begin{align}
	S_R^{(n,m)}(A:B) &= \frac{1}{1-n}\log \frac{\left< \sigma_{g_A}(p_1)\sigma_{g_Bg_A^{-1}}(p_2)\sigma_{g_B^{-1}}(p_3)\sigma_{g_B}(p_4)\sigma_{g_Ag_B^{-1}}(p_5)\sigma_{g_A^{-1}}(p_6) \right>_{\mathrm{BCFT}^{\bigotimes mn}}}{\left<\sigma_{g_m}(p_1)\sigma_{g_m^{-1}}(p_3)\sigma_{g_m}(p_4)\sigma_{g_m^{-1}}(p_6) \right>^n_{\mathrm{BCFT}^{\bigotimes m}}}\notag\\
	&= 2 \frac{1}{1-n}\log\left<\sigma_{g_Bg_A^{-1}}(p_2)\sigma_{g_Ag_B^{-1}}(q^b_2)\right>_{\mathrm{CFT}^{\bigotimes mn}}. \label{adjcase1h}
\end{align}
The reflected entropy may now be obtained from the replica limit of the above equation which is identical to that in
\cref{SRcaseb}.

\begin{figure}[H]
	\centering
	\includegraphics[scale=.9]{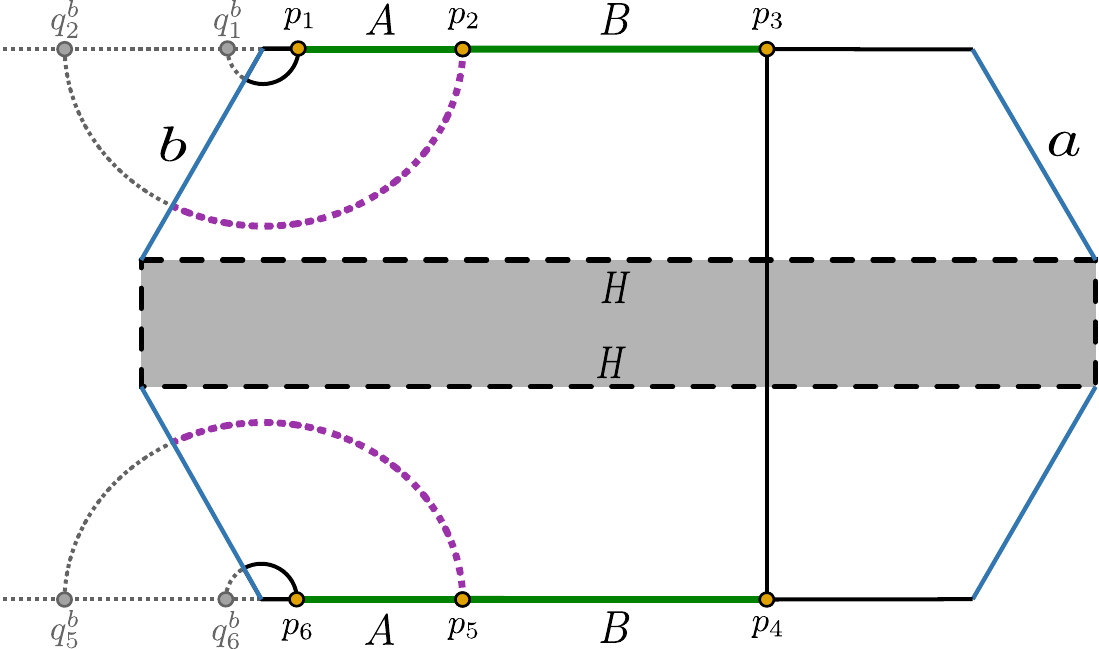}
	\caption{Here $q^b_{2}=2r_I-r_1$ and $q^b_{5}=2r_I-r_1$ are the mirror images of the points $p_2=r$ and $p_5=r$ respectively in the two $BCFT_2$ copies with respect to the $b$-boundary. The other points are identified similarly as in \cref{fig_adjcase1g}.}\label{fig_adjcase1h}
\end{figure}

Similar to the reflected entropy, the EWCS for this case is identical to the one discussed in configuration (b) as seen from the \cref{fig_adjcase1b,fig_adjcase1h}. Consequently the expression for the EWCS for this configuration may be given by the \cref{EWcaseb}.

\subsubsection*{Configuration (i):}
We now refer to a scenario discussed earlier in configuration (c) for the corresponding reflected entropy in this case. The present configuration involves a six-point correlation function in the reflected entropy expression which in the large central charge limit factorizes to the same two-point function as in \cref{adjcase1c}. This factorization of the six-point twist correlator is shown in \cref{adjcasei} of appendix \ref{app_adj}. The R\'enyi reflected entropy for the two adjacent intervals in this case may then be written as
\begin{align}
	S_R^{(n,m)}(A:B) &= \frac{1}{1-n}\log \frac{\left< \sigma_{g_A}(p_1)\sigma_{g_Bg_A^{-1}}(p_2)\sigma_{g_B^{-1}}(p_3)\sigma_{g_B}(p_4)\sigma_{g_Ag_B^{-1}}(p_5)\sigma_{g_A^{-1}}(p_6) \right>_{\mathrm{BCFT}^{\bigotimes mn}}}{\left<\sigma_{g_m}(p_1)\sigma_{g_m^{-1}}(p_3)\sigma_{g_m}(p_4)\sigma_{g_m^{-1}}(p_6) \right>^n_{\mathrm{BCFT}^{\bigotimes m}}}\notag\\
	& =\frac{1}{1-n}\log\left( \left<\sigma_{g_Bg_A^{-1}}(p_2)\sigma_{g_Ag_B^{-1}}(p_5)\right>_{\mathrm{CFT}^{\bigotimes mn}}\right). \label{adjcase2d}
\end{align}
Consequently, the reflected entropy $S_R(A:B)$ may be given by the \cref{SRcasec} after considering the replica limit of \cref{adjcase2d}.

\begin{figure}[H]
	\centering
	\includegraphics[scale=.9]{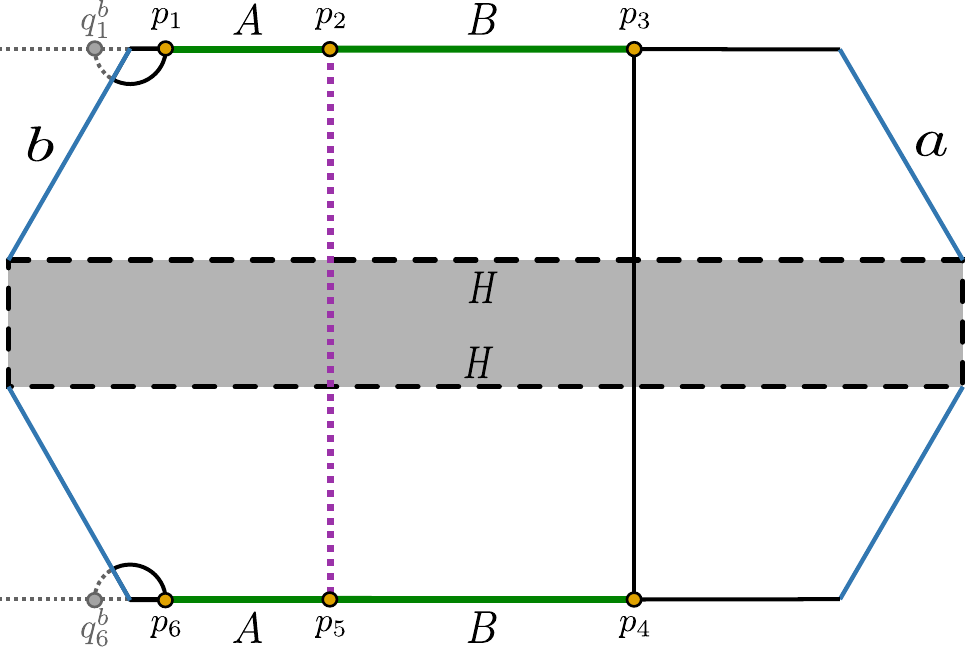}
	\caption{The points in the two copies of $BCFT_2$s in a TFD state are identified as $q^b_{1}=r_I-\epsilon$ and $q^b_{6}=r_I-\epsilon$ which are the mirror images of the points $p_1=r_I+\epsilon$ and $p_6=r_I+\epsilon$ respectively with respect to the $b$-boundary. The other points are identified similarly as in \cref{fig_adjcase1g}. }\label{fig_adjcase1i}
\end{figure}
We may note from the \cref{fig_adjcase1c,fig_adjcase1i}, that the EWCS for the entanglement wedge of $A\cup B$ in this scenario is identical to that discussed in configuration (c) where the EWCS was described by a HM surface. Hence the expression for the EWCS for this configuration can be given by the \cref{HM}.

\subsubsection*{Configuration (j):}
We proceed with a non-trivial contribution to the reflected entropy which arises from a four-point twist correlator after employing a factorization procedure in the large central charge limit as shown in \cref{adjcasej} of appendix \ref{app_adj}. The corresponding R\'enyi reflected entropy for the bipartite mixed state under consideration may be expressed as,
\begin{align}
	S_R^{(n,m)}(A:B) &= \frac{1}{1-n}\log \frac{\left< \sigma_{g_A}(p_1)\sigma_{g_Bg_A^{-1}}(p_2)\sigma_{g_B^{-1}}(p_3)\sigma_{g_B}(p_4)\sigma_{g_Ag_B^{-1}}(p_5)\sigma_{g_A^{-1}}(p_6) \right>_{\mathrm{BCFT}^{\bigotimes mn}}}{\left<\sigma_{g_m}(p_1)\sigma_{g_m^{-1}}(p_3)\sigma_{g_m}(p_4)\sigma_{g_m^{-1}}(p_6) \right>^n_{\mathrm{BCFT}^{\bigotimes m}}}\notag\\
	&= \frac{1}{1-n}\log \frac{\Big<\sigma_{g_Bg_A^{-1}}(p_2)\sigma_{g^{-1}_B}(p_3)\sigma_{g_B}(p_4)\sigma_{g_Ag_B^{-1}}(p_5) \Big>_{\mathrm{CFT}^{\bigotimes mn}}}{\left<\sigma_{g^{-1}_m}(p_3)\sigma_{g_m}(p_4) \right>^n_{\mathrm{CFT}^{\bigotimes m}}}. \label{adjcase2e}
\end{align}
For the four-point twist correlator in the numerator of the last line, we now use the doubling trick inversely such that it can be written as a two-point correlator in a $BCFT_2$. In an OPE channel, this $BCFT_2$ correlator is further equivalent to a three-point function on a flat $CFT_2$ \cite{Shao:2022gpg,Li:2021dmf}. Consequently, the above expression may be re-written as
\begin{align}
	S_R^{(n,m)}(A:B) &=\frac{1}{1-n}\log \frac{\Big<\sigma_{g_Bg_A^{-1}}(p_2)\sigma_{g^{-1}_B}(p_3)\Big>_{\mathrm{BCFT}^{\bigotimes mn}}}{\left<\sigma_{g^{-1}_m}(p_3)\sigma_{g_m}(p_4) \right>^n_{\mathrm{CFT}^{\bigotimes m}}}~~~ (\text{inverse doubling trick)}\notag\\
	&= \frac{1}{1-n}\log \frac{\Big<\sigma_{g_Bg_A^{-1}}(p_2)\sigma_{g^{-1}_B}(p_3)\sigma_{g_B}(q_3)\Big>_{\mathrm{CFT}^{\bigotimes mn}}}{\left<\sigma_{g^{-1}_m}(p_3)\sigma_{g_m}(p_4) \right>^n_{\mathrm{CFT}^{\bigotimes m}}}~~~(\text{OPE channel}).
\end{align}
We compute this OPE channel result following the methods discussed in \cite{Shao:2022gpg,Li:2021dmf} which after considering the replica limit reduces to
\begin{align}\label{SrCasej}
	S_R(A:B)= \frac{2c}{3} \log \frac{(r_1-r_2)~ \text{sech}\left(\frac{2 \pi  t}{\beta }\right) \sqrt{r_1^2+2 r_1 r_2 \cosh \left(\frac{4 \pi  t}{\beta }\right)+r_2^2}}{r_1 \epsilon }.
\end{align}

\begin{figure}[H]
	\centering
	\includegraphics[scale=.9]{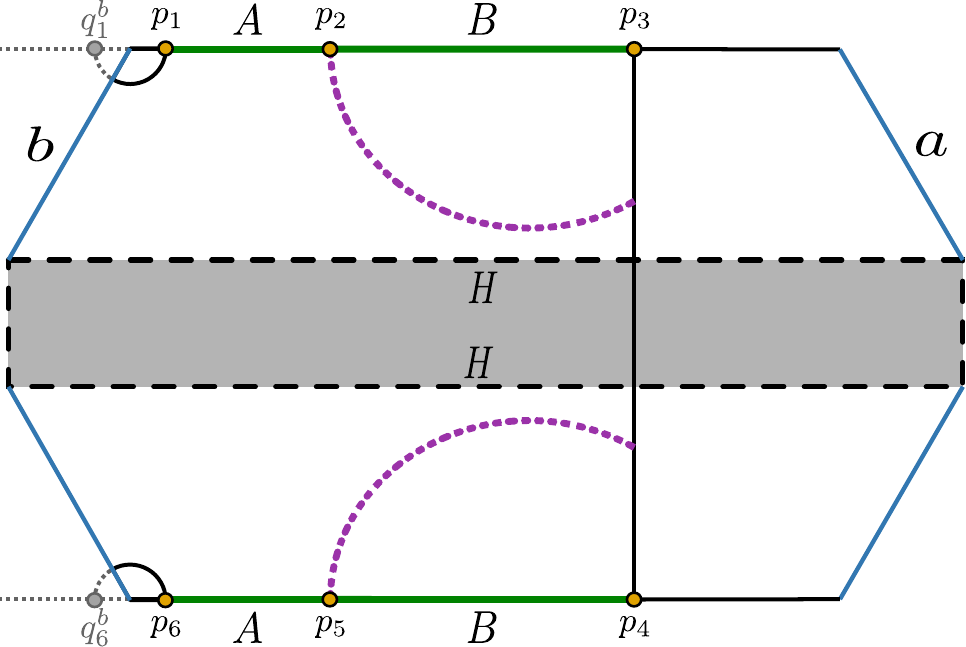}
	\caption{The points in the two copies of $BCFT_2$s in a TFD state are identified as $p_2=r_1$, $p_5=r_1$, $p_3=r_2$ and $p_4=r_2$.}\label{fig_adjcase1j}
\end{figure}
As depicted in \cref{fig_adjcase1j}, the EWCS for this configuration is described by a co-dimension two surface which starts from the common point of the two adjacent intervals and lands on the HM surface. We may now follow the procedure discussed in \cite{Shao:2022gpg} to compute this EWCS from a purely geometric perspective with a proper coordinate transformation given in eq. (7.34) of \cite{Shao:2022gpg} where we have considered $e^X=r$ in our work. The corresponding EWCS may then be expressed as
\begin{align}\label{EwCasej}
	E_W(A:B)= \frac{c}{3} \log \frac{(r_2-r_1)~ \text{sech}\left(\frac{2 \pi  t}{\beta }\right) \sqrt{r_1^2+2 r_1 r_2 \cosh \left(\frac{4 \pi  t}{\beta }\right)+r_2^2}}{r_2 \epsilon }.
\end{align}

\subsubsection*{Configuration (k):}
We now consider another possible contribution to the reflected entropy for the adjacent intervals $A$ and $B$ which produces an expression for $S_R(A:B)$ similar to the one obtained in the previous configuration. Note that after implementing the factorization to the six-point twist correlator in large central charge limit, the corresponding R\'enyi reflected entropy may be written as,
\begin{align}
	S_R^{(n,m)}(A:B) &= \frac{1}{1-n}\log \frac{\left< \sigma_{g_A}(p_1)\sigma_{g_Bg_A^{-1}}(p_2)\sigma_{g_B^{-1}}(p_3)\sigma_{g_B}(p_4)\sigma_{g_Ag_B^{-1}}(p_5)\sigma_{g_A^{-1}}(p_6) \right>_{\mathrm{BCFT}^{\bigotimes mn}}}{\left<\sigma_{g_m}(p_1)\sigma_{g_m^{-1}}(p_3)\sigma_{g_m}(p_4)\sigma_{g_m^{-1}}(p_6) \right>^n_{\mathrm{BCFT}^{\bigotimes m}}}\notag\\
	&= \frac{1}{1-n}\log \frac{\Big<\sigma_{g_Bg_A^{-1}}(p_2)\sigma_{g_A}(p_1)\sigma_{g^{-1}_A}(p_6)\sigma_{g_Ag_B^{-1}}(p_5) \Big>_{\mathrm{CFT}^{\bigotimes mn}}}{\left<\sigma_{g_m}(p_1)\sigma_{g^{-1}_m}(p_6) \right>^n_{\mathrm{CFT}^{\bigotimes m}}}
\end{align}	
The factorization procedure employed in the above expression is shown in \cref{adjcasek} of appendix \ref{app_adj}. Similar to the previous configuration, we now utilize the same techniques in this scenario to obtain the reflected entropy $S_R(A:B)$. Using the doubling trick inversely, the above expression may be re-expressed as
\begin{align}
	S_R^{(n,m)}(A:B) &= \frac{1}{1-n}\log \frac{\Big<\sigma_{g_Bg_A^{-1}}(p_2)\sigma_{g_A}(p_1)\Big>_{\mathrm{BCFT}^{\bigotimes mn}}}{\left<\sigma_{g_m}(p_1)\sigma_{g^{-1}_m}(p_6) \right>^n_{\mathrm{CFT}^{\bigotimes m}}}.
\end{align}	
However, in an OPE channel, this $BCFT_2$ two-point function is further equal to a three-point function on a flat $CFT_2$ as follows
\begin{align}
	S_R^{(n,m)}(A:B) &= \frac{1}{1-n}\log \frac{\Big<\sigma_{g_Bg_A^{-1}}(p_2)\sigma_{g_A}(p_1)\sigma_{g_A^{-1}}(q_1)\Big>_{\mathrm{CFT}^{\bigotimes mn}}}{\left<\sigma_{g_m}(p_1)\sigma_{g^{-1}_m}(p_6) \right>^n_{\mathrm{CFT}^{\bigotimes m}}}. \label{adjcase3a}
\end{align}
Finally considering the replica limit of the above expression, the reflected entropy for the adjacent intervals $A$ and $B$ is given by the \cref{SrCasej} with the point $r_2$ replaced by $r_I-\epsilon$ where $\epsilon$ is the UV cut off.

\begin{figure}[H]
	\centering
	\includegraphics[scale=.9]{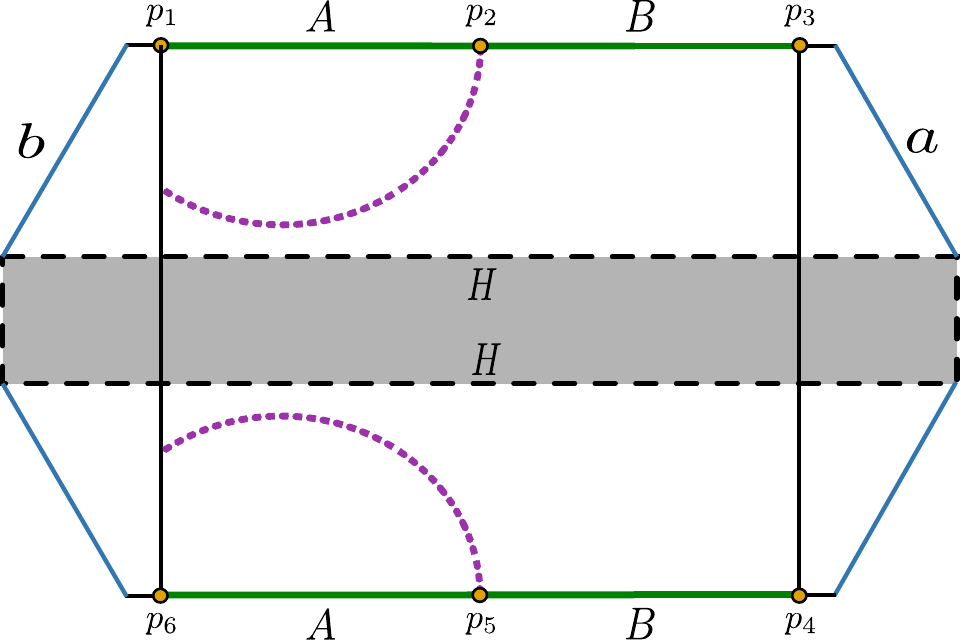}
	\caption{The points in the two $BCFT_2$ copies are identified as $p_1=r_I+\epsilon$, $p_6=r_I+\epsilon$ $p_2=r_1$, $p_5=r_1$, $p_3=r_O-\epsilon$ and $p_4=r_O-\epsilon$.}\label{fig_adjcase1k}
\end{figure}
Analogous to the reflected entropy, the EWCS for this configuration as depicted in \cref{fig_adjcase1k} is similar to the previous case. Consequently, the expression for the EWCS in this case is given by the \cref{EwCasej} with the point $r_2$ replaced by $r_I-\epsilon$.

\subsubsection*{Configuration (l):}
In this case, the contribution to the reflected entropy is given by the one obtained in configuration (c) since, the R\'enyi reflected entropy reduces to an identical expression after implementing a factorization of the six-point correlation function in the large central charge limit as follows
\begin{align}
	S_R^{(n,m)}(A:B) &= \frac{1}{1-n}\log \frac{\left< \sigma_{g_A}(p_1)\sigma_{g_Bg_A^{-1}}(p_2)\sigma_{g_B^{-1}}(p_3)\sigma_{g_B}(p_4)\sigma_{g_Ag_B^{-1}}(p_5)\sigma_{g_A^{-1}}(p_6) \right>_{\mathrm{BCFT}^{\bigotimes mn}}}{\left<\sigma_{g_m}(p_1)\sigma_{g_m^{-1}}(p_3)\sigma_{g_m}(p_4)\sigma_{g_m^{-1}}(p_6) \right>^n_{\mathrm{BCFT}^{\bigotimes m}}}\notag\\
	&= \frac{1}{1-n}\log\left( \left<\sigma_{g_Bg_A^{-1}}(p_2)\sigma_{g_Ag_B^{-1}}(p_5)\right>_{\mathrm{CFT}^{\bigotimes mn}}\right). \label{adjcase3b}
\end{align}
The corresponding factorization of the six-point twist correlator utilized in the above expression is shown in \cref{adjcasel} of appendix \ref{app_adj}. Consequently, the reflected entropy $S_R(A:B)$ for this configuration is given by the \cref{SRcasec}.

\begin{figure}[H]
	\centering
	\includegraphics[scale=.9]{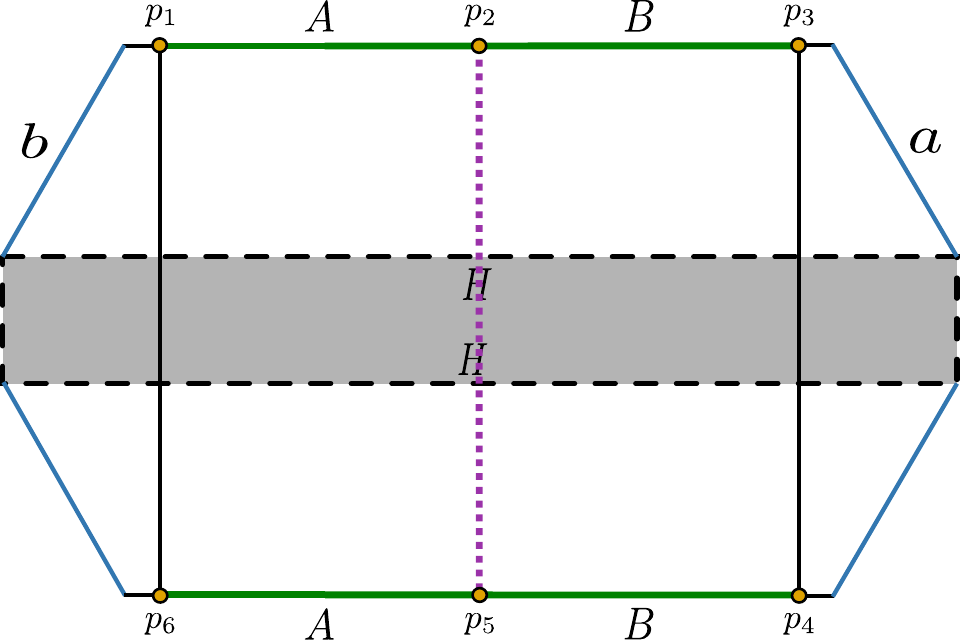}
	\caption{The points in the two $BCFT_2$ copies are identified as $p_2=r$ and $p_5=r$ and the other points are identified similarly as in the \cref{fig_adjcase1k}.}\label{fig_adjcase1l}
\end{figure}

Fig. \ref{fig_adjcase1l} depicts the corresponding EWCS for the entanglement wedge of the interval $A\cup B$ identical to that described in \cref{fig_adjcase1c} of configuration (c). Consequently, the expression for the EWCS in this scenario is given by the \cref{HM}.

\subsubsection*{Configuration (m):}
We refer to configuration (j) for the computation of reflected entropy in this scenario since the six-point twist correlator in the corresponding R\'enyi reflected entropy expression factorizes in the large central charge limit and produce an identical expression for the same. This factorization procedure is given in \cref{adjcasem} of appendix \ref{app_adj}. The R\'enyi reflected entropy may then be expressed as,
\begin{align}
	S_R^{(n,m)}(A:B) &= \frac{1}{1-n}\log \frac{\left< \sigma_{g_A}(p_1)\sigma_{g_Bg_A^{-1}}(p_2)\sigma_{g_B^{-1}}(p_3)\sigma_{g_B}(p_4)\sigma_{g_Ag_B^{-1}}(p_5)\sigma_{g_A^{-1}}(p_6) \right>_{\mathrm{BCFT}^{\bigotimes mn}}}{\left<\sigma_{g_m}(p_1)\sigma_{g_m^{-1}}(p_3)\sigma_{g_m}(p_4)\sigma_{g_m^{-1}}(p_6) \right>^n_{\mathrm{BCFT}^{\bigotimes m}}}\notag\\
	&= \frac{1}{1-n}\log \frac{\Big<\sigma_{g_Bg_A^{-1}}(p_2)\sigma_{g^{-1}_B}(p_3)\sigma_{g_B}(p_4)\sigma_{g_Ag_B^{-1}}(p_5) \Big>_{\mathrm{CFT}^{\bigotimes mn}}}{\left<\sigma_{g^{-1}_m}(p_3)\sigma_{g_m}(p_4) \right>^n_{\mathrm{CFT}^{\bigotimes m}}}.
\end{align}
As earlier, we first perform the inverse doubling trick as discussed in configuration (j) which transforms the above expression into 
\begin{align}
	S_R^{(n,m)}(A:B) &= \frac{1}{1-n}\log \frac{\Big<\sigma_{g_Bg_A^{-1}}(p_2)\sigma_{g^{-1}_B}(p_3)\Big>_{\mathrm{BCFT}^{\bigotimes mn}}}{\left<\sigma_{g^{-1}_m}(p_3)\sigma_{g_m}(p_4) \right>^n_{\mathrm{CFT}^{\bigotimes m}}}.
\end{align}
In an OPE channel, the $BCFT_2$ two-point function in the above expression is further equal to a three-point function on a flat $CFT_2$ as discussed in \cite{Shao:2022gpg,Li:2021dmf}. Hence, we obtain
\begin{align}
	S_R^{(n,m)}(A:B) &= \frac{1}{1-n}\log \frac{\Big<\sigma_{g_Bg_A^{-1}}(p_2)\sigma_{g^{-1}_B}(p_3)\sigma_{g_B}(q_3)\Big>_{\mathrm{CFT}^{\bigotimes mn}}}{\left<\sigma_{g^{-1}_m}(p_3)\sigma_{g_m}(p_4) \right>^n_{\mathrm{CFT}^{\bigotimes m}}}. \label{adjcase3c}
\end{align}
\noindent
The reflected entropy may then be obtained after considering the replica limit which produces an identical expression given in \cref{SrCasej}.

\begin{figure}[H]
	\centering
	\includegraphics[scale=.9]{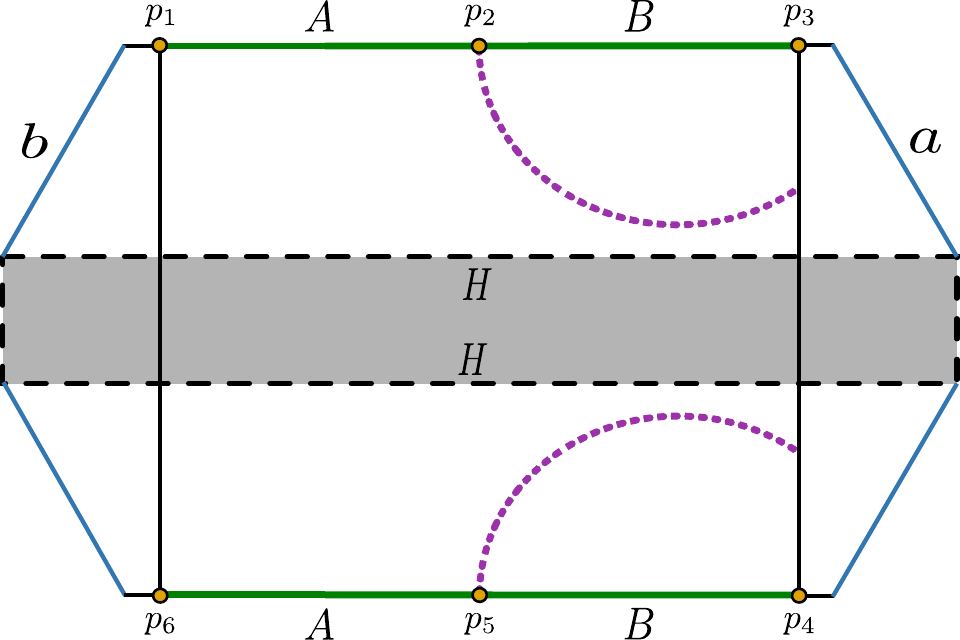}
	\caption{The points in the two $BCFT_2$ copies are identified similarly as in the \cref{fig_adjcase1j}.}\label{fig_adjcase1m}
\end{figure}
The EWCS for this configuration is shown by the dotted violet curve in \cref{fig_adjcase1m} which indicates an identical structure observed in \cref{fig_adjcase1j} of configuration (j). Therefore the corresponding expression for the EWCS in this scenario is given by the \cref{EwCasej}.

\subsubsection*{Configuration (n):}
Once again we refer to configuration (j) for the reflected entropy computation in this scenario. The six-point twist correlator involved in the R\'enyi reflected entropy factorizes in the large central charge limit as shown in \cref{adjcasen} of appendix \ref{app_adj} and produces an expression analogous to the one obtained in configuration (j) as follows
\begin{align}
	S_R^{(n,m)}(A:B) &= \frac{1}{1-n}\log \frac{\left< \sigma_{g_A}(p_1)\sigma_{g_Bg_A^{-1}}(p_2)\sigma_{g_B^{-1}}(p_3)\sigma_{g_B}(p_4)\sigma_{g_Ag_B^{-1}}(p_5)\sigma_{g_A^{-1}}(p_6) \right>_{\mathrm{BCFT}^{\bigotimes mn}}}{\left<\sigma_{g_m}(p_1)\sigma_{g_m^{-1}}(p_3)\sigma_{g_m}(p_4)\sigma_{g_m^{-1}}(p_6) \right>^n_{\mathrm{BCFT}^{\bigotimes m}}}\notag\\
	&= \frac{1}{1-n}\log \frac{\Big<\sigma_{g_Bg_A^{-1}}(p_2)\sigma_{g_A}(p_1)\sigma_{g^{-1}_A}(p_6)\sigma_{g_Ag_B^{-1}}(p_5) \Big>_{\mathrm{CFT}^{\bigotimes mn}}}{\left<\sigma_{g_m}(p_1)\sigma_{g^{-1}_m}(p_6) \right>^n_{\mathrm{CFT}^{\bigotimes m}}}.
\end{align}
Once again we utilize the same tricks discussed in configuration (j) to obtain the corresponding reflected entropy. Utilizing the inverse doubling trick, the above expression involving a four-point function in a $CFT_2$ may be re-expressed as
\begin{align}
	S_R^{(n,m)}(A:B) &= \frac{1}{1-n}\log \frac{\Big<\sigma_{g_Bg_A^{-1}}(p_2)\sigma_{g_A}(p_1)\Big>^{Flat}_{\mathrm{BCFT}^{\bigotimes mn}}}{\left<\sigma_{g_m}(p_1)\sigma_{g^{-1}_m}(p_6) \right>^n_{\mathrm{CFT}^{\bigotimes m}}},
\end{align}
which in an OPE channel further reduces to
\begin{align}\label{SRnmadjcasen}
	S_R^{(n,m)}(A:B) &= \frac{1}{1-n}\log \frac{\Big<\sigma_{g_Bg_A^{-1}}(p_2)\sigma_{g_A}(p_1)\sigma_{g_A^{-1}}(q_1)\Big>_{\mathrm{CFT}^{\bigotimes mn}}}{\left<\sigma_{g_m}(p_1)\sigma_{g^{-1}_m}(p_6) \right>^n_{\mathrm{CFT}^{\bigotimes m}}}.
\end{align}
Consequently, the reflected entropy $S_R(A:B)$ after considering the replica limit of \cref{SRnmadjcasen} is given by the \cref{SrCasej} with the point $r_2$ replaced by $r_I-\epsilon$ where $\epsilon$ is the UV cut off.

\begin{figure}[H]
	\centering
	\includegraphics[scale=.9]{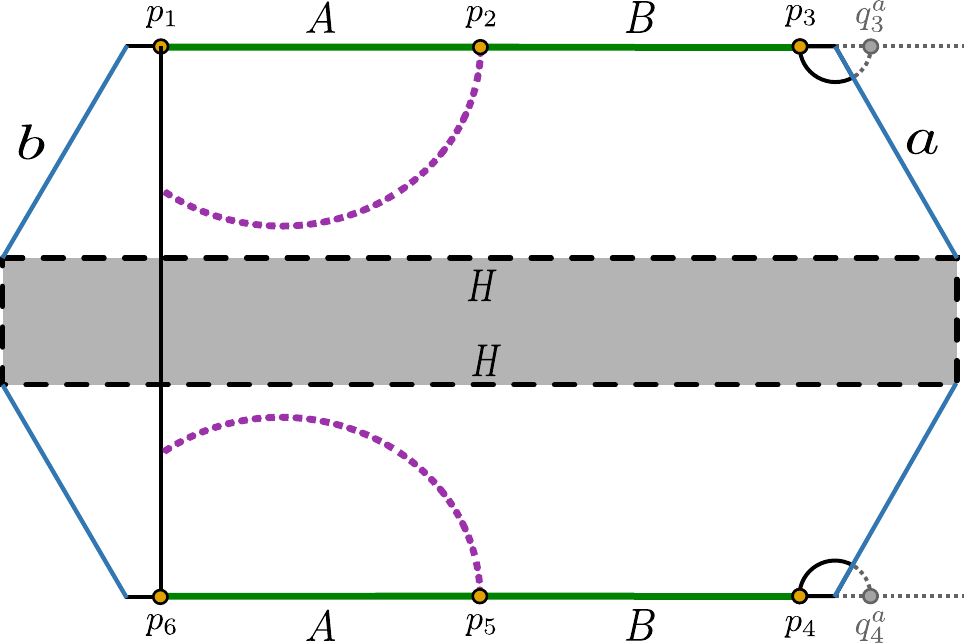}
	\caption{The points in the two $BCFT_2$ copies are identified similarly as in \cref{fig_adjcase1j} }\label{fig_adjcase1n}
\end{figure}
Similarly to the reflected entropy, the EWCS for this configuration (\cref{fig_adjcase1n}) is identical to the one described in \cref{fig_adjcase1j}. Therefore the corresponding expression for the EWCS in this case is given by the \cref{EwCasej} with the point $r_2$ replaced by $r_I-\epsilon$.

\subsubsection*{Configuration (o):}
This configuration corresponds to a dominant channel of the six-point twist correlator in the R\'enyi reflected entropy $S_R^{(n,m)}(A:B)$ which reduces to an expression identical to the one obtained in configuration (e). The R\'enyi reflected entropy for this case may be given as 
\begin{align}
	S_R^{(n,m)}(A:B) &= \frac{1}{1-n}\log \frac{\left< \sigma_{g_A}(p_1)\sigma_{g_Bg_A^{-1}}(p_2)\sigma_{g_B^{-1}}(p_3)\sigma_{g_B}(p_4)\sigma_{g_Ag_B^{-1}}(p_5)\sigma_{g_A^{-1}}(p_6) \right>_{\mathrm{BCFT}^{\bigotimes mn}}}{\left<\sigma_{g_m}(p_1)\sigma_{g_m^{-1}}(p_3)\sigma_{g_m}(p_4)\sigma_{g_m^{-1}}(p_6) \right>^n_{\mathrm{BCFT}^{\bigotimes m}}}\notag\\
	&= 2 \frac{1}{1-n}\log \frac{\Big< \sigma_{g_Bg_A^{-1}}(p_2)\sigma_{g_B^{-1}}(p_3)\sigma_{g_B}(q^a_3)\Big>_{\mathrm{CFT}^{\bigotimes mn}}}{\Big<\sigma_{g_m^{-1}}(p_3)\sigma_{g_m}(q^a_3)\Big>^n_{\mathrm{CFT}^{\bigotimes m}}}. \label{adjcase3e}
\end{align}
In the above expression we have employed a factorization of the six-point twist correlator in the large central charge limit as shown in \cref{adjcaseo} of appendix \ref{app_adj}. From the replica limit of \cref{adjcase3e}, the reflected entropy for this case may then be expressed as in \cref{SRcasee}. 

\begin{figure}[H]
	\centering
	\includegraphics[scale=.9]{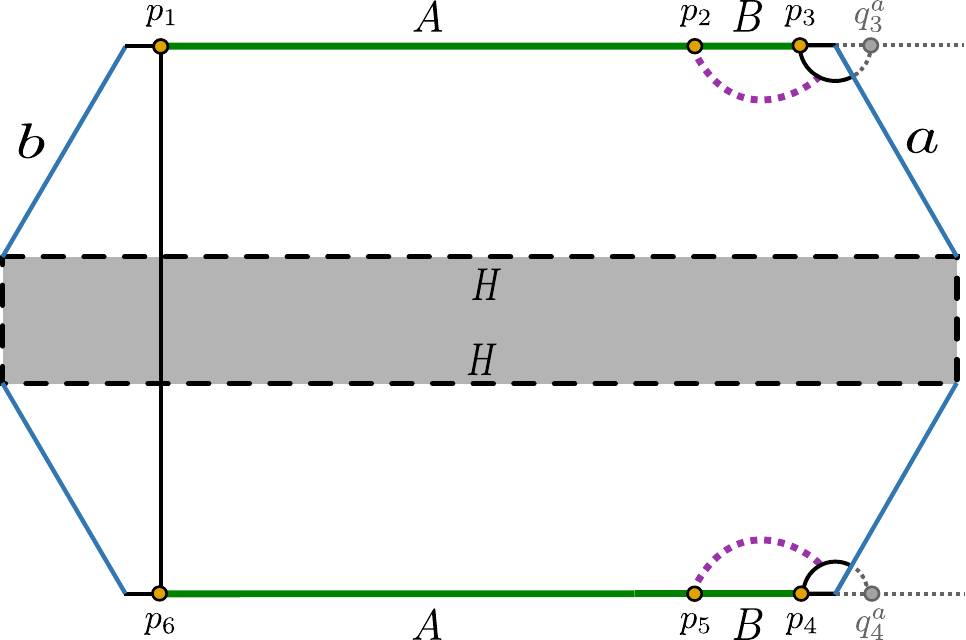}
	\caption{The points in the two $BCFT_2$ copies are identified similarly as in \cref{fig_adjcase1e}.}\label{fig_adjcase1o}
\end{figure}
The violet dotted curve as shown in \cref{fig_adjcase1o} corresponds to the EWCS for this present scenario which is identical to the one observed in \cref{fig_adjcase1e} of configuration (e). Consequently, the EWCS for this configuration is given by the \cref{EwCasee} with the cross-ratio in \cref{CrCasee}.

\subsubsection*{Configuration (p):}
In this scenario, the reflected entropy for the adjacent intervals $A$ and $B$ corresponds to an expressions for $S_R(A:B)$ which is identical to the one obtained in configuration (d). The six-point correlation function in the corresponding R\'enyi reflected entropy expression factorizes in the large central charge limit as shown in \cref{adjcasep} of appendix \ref{app_adj} and produces
\begin{align}
	S_R^{(n,m)}(A:B) &= \frac{1}{1-n}\log \frac{\left< \sigma_{g_A}(p_1)\sigma_{g_Bg_A^{-1}}(p_2)\sigma_{g_B^{-1}}(p_3)\sigma_{g_B}(p_4)\sigma_{g_Ag_B^{-1}}(p_5)\sigma_{g_A^{-1}}(p_6) \right>_{\mathrm{BCFT}^{\bigotimes mn}}}{\left<\sigma_{g_m}(p_1)\sigma_{g_m^{-1}}(p_3)\sigma_{g_m}(p_4)\sigma_{g_m^{-1}}(p_6) \right>^n_{\mathrm{BCFT}^{\bigotimes m}}}\notag\\
	&= 2\frac{1}{1-n}\log \left(\Big<\sigma_{g_Bg_A^{-1}}(p_2)\sigma_{g_Ag_B^{-1}}(q^a_2)\Big>_{\mathrm{CFT}^{\bigotimes mn}}\right). \label{adjcase3f}
\end{align}
Consequently the reflected entropy $S_R(A:B)$, from a replica limit of \cref{adjcase3f}, may be given by the \cref{SrCased}.

\begin{figure}[H]
	\centering
	\includegraphics[scale=.9]{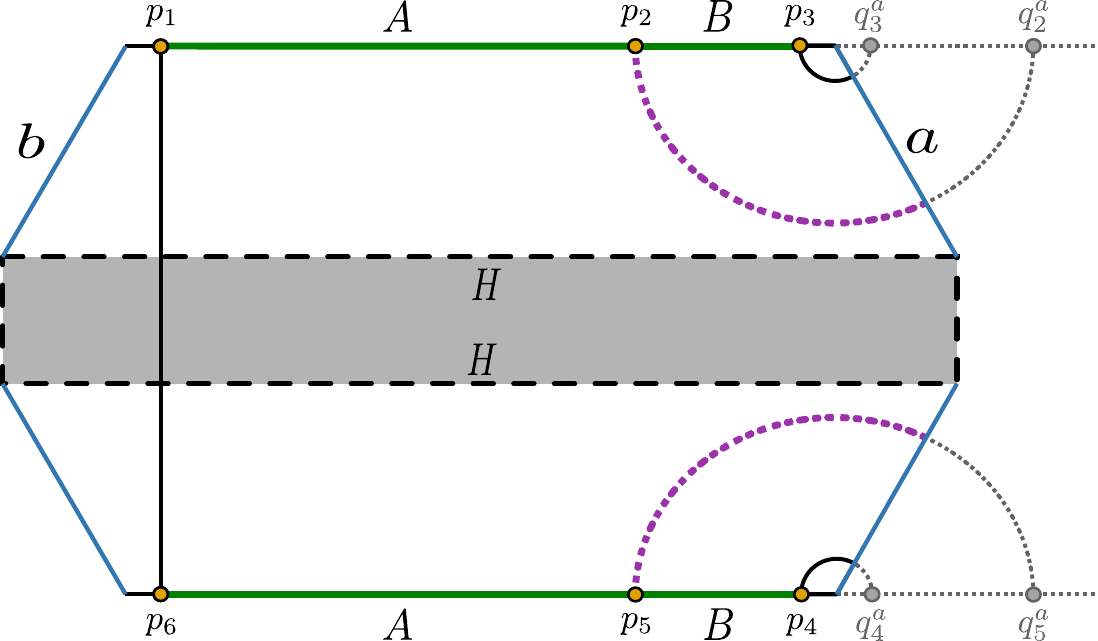}
	\caption{The points in the two $BCFT_2$ copies are identified similarly as in \cref{fig_adjcase1d}.}\label{fig_adjcase1p}
\end{figure}
The corresponding EWCS computation for this scenario follows an identical procedure as discussed in configuration (e). The expression for the same for this configuration may then be given by the \cref{EwCased}.

\subsubsection*{Configuration (q):}
Finally, the last contribution to the reflected entropy for the adjacent intervals $A$ and $B$ is similar to the $S_R(A:B)$ expression computed in configuration (c). Once again, we refer to \cref{adjcaseq} of appendix \ref{app_adj} where the factorization of the six-point twist correlator in the R\'enyi reflected entropy expression is shown in the large central charge limit. We then obtain a simplified expression for the R\'enyi reflected entropy as
\begin{align}
	S_R^{(n,m)}(A:B) &= \frac{1}{1-n}\log \frac{\left< \sigma_{g_A}(p_1)\sigma_{g_Bg_A^{-1}}(p_2)\sigma_{g_B^{-1}}(p_3)\sigma_{g_B}(p_4)\sigma_{g_Ag_B^{-1}}(p_5)\sigma_{g_A^{-1}}(p_6) \right>_{\mathrm{BCFT}^{\bigotimes mn}}}{\left<\sigma_{g_m}(p_1)\sigma_{g_m^{-1}}(p_3)\sigma_{g_m}(p_4)\sigma_{g_m^{-1}}(p_6) \right>^n_{\mathrm{BCFT}^{\bigotimes m}}}\notag\\
	&= \frac{1}{1-n}\log \left(\Big<\sigma_{g_Bg_A^{-1}}(p_2)\sigma_{g_Ag_B^{-1}}(p_5)\Big>_{\mathrm{CFT}^{\bigotimes mn}}\right). \label{adjcase1q}
\end{align}
Now implementing the replica limit to the above equation, the corresponding reflected entropy for the intervals $A$ and $B$ may be computed as in \cref{SRcasec}.
\begin{figure}[H]
	\centering
	\includegraphics[scale=.9]{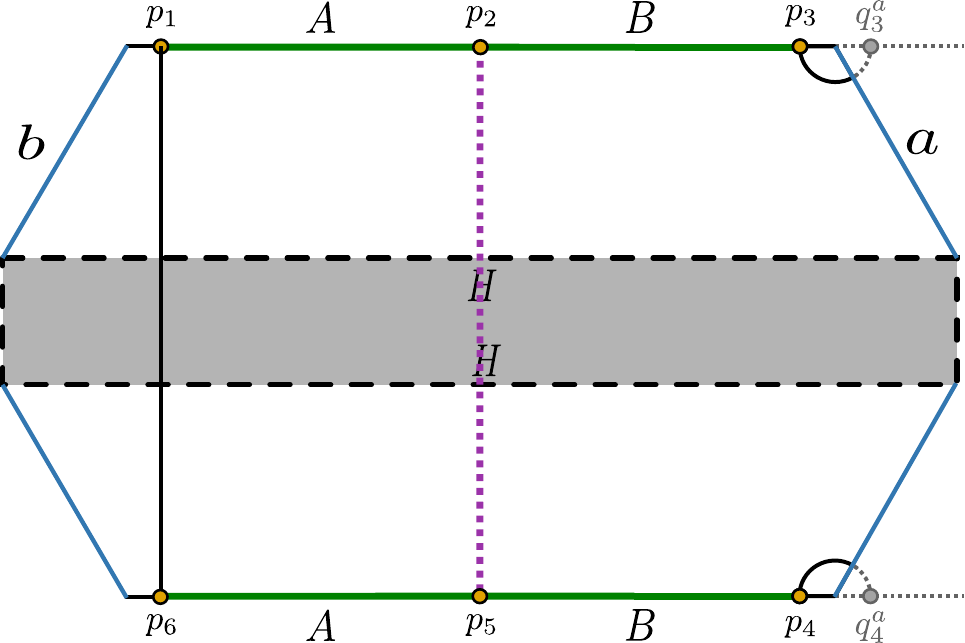}
	\caption{The points in the two $BCFT_2$ copies are identified similarly as in \cref{fig_adjcase1c}.}\label{fig_adjcase1q}
\end{figure}

Similar to the reflected entropy, the EWCS for this case is identical to configuration (c) and consequently the corresponding expression for the EWCS may be given by \cref{EwCased}.

\subsection{Disjoint intervals}
We now proceed to compute the reflected entropy for two disjoint intervals $A\equiv[p_1,p_2]\cup[p_8,p_7]$ and $B\equiv[p_3,p_4]\cup[p_6,p_5]$ in the bath $BCFT_2$s. Following this, we obtain the corresponding EWCSs for the dual bulk eternal BTZ black hole geometry to demonstrate the holographic duality between the reflected entropy and the
EWCS as described in \cref{Sr_Ew}. Once again the different contributions to the reflected entropy and the EWCS outlined below, occur 
due to the various configurations involving the sizes of the intervals similar to the case of adjacent intervals. 

\subsubsection*{Configuration (a):}
\begin{figure}[H]
	\centering
	\includegraphics[scale=.9]{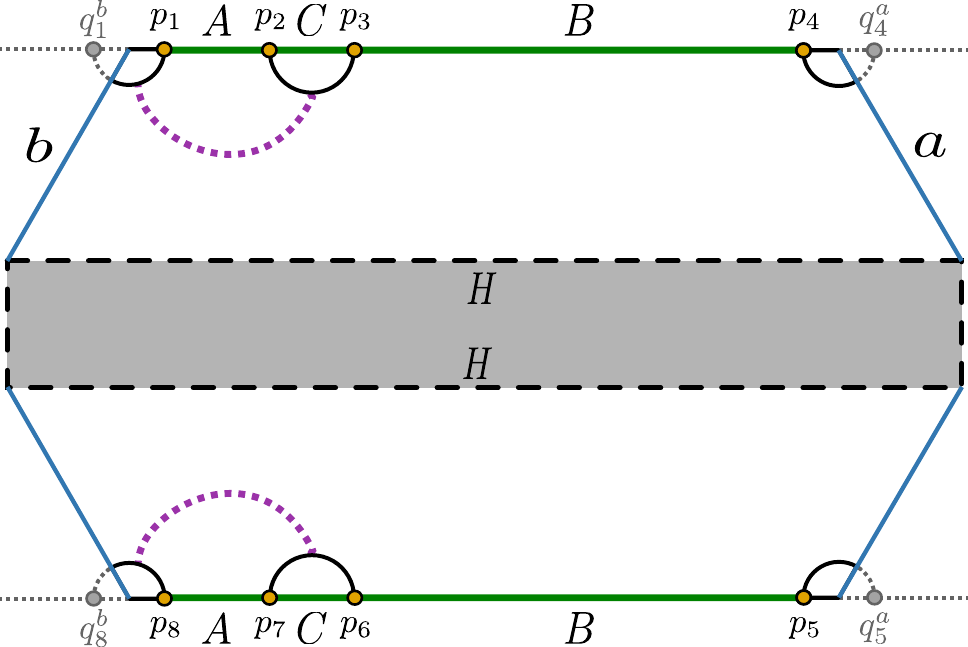}
	\caption{The endpoints of the intervals $A$ and $B$ in the two $BCFT_2$ copies are identified as $p_1=p_8=r_I+\epsilon$, $p_2=p_7=r_2$, $p_3=p_6=r_3$ and $p_4=p_5=r_O-\epsilon$. The other points $q_1^b=r_I-\epsilon$, $q_8^b=r_I-\epsilon$ are the image points of $p_1$ and $p_8$ respectively with respect to the $b$-boundary at $p_I=r_I$ and $q_4^a=r_O+\epsilon$, $q_5^b=r_O+\epsilon$ are the image points of $p_4$ and $p_5$ respectively with respect to the $a$-boundary at $p_O=r_O$. }\label{fig_discase1a}
\end{figure}

We begin with the first configuration where the R\'enyi reflected entropy involves four-point twist correlators in a $BCFT_2$. In a dominant channel, this $BCFT_2$ four-point twist correlator factorizes into three-point and one-point functions in the large central charge limit which upon utilizing the doubling trick results in $CFT_2$ four-point and two-point functions respectively . This intermediate factorization procedure is described in \cref{discasea} of appendix \ref{app_disj}. The corresponding R\'enyi reflected entropy is given as 
\begin{align}
	S_R^{(n,m)}(A:B) &= 2\frac{1}{1-n}\log \frac{\left< \sigma_{g_A}(p_1)\sigma_{g_A^{-1}}(p_2)\sigma_{g_B}(p_3)\sigma_{g_B^{-1}}(p_4) \right>_{\mathrm{BCFT}^{\bigotimes mn}}}{\left<\sigma_{g_m}(p_1)\sigma_{g_m^{-1}}(p_2)\sigma_{g_m}(p_3)\sigma_{g_m^{-1}}(p_4) \right>^n_{\mathrm{BCFT}^{\bigotimes m}}}\notag\\
	&= 2\frac{1}{1-n}\log \frac{\left< \sigma_{g_A^{-1}}(q^b_1)\sigma_{g_A}(p_1)\sigma_{g_A^{-1}}(p_2)\sigma_{g_B}(p_3)\right>_{\mathrm{CFT}^{\bigotimes mn}}}{\left<\sigma_{g_m^{-1}}(q^b_1)\sigma_{g_m}(p_1)\sigma_{g_m^{-1}}(p_2)\sigma_{g_m}(p_3)\right>^n_{\mathrm{CFT}^{\bigotimes m}}},\label{discase1a}
\end{align}
where the factor two arises due to the symmetry of the configuration as depicted in \cref{fig_discase1a}. Implementing the replica limit and following the procedure discussed in \cite{Jeong:2019xdr}, the reflected entropy for the disjoint intervals $A$ and $B$ in this case may be obtained as
\begin{equation}\label{SRdiscasea}
	S_R(A:B)=\frac{2c}{3}  \log \left(  \frac{1+\sqrt{\tilde x}}{1-\sqrt{\tilde x}}\right) ,
\end{equation}
with $\tilde x$ being the cross-ratio at a finite temperature given by
\begin{equation}\label{CRdiscase1a}
	\tilde x=\frac{\sinh \left(\frac{\pi  (p_1-p_2)}{\beta }\right) \sinh \left(\frac{\pi  (p_3-q_1^b)}{\beta }\right)}{\sinh \left(\frac{\pi  (p_1-p_3)}{\beta }\right) \sinh \left(\frac{\pi  (p_2-q_1^b) }{\beta }\right)}.
\end{equation}

The corresponding EWCS for this configuration is described by a codimension two surface as depicted by the violet dotted curve in \cref{fig_discase1a}. We follow the procedure described in \cref{EWCS} to obtain the EWCS in this scenario which is given by twice of the contribution in \cref{EW} with the finite temperature cross-ratio in \cref{CrRatio}. However, in this configuration we have identified the points as $a_1=p_1$, $a_2=p_2$, $b_1=p_3$ and $b_2=q_1^b$ with $q_1^b$ being the image point of $p_1$ with respect to the $b$-boundary in the $BCFT_2$. Although, \cref{SRdiscasea,EW} appears to be different with the corresponding cross-ratios being distinct, we can always transform one expression to another with identical cross-ratios utilizing a transformation $\tilde x \rightarrow \frac{z}{1+z}$.

\subsubsection*{Configuration (b):}
\begin{figure}[H]
	\centering
	\includegraphics[scale=.9]{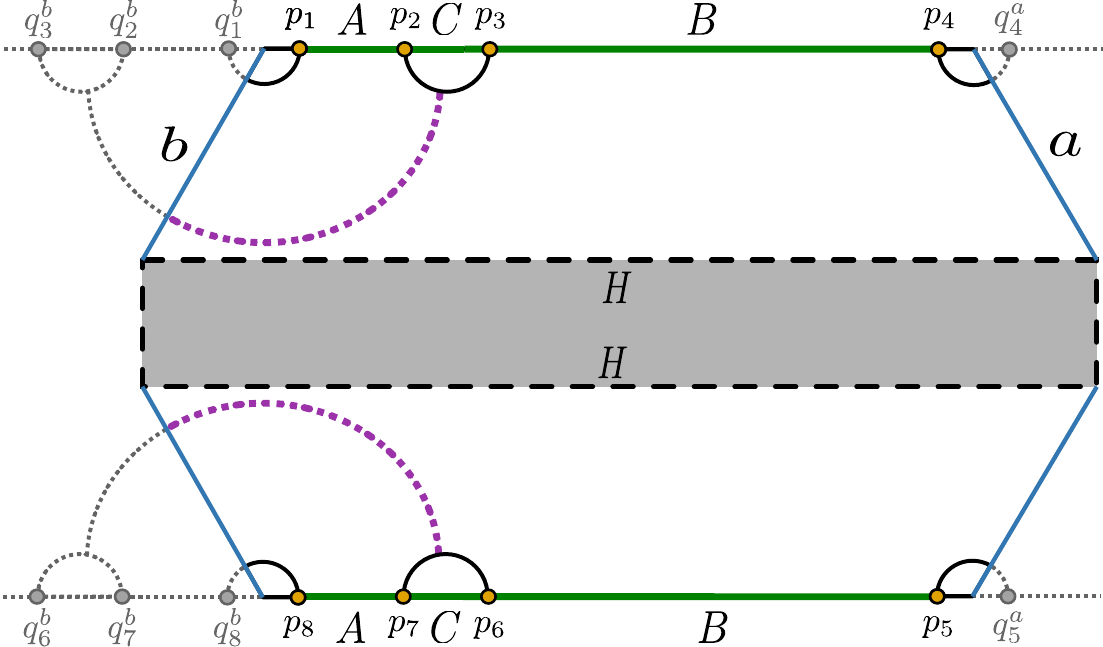}
	\caption{The endpoints of the intervals $A$ and $B$ and the image points in the two $BCFT_2$ copies are identified similarly as in \cref{fig_discase1a}. The other points	$q_3^b=2r_I-r_3$, $q_2^b=2r_I-r_2$ are the image points of $p_2$ and $p_3$ respectively with respect to the $b$-boundary. Also $q_6^b=2r_I-r_3$, $q_7^b=2r_I-r_2$ are the image points of $p_6$ and $p_7$ respectively with respect to the $b$-boundary. }\label{fig_discase1b}
\end{figure}

The second configuration corresponds to a different channel for the four-point twist correlator which results in another contribution to the reflected entropy for the disjoint intervals $A$ and $B$ in the $BCFT_2$s. The R\'enyi reflected entropy for this case may be written as
\begin{align}
	S_R^{(n,m)}(A:B) &= 2\frac{1}{1-n}\log \frac{\left< \sigma_{g_A}(p_1)\sigma_{g_A^{-1}}(p_2)\sigma_{g_B}(p_3)\sigma_{g_B^{-1}}(p_4) \right>_{\mathrm{BCFT}^{\bigotimes mn}}}{\left<\sigma_{g_m}(p_1)\sigma_{g_m^{-1}}(p_2)\sigma_{g_m}(p_3)\sigma_{g_m^{-1}}(p_4) \right>^n_{\mathrm{BCFT}^{\bigotimes m}}}\notag\\
	&= 2 \frac{1}{1-n}\log \frac{\left<\sigma_{g_A}(q^b_2)\sigma_{g_A^{-1}}(p_2)\sigma_{g_B}(p_3)\sigma_{g_B^{-1}}(q^b_3)\right>_{\mathrm{CFT}^{\bigotimes mn}}}{\left<\sigma_{g_m}(q^b_2)\sigma_{g_m^{-1}}(p_2)\sigma_{g_m}(p_3)\sigma_{g_m^{-1}}(q^b_3)\right>^n_{\mathrm{CFT}^{\bigotimes m}}},\label{discase1b}
\end{align}
where once again the factor two arises from the symmetry of the configuration as depicted in \cref{fig_discase1b}. The $BCFT_2$ four-point function in the R\'enyi reflected entropy factorizes into two one-point and a two-point functions in the large central charge limit which as earlier upon utilizing the doubling trick may be expressed in terms of appropriate $CFT_2$ twist correlators. These intermediate procedures are described in \cref{discaseb} of appendix \ref{app_disj}. Note from \cref{fig_discase1b} that $q_2^b$ and $q_3^a$ are the image points for $p_2$ and $p_3$ respectively with respect to the $b$ and $a$-boundaries in the $BCFT_2$. 

A similar configuration for the reflected entropy of two disjoint intervals in a $BCFT_2$ was demonstrated in \cite{Malvimat} the final expression for the reflected entropy was computed as
\begin{align}\label{Srcaseb}
	S_R(A:B)=\frac{2c}{3}\log\left(\frac{1+\sqrt{1-\tilde{x}}}{\sqrt{\tilde{x}}}\right)+4S_{\text{bdy}}\,,
\end{align}
where the $\tilde{x}$ is the finite temperature cross-ratio given by
\begin{equation}\label{CRdiscase1b}
	\tilde{x}=\frac{\sinh \left(\frac{\pi  (p_2-p_3)}{\beta }\right) \sinh \left(\frac{\pi  (q_2^b-q_3^b)}{\beta }\right)}{\sinh \left(\frac{\pi  (q_2^b-p_3)}{\beta }\right) \sinh \left(\frac{\pi  (p_2-q_3^b) }{\beta }\right)}.
\end{equation}
The above expression in \cref{Srcaseb} upon utilization $\tilde{x} \rightarrow \frac{1}{z+1}$ can be further re-expressed as
\begin{equation}\label{srcaseb}
	S_R(A:B)=\frac{c}{3}  \log \left(1+2 z+2 \sqrt{z (z+1)}\right)+4S_{\text{bdy}}\,,
\end{equation}
where $z$ is the finite temperature cross ratio defined as in \cref{CrRatio} with $a_1$, $a_2$, $b_1$, $b_2$ replaced by the points $q_2^b$, $p_2$, $p_3$, $q_3^b$ respectively.

The EWCS for this configuration corresponds to the violet dotted curve shown in \cref{fig_discase1b} which starts from the tip of the dome-type RT surfaces admitted by the interval $C$ and lands on the $b$-brane in the dual bulk $AdS_3$ geometry. The computation for the EWCS in this scenario involves half of the contribution given in \cref{EW} for two disjoint intervals $[q_2^b,p_2]$ and $[p_3,q_3^b]$ with an additive $S_{\text{bdy}}$ term. Finally, considering the two copies of the $BCFT_2$s in a TFD state we obtain the final expression for the EWCS as
\begin{equation}\label{Ewcaseb}
		E_W(A:B)=\frac{c}{6}  \log \left(1+2 z+2 \sqrt{z (z+1)}\right)+2S_{\text{bdy}}\,,
\end{equation}
where $z$ being the finite temperature cross-ratio given in \cref{CrRatio} with $a_1$, $a_2$, $b_1$, $b_2$ replaced by the points $q_2^b$, $p_2$, $p_3$ and $q_3^b$ respectively.

\subsubsection*{Configuration (c):}
We now consider the configuration where the correlation function in the reflected entropy expression involves eight twist operators located at the end points of the intervals in the two copies of the $BCFT_2$s. However, after utilizing a factorization procedure (as shown in \cref{discasec} of appendix \ref{app_disj}), this eight-point function reduces to a four-point function in the $BCFT_2$ which in an appropriate (bulk) OPE channel reduces to a four-point function in a $CFT_2$.

\begin{align}
	S_R^{(n,m)}(A:B) &= \frac{1}{1-n}\log \frac{\left< \sigma_{g_A}(p_1)\sigma_{g_A^{-1}}(p_2)\sigma_{g_B}(p_3)\sigma_{g_B^{-1}}(p_4)\sigma_{g_B}(p_5)\sigma_{g_B^{-1}}(p_6)\sigma_{g_A}(p_7)\sigma_{g_A^{-1}}(p_8) \right>_{\mathrm{BCFT}^{\bigotimes mn}}}{\left<\sigma_{g_m}(p_1)\sigma_{g_m^{-1}}(p_2)\sigma_{g_m}(p_3)\sigma_{g_m^{-1}}(p_4)\sigma_{g_m}(p_5)\sigma_{g_m^{-1}}(p_6)\sigma_{g_m}(p_7)\sigma_{g_m^{-1}}(p_8) \right>^n_{\mathrm{BCFT}^{\bigotimes m}}}\notag\\
	&= \frac{1}{1-n}\log \frac{\left< \sigma_{g_A^{-1}}(p_2)\sigma_{g_B}(p_3)\sigma_{g_B^{-1}}(p_6)\sigma_{g_A}(p_7)\right>_{\mathrm{CFT}^{\bigotimes mn}}}{\left<\sigma_{g_m^{-1}}(p_2)\sigma_{g_m}(p_3)\sigma_{g_m^{-1}}(p_6)\sigma_{g_m}(p_7) \right>^n_{\mathrm{CFT}^{\bigotimes m}}}. \label{discase1c}
\end{align}
Following the discussions in \cite{Jeong:2019xdr} upon implementing the replica limit of the above \cref{discase1c}, we obtain the reflected entropy for the two disjoint intervals $A$ and $B$, which is equivalent to the reflected entropy for the intervals $[p_7,p_2]$ and $[p_3,p_6]$, as
\begin{equation}\label{SRdisjc}
	S_R(A:B)= \frac{c}{3} \log \left(  \frac{1+\sqrt{\tilde{x}}}{1-\sqrt{\tilde{x}}}\right) ,
\end{equation}
where $\tilde{x}=\frac{\tilde{x}_{12}\tilde{x}_{34}}{\tilde{x}_{13}\tilde{x}_{24}}$ is the cross-ratio at a finite temperature $\beta$. In the proximity limit, the reflected entropy for the two disjoint intervals may be expressed following the similar analysis demonstrated in \cite{Shao:2022gpg} as
\begin{align}\label{SRdisjcc}
	S_R(A:B)&=  \frac{c}{3} \log 				\left(\frac{4}{1-\tilde{x}}\right)\notag\\
			&=2\frac{c}{3} \log \frac{\left( r_2+r_3\right)  \cosh \left( \frac{2\pi t}{\beta}\right) }{\left( r_2-r_3\right) }.
\end{align}

\begin{figure}[H]
	\centering
	\includegraphics[scale=.9]{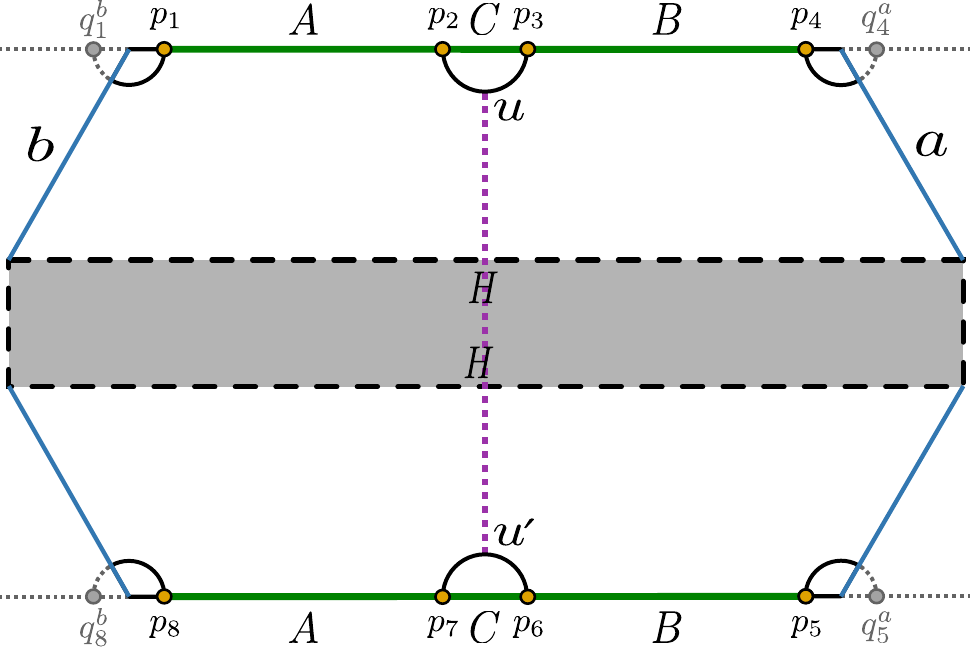}
	\caption{The endpoints of the intervals $A$ and $B$ and the image points in the two $BCFT_2$ copies are identified similarly as in \cref{fig_discase1a}.}\label{fig_discase1c}
\end{figure}

The EWCS for this configuration stretches between the tip of the dome-type RT surfaces for the interval $C$ in the two $BCFT_2$ copies as depicted in \cref{fig_discase1c}. We compute the length of the EWCS in the dual bulk $AdS_3$ geometry in the Poincar\'e coordinates since the dual bulk BTZ black hole space time can always be mapped to a Poincar\'e patch with an appropriate coordinate transformation described in \cite{Geng:2021iyq}. Consequently, the length of one violet dotted curve in the the Poincar\'e patch of the dual bulk BTZ black hole geometry may be obtained as
\begin{equation}\label{LengthEW1}
	{d}=L\, \cosh^{-1}\left(\frac{(\tau-\tau')^2+(x-x')^2+z^2+z'^2}{2 z\, z'}\right)~,
\end{equation}
where $L$ is the length scale of the dual $AdS_3$ geometry and $(\tau,x,z)$, $(\tau',x',z')$ are the Poincar\'e coordinates of some arbitrary points $u$ and $u'$ as shown in \cref{fig_discase1c}. We then extremize \cref{LengthEW1} with respect to the points $u$ and $u'$ and divide it by $4G_N$. Finally using the Brown-Henneaux formula $c=\frac{3L}{2G_N}$ and considering a proper coordinate transformation given in eq. (7.34) of \cite{Shao:2022gpg} where we have considered $e^X=r$ in our work, we obtain the corresponding expression for the EWCS as
\begin{equation}\label{SRdiscase1c}
	E_W(A:B)= \frac{c}{3} \log \frac{\left( r_2+r_3\right)  \cosh \left( \frac{2\pi t}{\beta}\right) }{\left( r_2-r_3\right) }.
\end{equation}

\subsubsection*{Configuration (d):}
\begin{figure}[H]
	\centering
	\includegraphics[scale=.9]{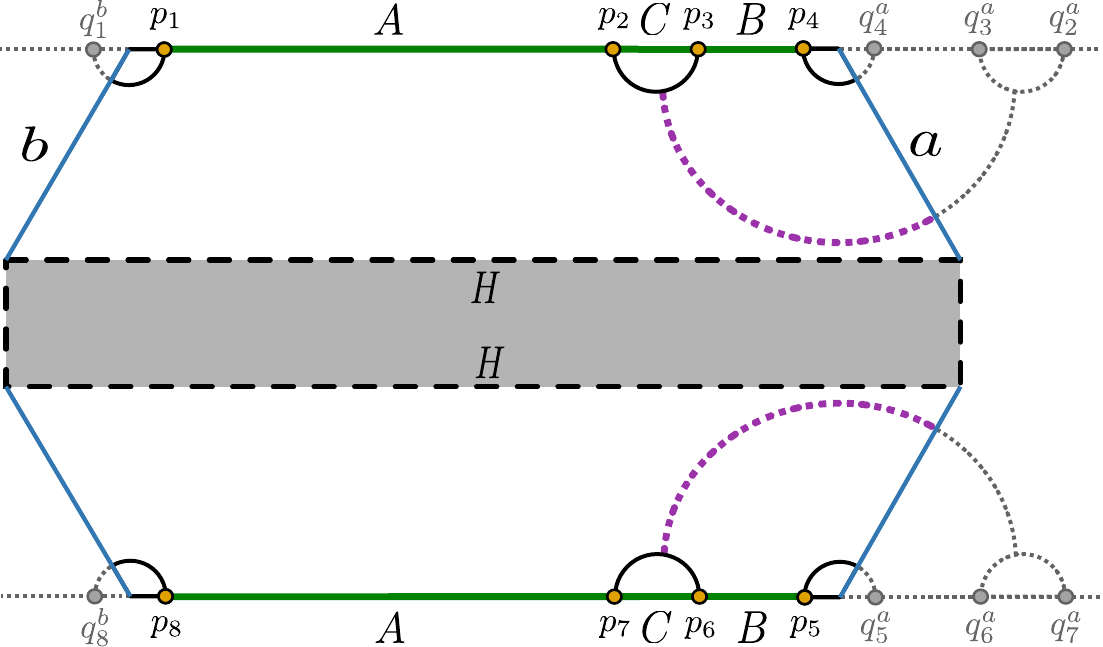}
	\caption{The endpoints of the intervals $A$ and $B$ and the image points in the two $BCFT_2$ copies are identified similarly as in \cref{fig_discase1a}. The other points	$q_2^a=2r_O-r_2$, $q_3^a=2r_O-r_3$  are the image points of $p_2$ and $p_3$ respectively with respect to the $a$-boundary. Also $q_6^a=2r_I-r_3$, $q_7^a=2r_I-r_2$ are the image points of $p_6$ and $p_7$ respectively with respect to the $a$-boundary.}\label{fig_discase1d}
\end{figure}

This case is similar to configuration (b) which involves four-point functions in the corresponding reflected entropy for the disjoint intervals $A$ and $B$. In the large central charge limit each four-point function factorizes into two one-point and one single-point function as shown in \cref{discased} of appendix \ref{app_disj}. finally we uilize the doubling trick to obtain the R\'enyi reflected entropy as 
\begin{align}
	S_R^{(n,m)}(A:B) &= 2\frac{1}{1-n}\log \frac{\left< \sigma_{g_A}(p_1)\sigma_{g_A^{-1}}(p_2)\sigma_{g_B}(p_3)\sigma_{g_B^{-1}}(p_4)\right>_{\mathrm{BCFT}^{\bigotimes mn}}}{\left<\sigma_{g_m}(p_1)\sigma_{g_m^{-1}}(p_2)\sigma_{g_m}(p_3)\sigma_{g_m^{-1}}(p_4) \right>^n_{\mathrm{BCFT}^{\bigotimes m}}}\notag\\
	&= 2 \frac{1}{1-n}\log \frac{\left<\sigma_{g_A}(q^a_2)\sigma_{g_A^{-1}}(p_2)\sigma_{g_B}(p_3)\sigma_{g_B^{-1}}(q^a_3)\right>_{\mathrm{CFT}^{\bigotimes mn}}}{\left<\sigma_{g_m}(q^a_2)\sigma_{g_m^{-1}}(p_2)\sigma_{g_m}(p_3)\sigma_{g_m^{-1}}(q^a_3)\right>^n_{\mathrm{CFT}^{\bigotimes m}}}. \label{discase1d}
\end{align}
However, in this configuration the doubling tricks for the points $p_2$ and $p_3$ are performed with respect to the $a$-boundary in the $BCFT_2$ as depicted in \cref{fig_discase1d}. 

The computation for the reflected entropy in this scenario is similar to the one demonstrated in configuration (b). As a consequence, the reflected entropy for the two disjoint intervals $A$ and $B$ in this configuration may be expressed as in \cref{srcaseb} where the finite temperature cross-ratio $z$ is given by the \cref{CrRatio} with $a_1$, $a_2$, $b_1$, $b_2$ replaced by the points $p_3$, $q_3^a$, $q_2^a$ and $p_2$ respectively.

The corresponding EWCS for this configuration as shown in \cref{fig_discase1b} starts from the tip of the dome-type RT surface supported by the interval $C$ and intersects $b$-brane in the dual bulk BTZ black hole geometry. Once again, the expression for the EWCS in this configuration may be obtained as in \cref{Ewcaseb} where $z$ is the finite temperature cross-ratio given by the \cref{CrRatio} with $a_1$, $a_2$, $b_1$, $b_2$ replaced by the points $p_3$, $q_3^a$, $q_2^a$ and $p_2$ respectively.

\subsubsection*{Configuration (e):}
This case is similar to the one described in configuration (a) which involves a four-point twist correlator in the expression for the reflected entropy with a symmetry factor two. However, after a factorization of the four-point function in the dominant channel, we obtain a three-point twist correlator involving the points $p_2$, $p_3$, $p_4$ and a one-point function with a twist operator at $p_1$ in the $BCFT_2$. This factorization procedure is shown in \cref{discasee} of appendix \ref{app_disj}. Finally upon utilization of the doubling trick this results in a
four-point function in a $CFT_2$. The R\'enyi reflected entropy may then be expressed as
\begin{align}
	S_R^{(n,m)}(A:B) &= 2\frac{1}{1-n}\log \frac{\left< \sigma_{g_A}(p_1)\sigma_{g_A^{-1}}(p_2)\sigma_{g_B}(p_3)\sigma_{g_B^{-1}}(p_4)\right>_{\mathrm{BCFT}^{\bigotimes mn}}}{\left<\sigma_{g_m}(p_1)\sigma_{g_m^{-1}}(p_2)\sigma_{g_m}(p_3)\sigma_{g_m^{-1}}(p_4) \right>^n_{\mathrm{BCFT}^{\bigotimes m}}}\notag\\
	&= 2\frac{1}{1-n}\log \frac{\left< \sigma_{g_A^{-1}}(p_2)\sigma_{g_B}(p_3)\sigma_{g_B^{-1}}(p_4)\sigma_{g_B}(q^a_4)\right>_{\mathrm{CFT}^{\bigotimes mn}}}{\left<\sigma_{g_m^{-1}}(p_2)\sigma_{g_m}(p_3)\sigma_{g_m^{-1}}(p_4)\sigma_{g_m}(q^a_4) \right>^n_{\mathrm{CFT}^{\bigotimes m}}}. \label{discase1e}
\end{align}
Similar to the configuration (c), we follow the procedure discussed in \cite{Dutta:2019gen} to obtain the reflected entropy which is given by the \cref{SRdiscasea} with the finite temperature cross-ratio $z$ expressed as
\begin{equation}\label{CRdcase1e}
	z=\frac{\sinh \left(\frac{\pi  (p_3-p_4)}{\beta }\right) \sinh \left(\frac{\pi  (q_4^a-p_2}{\beta }\right)}{\sinh \left(\frac{\pi  (p_4-q_4^q)}{\beta }\right) \sinh \left(\frac{\pi  (p_2-p_3) }{\beta }\right)}.
\end{equation}

\begin{figure}[H]
	\centering
	\includegraphics[scale=.9]{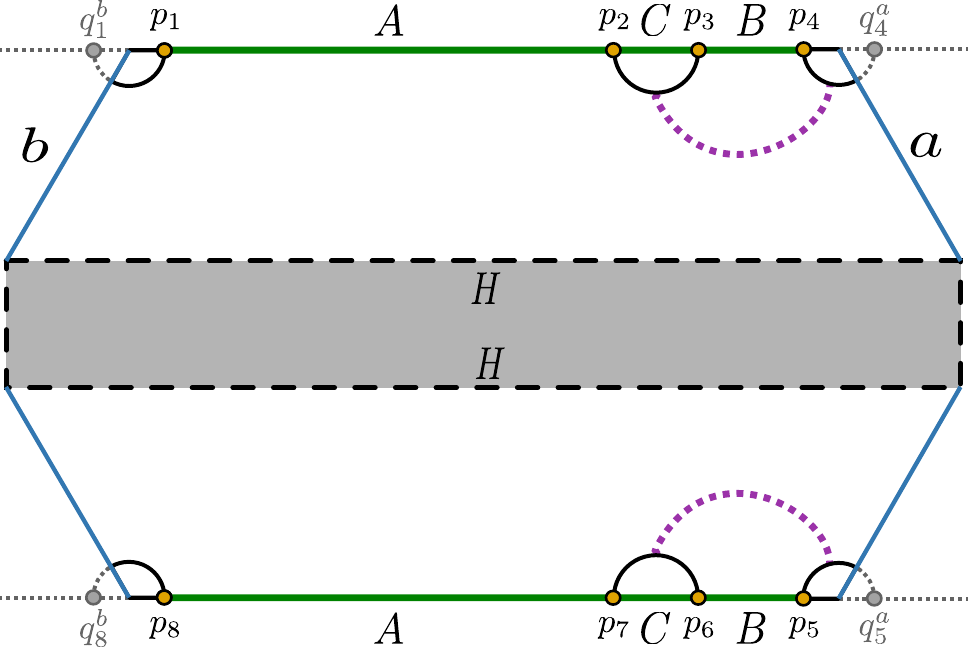}
	\caption{The endpoints of the intervals $A$ and $B$ and the image points in the two $BCFT_2$ copies are identified similarly as in \cref{fig_discase1a}.}\label{fig_discase1e}
\end{figure}
The EWCS for this configuration is described by the violet dotted curve in \cref{fig_discase1e} which is analogous to the one observed in \cref{fig_discase1e} of configuration (a). Consequently the EWCS for this case may be expressed by the \cref{EW} with the finite temperature cross-ration in \cref{CrRatio}. However the points in \cref{CrRatio} are identified as $a_1=p_3$, $a_2=p_4$, $b_1=p^a_4$ and $b_2=p_4$ with $q_4^a$ being the image point of $p_4$ with respect to the $a$-boundary in the $BCFT_2$. As earlier, \cref{SRdiscasea,EW} appear to be different with distinct cross-ratios, however one can always transform one expression to another with identical cross-ratios utilizing a transformation $\tilde x \rightarrow \frac{z}{1+z}$.

\subsubsection*{Configuration (f):}
\begin{figure}[H]
	\centering
	\includegraphics[scale=.9]{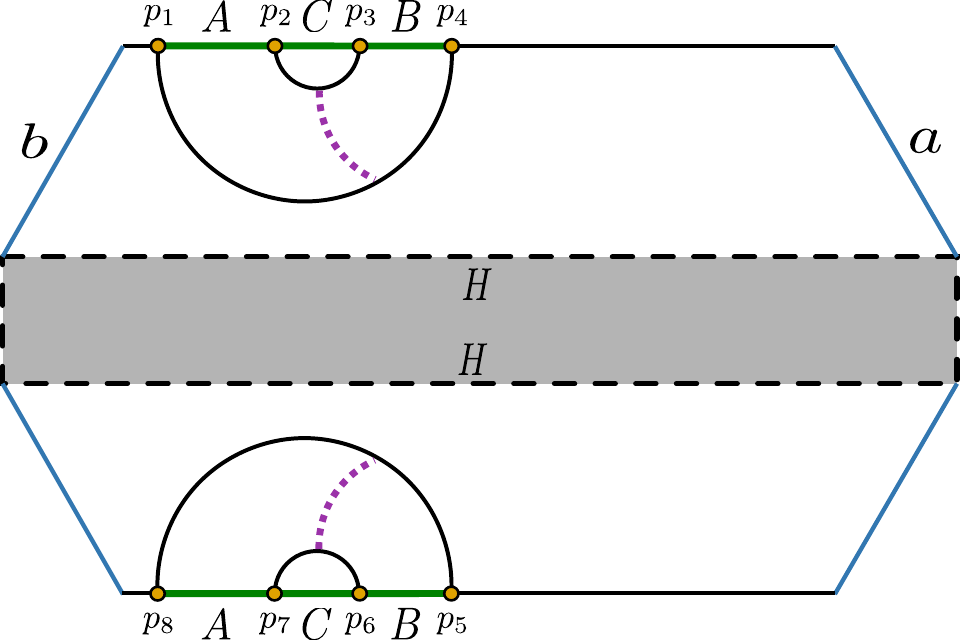}
	\caption{The endpoints of the intervals $A$ and $B$ in the two $BCFT_2$ copies are identified as  $p_1=p_8=r_I+\epsilon$, $p_2=p_7=r_2$, $p_3=p_6=r_3$ and $p_4=p_5=r_4$.}\label{fig_discase1f}
\end{figure}

Next we consider the configuration depicted in \cref{fig_discase1f} and compute the reflected entropy for the disjoint intervals $A$ and $B$. As seen from the figure, theory of $BCFT_2$ does not play any significant role for the reflected entropy computation in this scenario  since the $BCFT_2$ four-point twist correlators in the R\'enyi reflected entropy expression reduce to $CFT_2$ four-point correlators in the an appropriate (bulk) OPE channel. Consequently this configuration corresponds to a trivial one where the reflected entropy for the two disjoint intervals may be obtained following the usual procedure in a $CFT_2$ as discussed in \cite{Dutta:2019gen}. The R\'enyi reflected entropy for this configuration may be obtained as
\begin{align}
	S_R^{(n,m)}(A:B) &= 2\frac{1}{1-n}\log \frac{\left< \sigma_{g_A}(p_1)\sigma_{g_A^{-1}}(p_2)\sigma_{g_B}(p_3)\sigma_{g_B^{-1}}(p_4) \right>_{\mathrm{BCFT}^{\bigotimes mn}}}{\left<\sigma_{g_m}(p_1)\sigma_{g_m^{-1}}(p_2)\sigma_{g_m}(p_3)\sigma_{g_m^{-1}}(p_4) \right>^n_{\mathrm{BCFT}^{\bigotimes m}}}\notag\\
	&= 2\frac{1}{1-n}\log \frac{\left< \sigma_{g_A}(p_1)\sigma_{g_A^{-1}}(p_2)\sigma_{g_B}(p_3)\sigma_{g_B^{-1}}(p_4) \right>_{\mathrm{CFT}^{\bigotimes mn}}}{\left<\sigma_{g_m}(p_1)\sigma_{g_m^{-1}}(p_2)\sigma_{g_m}(p_3)\sigma_{g_m^{-1}}(p_4) \right>^n_{\mathrm{CFT}^{\bigotimes m}}}.\label{discase1f}
\end{align}
As earlier, the factor two in the above expression arises from the symmetry of the configuration as seen from \cref{fig_discase1f}. Now implementing the replica limit for the \cref{discase1f}, we obtain the reflected entropy for this configuration as
\begin{equation}\label{SRdiscasef}
	S_R(A:B)=\frac{2c}{3}  \log \left(  \frac{1+\sqrt{\tilde x}}{1-\sqrt{\tilde x}}\right) .
\end{equation}
Here $\tilde x$ is the cross ratio at a finite temperature which is given by
\begin{equation}\label{CRSRdcase1f}
	\tilde x=\frac{\sinh \left(\frac{\pi  (p_1-p_2)}{\beta }\right) \sinh \left(\frac{\pi  (p_3-p_4)}{\beta }\right)}{\sinh \left(\frac{\pi  (p_1-p_3)}{\beta }\right) \sinh \left(\frac{\pi  (p_2-p_4) }{\beta }\right)}.
\end{equation}

The corresponding EWCS computation for this configuration is also a trivial one since it refers to the case of two disjoint intervals in the context of the usual $AdS_3/CFT_2$ scenario as seen from the \cref{fig_discase1f}. Therefore, the expression for the EWCS in this case is given by the \cref{EW} with the finite temperature cross-ratio in \cref{CrRatio} where the points $a_1$, $a_2$, $b_1$ and $b_2$ are replaced by $p_1$, $p_2$, $p_3$ and $p_4$ respectively. Once again, \cref{SRdiscasef,EW} appear to be different with distinct cross-ratios, however one expression can always be transformed to the other one with identical cross-ratios utilizing a transformation $\tilde x \rightarrow \frac{z}{1+z}$.

\subsubsection*{Configuration (g):}
Another possible contribution to the reflected entropy is considered in this case which results into an expression for $S_R(A:B)$ analogous to the one obtained in configuration (a). However, in this case the R\'enyi reflected entropy involves an eight-point function of the twist operators located at the endpoints of the intervals in the two copies of the $BCFT_2$s. Nevertheless, after using a factorization of the eight-point function in the large central charge limit, we obtain an identical expression for the R\'enyi reflected entropy as in configuration (a),
\begin{align}
	S_R^{(n,m)}(A:B) &= \frac{1}{1-n}\log \frac{\left< \sigma_{g_A}(p_1)\sigma_{g_A^{-1}}(p_2)\sigma_{g_B}(p_3)\sigma_{g_B^{-1}}(p_4)\sigma_{g_B}(p_5)\sigma_{g_B^{-1}}(p_6)\sigma_{g_A}(p_7)\sigma_{g_A^{-1}}(p_8) \right>_{\mathrm{BCFT}^{\bigotimes mn}}}{\left<\sigma_{g_m}(p_1)\sigma_{g_m^{-1}}(p_2)\sigma_{g_m}(p_3)\sigma_{g_m^{-1}}(p_4)\sigma_{g_m}(p_5)\sigma_{g_m^{-1}}(p_6)\sigma_{g_m}(p_7)\sigma_{g_m^{-1}}(p_8) \right>^n_{\mathrm{BCFT}^{\bigotimes m}}}\notag\\
	&= 2\frac{1}{1-n}\log \frac{\left< \sigma_{g_A^{-1}}(q^b_1)\sigma_{g_A}(p_1)\sigma_{g_A^{-1}}(p_2)\sigma_{g_B}(p_3)\right>_{\mathrm{CFT}^{\bigotimes mn}}}{\left<\sigma_{g_m^{-1}}(q^b_1)\sigma_{g_m}(p_1)\sigma_{g_m^{-1}}(p_2)\sigma_{g_m}(p_3)\right>^n_{\mathrm{CFT}^{\bigotimes m}}}. \label{discase1g}
\end{align}
The factorization procedure utilized in the above expression is shown in \cref{discaseg} of appendix \ref{app_disj}. Now considering the replica limit of \cref{discase1g}, the reflected entropy $S_R(A:B)$ is given by the \cref{SRdiscasea} with the finite temperature cross-ratio in \cref{CRdiscase1a}.

\begin{figure}[H]
	\centering
	\includegraphics[scale=.9]{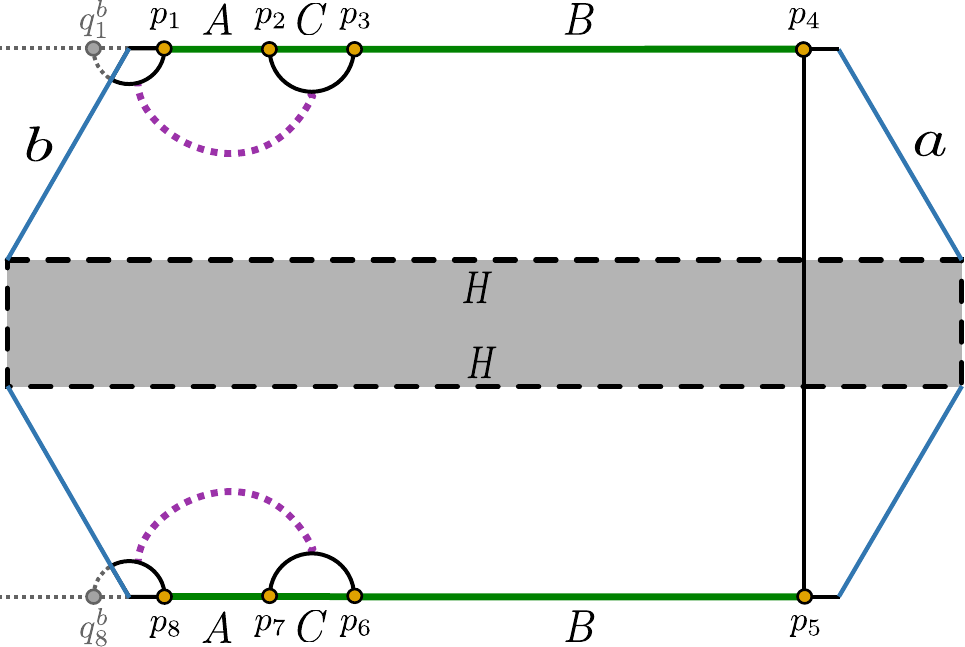}
	\caption{The endpoints of the intervals $A$ and $B$ add the image points in the two $BCFT_2$ copies are identified similarly as in \cref{fig_discase1a}.}\label{fig_discase1g}
\end{figure}

The corresponding EWCS for this case is identical to that in configuration (a) as seen from the \cref{fig_discase1a,fig_discase1g}. Consequently the EWCS for this configuration may be expressed as in \cref{EW} with the cross-ratio at a finite temperature given in \cref{CrRatio}. Similar to the configuration (a), we have identified the points in this scenario as $a_1=p_1$, $a_2=p_2$, $b_1=p_3$ and $b_2=q_1^b$ with $q_1^b$ being the image point of $p_1$ with respect to the $b$-boundary in the $BCFT_2$. As explained earlier, the \cref{SRdiscasea,EW} appear to be different with distinct cross-ratios, however one can always transform one expression to another with identical cross-ratios utilizing a transformation $\tilde x \rightarrow \frac{z}{1+z}$.

\subsubsection*{Configuration (h):}
In this configuration, we discuss another contribution to the reflected entropy for the disjoint intervals under consideration which is similar to the one described in configuration (b). The R\'enyi reflected entropy for this case involves eight-point correlation functions with the twist operators located at the end points of the intervals in the bath $BCFT_2$s. However after employing a factorization of the correlation function in the large central charge limit (as described in \cref{discaseh} of appendix \ref{app_disj}), we obtain an expression for the $S_R^{(n,m)}(A:B)$ similar to the one in configuration (b) as follows
\begin{align}
	S_R^{(n,m)}(A:B) &= \frac{1}{1-n}\log \frac{\left< \sigma_{g_A}(p_1)\sigma_{g_A^{-1}}(p_2)\sigma_{g_B}(p_3)\sigma_{g_B^{-1}}(p_4)\sigma_{g_B}(p_5)\sigma_{g_B^{-1}}(p_6)\sigma_{g_A}(p_7)\sigma_{g_A^{-1}}(p_8) \right>_{\mathrm{BCFT}^{\bigotimes mn}}}{\left<\sigma_{g_m}(p_1)\sigma_{g_m^{-1}}(p_2)\sigma_{g_m}(p_3)\sigma_{g_m^{-1}}(p_4)\sigma_{g_m}(p_5)\sigma_{g_m^{-1}}(p_6)\sigma_{g_m}(p_7)\sigma_{g_m^{-1}}(p_8) \right>^n_{\mathrm{BCFT}^{\bigotimes m}}}\notag\\
	&= 2 \frac{1}{1-n}\log \frac{\left<\sigma_{g_A}(q^b_2)\sigma_{g_A^{-1}}(p_2)\sigma_{g_B}(p_3)\sigma_{g_B^{-1}}(q^b_3)\right>_{\mathrm{CFT}^{\bigotimes mn}}}{\left<\sigma_{g_m}(q^b_2)\sigma_{g_m^{-1}}(p_2)\sigma_{g_m}(p_3)\sigma_{g_m^{-1}}(q^b_3)\right>^n_{\mathrm{CFT}^{\bigotimes m}}}. \label{discase1h}
\end{align}
Considering the replica limit to the above equation and following the same procedure discussed in configuration (b), $S_R(A:B)$ in this case may be given by the \cref{srcaseb} where the finite temperature cross-ratio $z$ defined as in \cref{CrRatio} with $a_1$, $a_2$, $b_1$, $b_2$ replaced by the points $q_2^b$, $p_2$, $p_3$, $q_3^b$ respectively.

\begin{figure}[H]
	\centering
	\includegraphics[scale=.9]{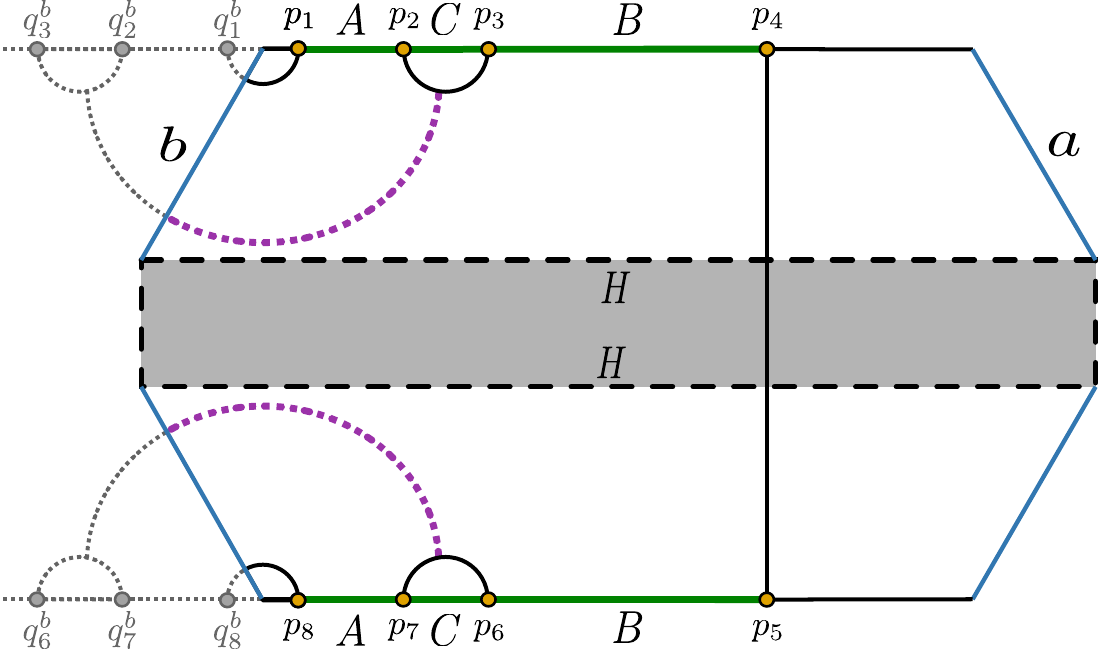}
	\caption{The endpoints of the intervals $A$ and $B$ and the image points in the two $BCFT_2$ copies are identified similarly as in \cref{fig_discase1a,fig_discase1b} with $p_4=p_5=r_4$.}\label{fig_discase1h}
\end{figure}

Similar to the reflected entropy, the corresponding EWCS in the bulk dual geometry is identical to that discussed in configuration (b) as seen from the \cref{fig_discase1b,fig_discase1h}. Therefore the expression for the EWCS for this configuration may be given by the \cref{Ewcaseb} where once again the finite temperature cross-ratio $z$ is defined as in \cref{CrRatio} with $a_1$, $a_2$, $b_1$, $b_2$ replaced by the points $q_2^b$, $p_2$, $p_3$, $q_3^b$ respectively.

\subsubsection*{Configuration (i):}
For this scenario, we may refer to configuration (c) for the computation of the reflected entropy since, in the large central charge limit, after employing a factorization procedure to the eight-point correlation functions involved in the R\'enyi reflected entropy, we obtain an identical expression as follows
\begin{align}
	S_R^{(n,m)}(A:B) &= \frac{1}{1-n}\log \frac{\left< \sigma_{g_A}(p_1)\sigma_{g_A^{-1}}(p_2)\sigma_{g_B}(p_3)\sigma_{g_B^{-1}}(p_4)\sigma_{g_B}(p_5)\sigma_{g_B^{-1}}(p_6)\sigma_{g_A}(p_7)\sigma_{g_A^{-1}}(p_8) \right>_{\mathrm{BCFT}^{\bigotimes mn}}}{\left<\sigma_{g_m}(p_1)\sigma_{g_m^{-1}}(p_2)\sigma_{g_m}(p_3)\sigma_{g_m^{-1}}(p_4)\sigma_{g_m}(p_5)\sigma_{g_m^{-1}}(p_6)\sigma_{g_m}(p_7)\sigma_{g_m^{-1}}(p_8) \right>^n_{\mathrm{BCFT}^{\bigotimes m}}}\notag\\
	&= \frac{1}{1-n}\log \frac{\left< \sigma_{g_A^{-1}}(p_2)\sigma_{g_B}(p_3)\sigma_{g_B^{-1}}(p_6)\sigma_{g_A}(p_7)\right>_{\mathrm{CFT}^{\bigotimes mn}}}{\left<\sigma_{g_m^{-1}}(p_2)\sigma_{g_m}(p_3)\sigma_{g_m^{-1}}(p_6)\sigma_{g_m}(p_7) \right>^n_{\mathrm{CFT}^{\bigotimes m}}}. \label{discase1i}
\end{align}
The factorization of the eight-point functions employed in the above expression is described in \cref{discasei} of appendix \ref{app_disj}. Consequently, the reflected entropy for the disjoint intervals $A$ and $B$ for this configuration is given by the \cref{SRdisjcc}.

\begin{figure}[H]
	\centering
	\includegraphics[scale=.9]{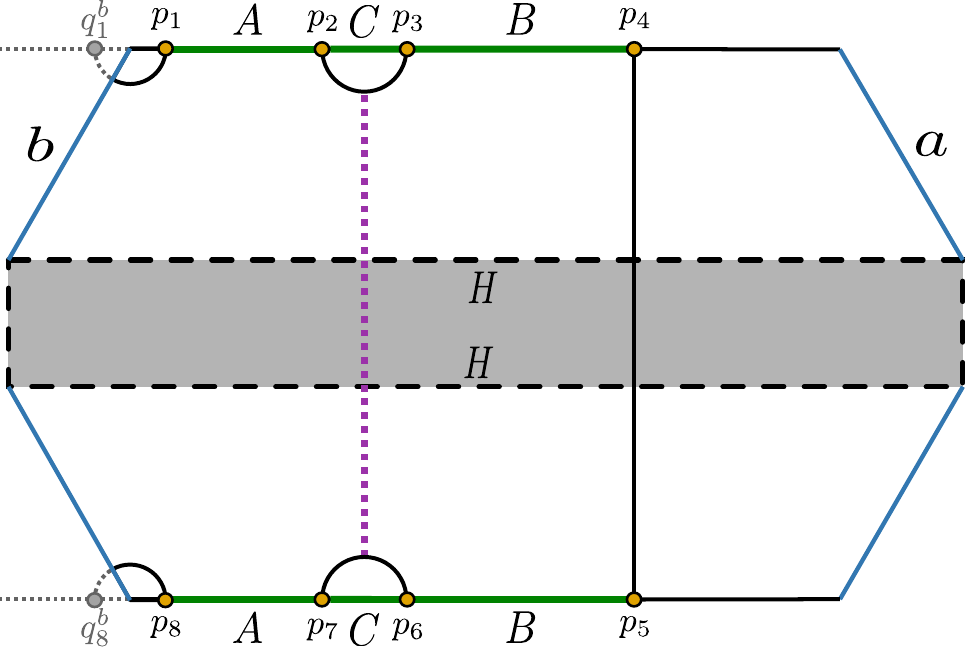}
	\caption{The endpoints of the intervals $A$ and $B$ and the image points in the two $BCFT_2$ copies are identified similarly as in \cref{fig_discase1a} with $p_4=p_5=r_4$.}\label{fig_discase1i}
\end{figure}
As shown in the \cref{fig_discase1i}, the violet dotted curve describes the corresponding EWCS in the bulk dual BTZ black hole space time which is analogous to the one observed in \cref{fig_discase1c} of configuration (c). Consequently the EWCS for this case may be given by the \cref{SRdiscase1c}.

\subsubsection*{Configuration (j):}
We now discuss a non-trivial contribution to the reflected entropy for the two disjoint intervals $A$ and $B$ in the two copies of the $BCFT_2$s. This scenario is similar to the one discussed in configuration (j) of the two adjacent intervals. The corresponding computation initially involves eight-point correlation functions in the R\'enyi reflected entropy expression which factorizes in the large central charge limit (as described in \cref{discasej} of appendix \ref{app_disj}) into two one-point and one six-point functions in a $BCFT_2$. Finally this reduces to an expression for the R\'enyi reflected entropy which include six-point functions in a $CFT_2$ as follows
\begin{align}
	S_R^{(n,m)}(A:B) &= \frac{1}{1-n}\log \frac{\left< \sigma_{g_A}(p_1)\sigma_{g_A^{-1}}(p_2)\sigma_{g_B}(p_3)\sigma_{g_B^{-1}}(p_4)\sigma_{g_B}(p_5)\sigma_{g_B^{-1}}(p_6)\sigma_{g_A}(p_7)\sigma_{g_A^{-1}}(p_8) \right>_{\mathrm{BCFT}^{\bigotimes mn}}}{\left<\sigma_{g_m}(p_1)\sigma_{g_m^{-1}}(p_2)\sigma_{g_m}(p_3)\sigma_{g_m^{-1}}(p_4)\sigma_{g_m}(p_5)\sigma_{g_m^{-1}}(p_6)\sigma_{g_m}(p_7)\sigma_{g_m^{-1}}(p_8) \right>^n_{\mathrm{BCFT}^{\bigotimes m}}}\notag\\
	&= \frac{1}{1-n}\log \frac{\left< \sigma_{g_A^{-1}}(p_2)\sigma_{g_B}(p_3)\sigma_{g_B^{-1}}(p_4)\sigma_{g_B}(p_5)\sigma_{g_B^{-1}}(p_6)\sigma_{g_A}(p_7)\right>_{\mathrm{CFT}^{\bigotimes mn}}}{\left<\sigma_{g_m^{-1}}(p_2)\sigma_{g_m}(p_3)\sigma_{g_m^{-1}}(p_4)\sigma_{g_m}(p_5)\sigma_{g_m^{-1}}(p_6)\sigma_{g_m}(p_7)\right>^n_{\mathrm{CFT}^{\bigotimes m}}}. \label{discase1j}
\end{align}
We now utilize a technique that has already been discussed for the adjacent intervals termed as inverse doubling trick which transforms the above $CFT_2$ six-point twist correlator into a $BCFT_2$ three-point correlator \cite{Shao:2022gpg,Li:2021dmf}. Finally in an appropriate OPE channel and implementing the replica limit, we obtain the reflected entropy for this configuration in a proximity limit as
\begin{align}\label{SRdiscasej}
	S_R(A:B) &= \frac{2c}{3}\log \left(\frac{2 \sech(\frac{2\pi t}{\beta}) (p_4-p_2) \sqrt{p_4^2+p_2^2+2 p_4 p_2 \cosh (\frac{4\pi t}{\beta})}}{p_4 (p_1-p_2)}\right).
\end{align}

\begin{figure}[H]
	\centering
	\includegraphics[scale=.9]{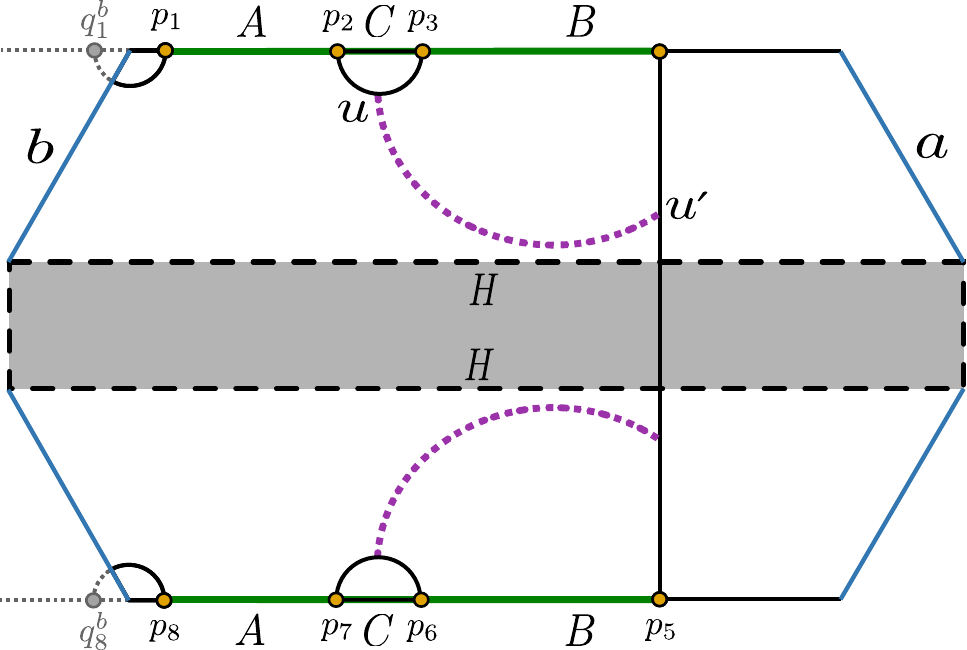}
	\caption{The endpoints of the intervals $A$ and $B$ and the image points in the two $BCFT_2$ copies are identified similarly as in \cref{fig_discase1a} with $p_4=p_5=r_4$.}\label{fig_discase1j}
\end{figure}

In the bulk dual geometry, the EWCS is described by a codimension two surface that starts from the tip of the dome-type RT surface supported by the interval $C$ and lands on the HM surface as depicted in \cref{fig_discase1j}. We compute the length for such EWCS from a purely geometric perspective. First we consider two arbitrary points $u$ and $u'$ in the bulk dual geometry where the EWCS intersects the dome-type RT surface supported by the interval $C$ and the HM surface respectively. We then compute the corresponding length of the EWCS in the Poincar\'e coordinates since the dual bulk BTZ black hole space time can always be mapped to a Poincar\'e patch with an appropriate coordinate transformation described in \cite{Geng:2021iyq}. Consequently, the length of one violet dotted curve in the the Poincar\'e patch of the dual bulk BTZ black hole geometry may be obtained as
\begin{equation}\label{LengthEW}
{d}=L\, \cosh^{-1}\left(\frac{(\tau-\tau')^2+(x-x')^2+z^2+z'^2}{2 z\, z'}\right)~,
\end{equation}
where $L$ is the length scale of the dual $AdS_3$ geometry and $(\tau,x,z)$, $(\tau',x',z')$ are the Poincar\'e coordinates of the points $u$ and $u'$. Finally, minimizing \cref{LengthEW} with respect to the points $u$ and $u'$ and multiplying by $\frac{2}{4G_N}$ in a proximity limit ($p_2\rightarrow p_3$) we obtain the expression for the EWCS as
\begin{align}\label{EWdiscasej}
	E_W(A:B)&=\frac{c}{3}\log \left(\frac{2 \sech(\frac{2\pi t}{\beta}) (p_4-p_2) \sqrt{p_4^2+p_2^2+2 p_4 p_2 \cosh (\frac{4\pi t}{\beta})}}{p_4 (p_1-p_2)}\right).
\end{align}
In the above expression we have used the Brown-Henneaux formula $c=\frac{3L}{2G_N}$ to express it in terms of the central charge $c$ of the boundary $CFT_2$ and a proper coordinate transformation given in eq. (7.34) of \cite{Shao:2022gpg} where we have considered $e^X=r$ in our work.

\subsubsection*{Configuration (k):}
This configuration is similar to the previous one however with a different factorization procedure employed in the large central charge limit for the eight-point function in the R\'enyi reflected entropy expression. This factorization of the eight-point twist correlator is described in \cref{discasek} of appendix \ref{app_disj}. The final expression for the $S_R^{(n,m)}(A:B)$ involving six-point functions is given as follows
\begin{align}
	S_R^{(n,m)}(A:B) &= \frac{1}{1-n}\log \frac{\left< \sigma_{g_A}(p_1)\sigma_{g_A^{-1}}(p_2)\sigma_{g_B}(p_3)\sigma_{g_B^{-1}}(p_4)\sigma_{g_B}(p_5)\sigma_{g_B^{-1}}(p_6)\sigma_{g_A}(p_7)\sigma_{g_A^{-1}}(p_8) \right>_{\mathrm{BCFT}^{\bigotimes mn}}}{\left<\sigma_{g_m}(p_1)\sigma_{g_m^{-1}}(p_2)\sigma_{g_m}(p_3)\sigma_{g_m^{-1}}(p_4)\sigma_{g_m}(p_5)\sigma_{g_m^{-1}}(p_6)\sigma_{g_m}(p_7)\sigma_{g_m^{-1}}(p_8) \right>^n_{\mathrm{BCFT}^{\bigotimes m}}}\notag\\
	&= \frac{1}{1-n}\log \frac{\left< \sigma_{g_A}(p_1)\sigma_{g_A^{-1}}(p_2)\sigma_{g_B}(p_3)\sigma_{g_B^{-1}}(p_6)\sigma_{g_A}(p_7)\sigma_{g_A^{-1}}(p_8) \right>_{\mathrm{CFT}^{\bigotimes mn}}}{\left<\sigma_{g_m}(p_1)\sigma_{g_m^{-1}}(p_2)\sigma_{g_m}(p_3)\sigma_{g_m^{-1}}(p_6)\sigma_{g_m}(p_7)\sigma_{g_m^{-1}}(p_8) \right>^n_{\mathrm{CFT}^{\bigotimes m}}}, \label{discase1k}
\end{align}
Similar to the previous configuration, we now utilize the same tricks discussed in \cite{Shao:2022gpg,Li:2021dmf} to obtain the final result for the reflected entropy which may be given by the \cref{SRdiscasej} with the points $p_4$ and $p_5$ interchanged with $p_1$ and $p_8$ respectively.

\begin{figure}[H]
	\centering
	\includegraphics[scale=.9]{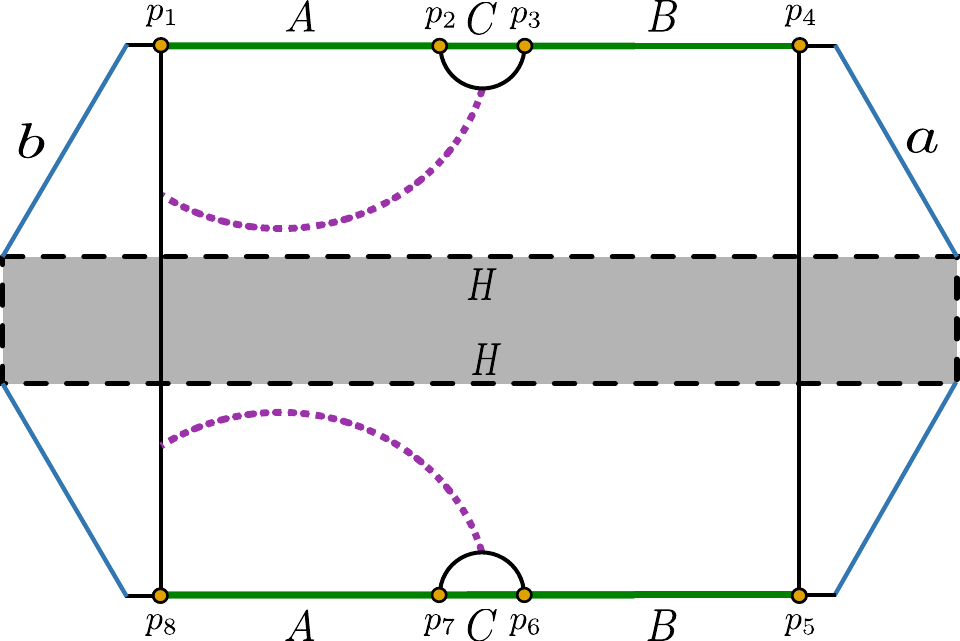}
	\caption{The endpoints of the intervals $A$ and $B$ in the two $BCFT_2$ copies are identified similarly as in \cref{fig_discase1a}.}\label{fig_discase1k}
\end{figure}

The EWCS for this scenario is analogous to the previous configuration as seen from the \cref{fig_discase1j,fig_discase1k}. Consequently the EWCS for this configuration may be expressed as in \cref{EWdiscasej} with the points $p_4$ and $p_5$ being interchanged by $p_1$ and $p_8$ respectively.

\subsubsection*{Configuration (l):}
We refer to configuration (c) to obtain the reflected entropy for the disjoint intervals $A$ and $B$ in this scenario. The R\'enyi reflected entropy for this configuration reduces to an expression identical to the one in \cref{discase1c} after employing a factorization procedure for the eight-point twist correlator in the large central charge limit as described in \cref{discasel} of appendix \ref{app_disj}. The corresponding R\'enyi reflected entropy may then be given as follows
\begin{align}
	S_R^{(n,m)}(A:B) &= \frac{1}{1-n}\log \frac{\left< \sigma_{g_A}(p_1)\sigma_{g_A^{-1}}(p_2)\sigma_{g_B}(p_3)\sigma_{g_B^{-1}}(p_4)\sigma_{g_B}(p_5)\sigma_{g_B^{-1}}(p_6)\sigma_{g_A}(p_7)\sigma_{g_A^{-1}}(p_8) \right>_{\mathrm{BCFT}^{\bigotimes mn}}}{\left<\sigma_{g_m}(p_1)\sigma_{g_m^{-1}}(p_2)\sigma_{g_m}(p_3)\sigma_{g_m^{-1}}(p_4)\sigma_{g_m}(p_5)\sigma_{g_m^{-1}}(p_6)\sigma_{g_m}(p_7)\sigma_{g_m^{-1}}(p_8) \right>^n_{\mathrm{BCFT}^{\bigotimes m}}}\notag\\
	&= \frac{1}{1-n}\log \frac{\left< \sigma_{g_A^{-1}}(p_2)\sigma_{g_B}(p_3)\sigma_{g_B^{-1}}(p_6)\sigma_{g_A}(p_7)\right>_{\mathrm{CFT}^{\bigotimes mn}}}{\left<\sigma_{g_m^{-1}}(p_2)\sigma_{g_m}(p_3)\sigma_{g_m^{-1}}(p_6)\sigma_{g_m}(p_7) \right>^n_{\mathrm{CFT}^{\bigotimes m}}}. \label{discase1l}
\end{align}
Consequently, after implementing the replica limit to the above equation, the reflected entropy $S_R(A:B)$ may be given by the \cref{SRdisjcc}.

\begin{figure}[H]
	\centering
	\includegraphics[scale=.9]{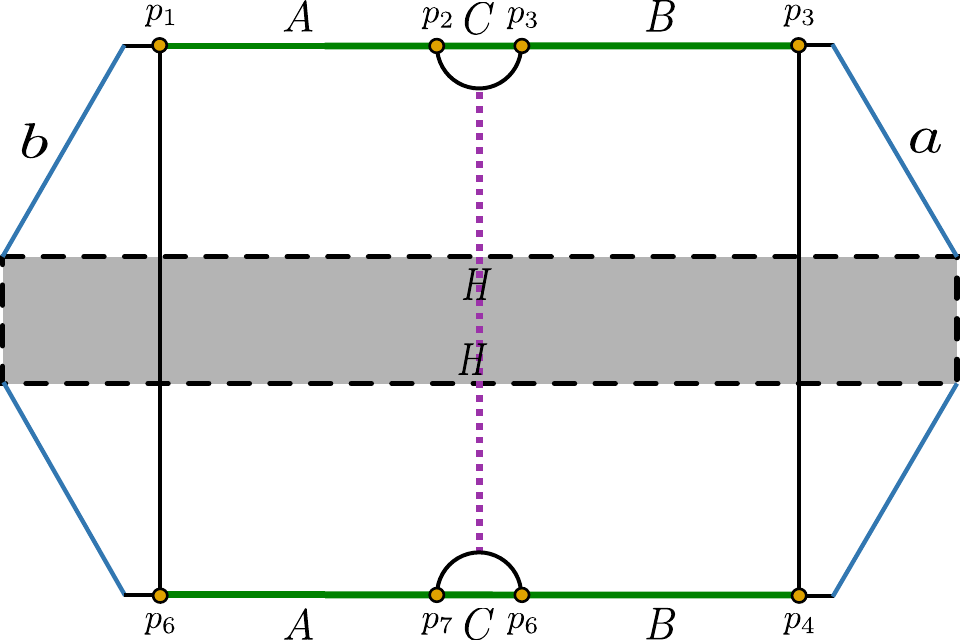}
	\caption{The endpoints of the intervals $A$ and $B$ in the two $BCFT_2$ copies are identified similarly as in \cref{fig_discase1a}.}\label{fig_discase1l}
\end{figure}

As seen from the \cref{fig_discase1c,fig_discase1l}, the EWCS for this scenario is identical to the one described in configuration (c). Therefore we follow the same procedure discussed in configuration (c) to obtain the EWCS for this case and the corresponding expression may then be given by the \cref{SRdiscase1c}.

\subsubsection*{Configuration (m):}
The reflected entropy for this case is identical to the one computed in configuration (j) since factorizing the eight-point function in the corresponding expression in a dominant channel (as shown in \cref{discasem} of appendix \ref{app_disj}) we obtain the same six-point function as
\begin{align}
	S_R^{(n,m)}(A:B) &= \frac{1}{1-n}\log \frac{\left< \sigma_{g_A}(p_1)\sigma_{g_A^{-1}}(p_2)\sigma_{g_B}(p_3)\sigma_{g_B^{-1}}(p_4)\sigma_{g_B}(p_5)\sigma_{g_B^{-1}}(p_6)\sigma_{g_A}(p_7)\sigma_{g_A^{-1}}(p_8) \right>_{\mathrm{BCFT}^{\bigotimes mn}}}{\left<\sigma_{g_m}(p_1)\sigma_{g_m^{-1}}(p_2)\sigma_{g_m}(p_3)\sigma_{g_m^{-1}}(p_4)\sigma_{g_m}(p_5)\sigma_{g_m^{-1}}(p_6)\sigma_{g_m}(p_7)\sigma_{g_m^{-1}}(p_8) \right>^n_{\mathrm{BCFT}^{\bigotimes m}}}\notag\\
	&= \frac{1}{1-n}\log \frac{\left< \sigma_{g_A^{-1}}(p_2)\sigma_{g_B}(p_3)\sigma_{g_B^{-1}}(p_4)\sigma_{g_B}(p_5)\sigma_{g_B^{-1}}(p_6)\sigma_{g_A}(p_7)\right>_{\mathrm{CFT}^{\bigotimes mn}}}{\left<\sigma_{g_m^{-1}}(p_2)\sigma_{g_m}(p_3)\sigma_{g_m^{-1}}(p_4)\sigma_{g_m}(p_5)\sigma_{g_m^{-1}}(p_6)\sigma_{g_m}(p_7)\right>^n_{\mathrm{CFT}^{\bigotimes m}}}. \label{discase1m}
\end{align}
Once again, employing the same techniques as discussed in configuration (j), the reflected entropy $S_R(A:B)$ in this scenario is given by the \cref{SRdiscasej}.

\begin{figure}[H]
	\centering
	\includegraphics[scale=.9]{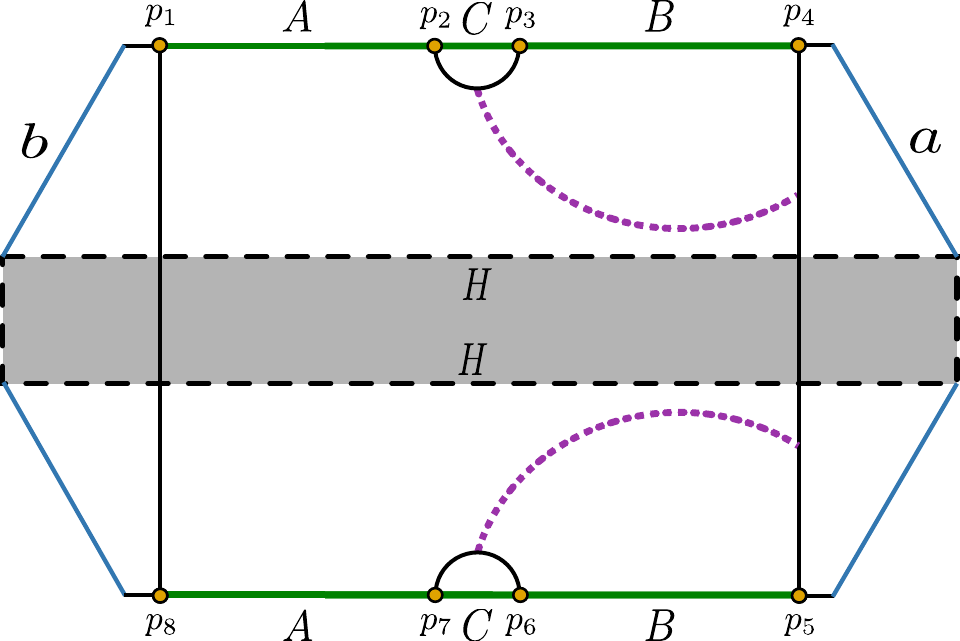}
	\caption{The endpoints of the intervals $A$ and $B$ in the two $BCFT_2$ copies are identified similarly as in \cref{fig_discase1a}. }\label{fig_discase1m}
\end{figure}
As seen from the \cref{fig_discase1j,fig_discase1m}, the EWCS for the entanglement wedge of $A\cup B$ for this case is identical to that of in configuration (j). Consequently the corresponding expression for the EWCS for this configuration may be given by the \cref{EWdiscasej}.

\subsubsection*{Configuration (n):}
Again we refer to configuration (k), the R\'enyi reflected entropy for the two disjoint intervals under consideration for this case reduces to an identical expression after the possible factorization of the eight-point function involved in the corresponding expression in a dominant channel as shown in \cref{discasen} of appendix \ref{app_disj}. The R\'enyi reflected entropy may then be given by
\begin{align}
	S_R^{(n,m)}(A:B) &= \frac{1}{1-n}\log \frac{\left< \sigma_{g_A}(p_1)\sigma_{g_A^{-1}}(p_2)\sigma_{g_B}(p_3)\sigma_{g_B^{-1}}(p_4)\sigma_{g_B}(p_5)\sigma_{g_B^{-1}}(p_6)\sigma_{g_A}(p_7)\sigma_{g_A^{-1}}(p_8) \right>_{\mathrm{BCFT}^{\bigotimes mn}}}{\left<\sigma_{g_m}(p_1)\sigma_{g_m^{-1}}(p_2)\sigma_{g_m}(p_3)\sigma_{g_m^{-1}}(p_4)\sigma_{g_m}(p_5)\sigma_{g_m^{-1}}(p_6)\sigma_{g_m}(p_7)\sigma_{g_m^{-1}}(p_8) \right>^n_{\mathrm{BCFT}^{\bigotimes m}}}\notag\\
	&= \frac{1}{1-n}\log \frac{\left< \sigma_{g_A}(p_1)\sigma_{g_A^{-1}}(p_2)\sigma_{g_B}(p_3)\sigma_{g_B^{-1}}(p_6)\sigma_{g_A}(p_7)\sigma_{g_A^{-1}}(p_8) \right>_{\mathrm{CFT}^{\bigotimes mn}}}{\left<\sigma_{g_m}(p_1)\sigma_{g_m^{-1}}(p_2)\sigma_{g_m}(p_3)\sigma_{g_m^{-1}}(p_6)\sigma_{g_m}(p_7)\sigma_{g_m^{-1}}(p_8) \right>^n_{\mathrm{CFT}^{\bigotimes m}}}, \label{discase1n}
\end{align}
As earlier we utilize the inverse doubling trick for which the above six-point function in a $BCFT_2$ transforms into an three-point function in a $CFT_2$. Finally in an OPE channel followed by the replica limit we obtain the corresponding reflected entropy $S_R(A:B)$ as in \cref{SRdiscasej} with the points $p_4$ and $p_5$ replaced by $p_1$ and $p_8$ respectively.

\begin{figure}[H]
	\centering
	\includegraphics[scale=.9]{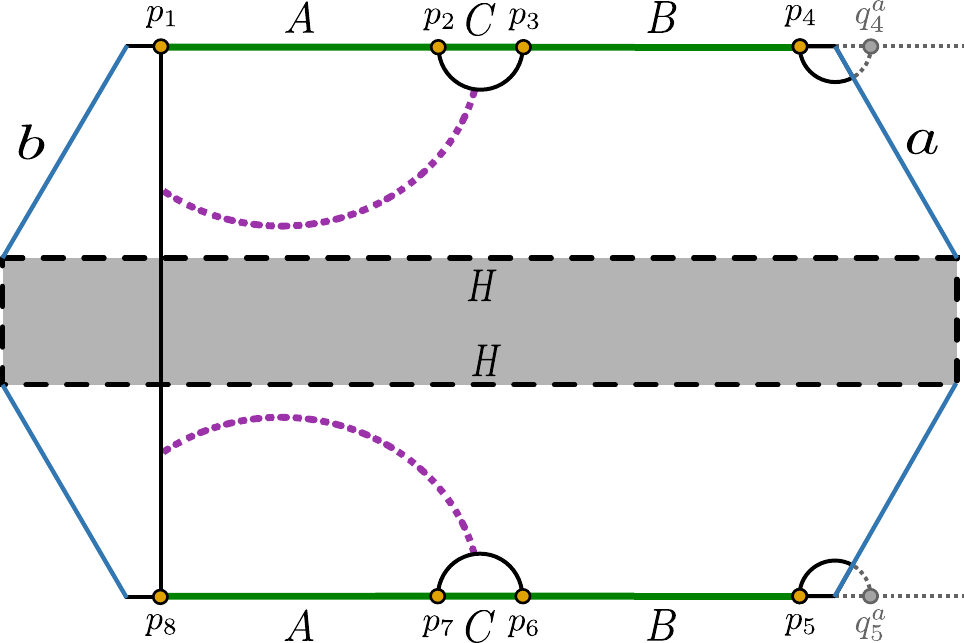}
	\caption{The endpoints of the intervals $A$ and $B$ and the image points in the two $BCFT_2$ copies are identified similarly as in \cref{fig_discase1a}.}\label{fig_discase1n}
\end{figure}

Similar to the reflected entropy, the EWCS for this scenario is analogous to the one described in configuration (j) as seen from the \cref{fig_discase1j,fig_discase1n}. Therefore, the EWCS for this configuration may be given by the \cref{EWdiscasej} with the points $p_4$ and $p_5$ replaced by $p_1$ and $p_8$ respectively.

\subsubsection*{Configuration (o):}
We proceed with another possible contribution to the reflected entropy for the disjoint intervals $A$ and $B$ as depicted in \cref{discase1o}. However in this scenario the R\'enyi reflected entropy reduces to the same expression of configuration (e) after a possible factorization of the correlation function as
\begin{align}
	S_R^{(n,m)}(A:B) &= \frac{1}{1-n}\log \frac{\left< \sigma_{g_A}(p_1)\sigma_{g_A^{-1}}(p_2)\sigma_{g_B}(p_3)\sigma_{g_B^{-1}}(p_4)\sigma_{g_B}(p_5)\sigma_{g_B^{-1}}(p_6)\sigma_{g_A}(p_7)\sigma_{g_A^{-1}}(p_8) \right>_{\mathrm{BCFT}^{\bigotimes mn}}}{\left<\sigma_{g_m}(p_1)\sigma_{g_m^{-1}}(p_2)\sigma_{g_m}(p_3)\sigma_{g_m^{-1}}(p_4)\sigma_{g_m}(p_5)\sigma_{g_m^{-1}}(p_6)\sigma_{g_m}(p_7)\sigma_{g_m^{-1}}(p_8) \right>^n_{\mathrm{BCFT}^{\bigotimes m}}}\notag\\
	&= 2\frac{1}{1-n}\log \frac{\left< \sigma_{g_A^{-1}}(p_2)\sigma_{g_B}(p_3)\sigma_{g_B^{-1}}(p_4)\sigma_{g_B}(q^a_4)\right>_{\mathrm{CFT}^{\bigotimes mn}}}{\left<\sigma_{g_m^{-1}}(p_2)\sigma_{g_m}(p_3)\sigma_{g_m^{-1}}(p_4)\sigma_{g_m}(q^a_4) \right>^n_{\mathrm{CFT}^{\bigotimes m}}}, \label{discase1o}
\end{align}
The factorization procedure of the eight-point function utilized above in a dominant channel is described in \cref{discaseo} of appendix \ref{app_disj}. Consequently, the reflected entropy $S_R(A:B)$ after considering the replica limit of the above equation is then given by the \cref{SRdiscasea} with the finite temperature cross-ratio as in \cref{CRdcase1e}.
\begin{figure}[H]
	\centering
	\includegraphics[scale=.9]{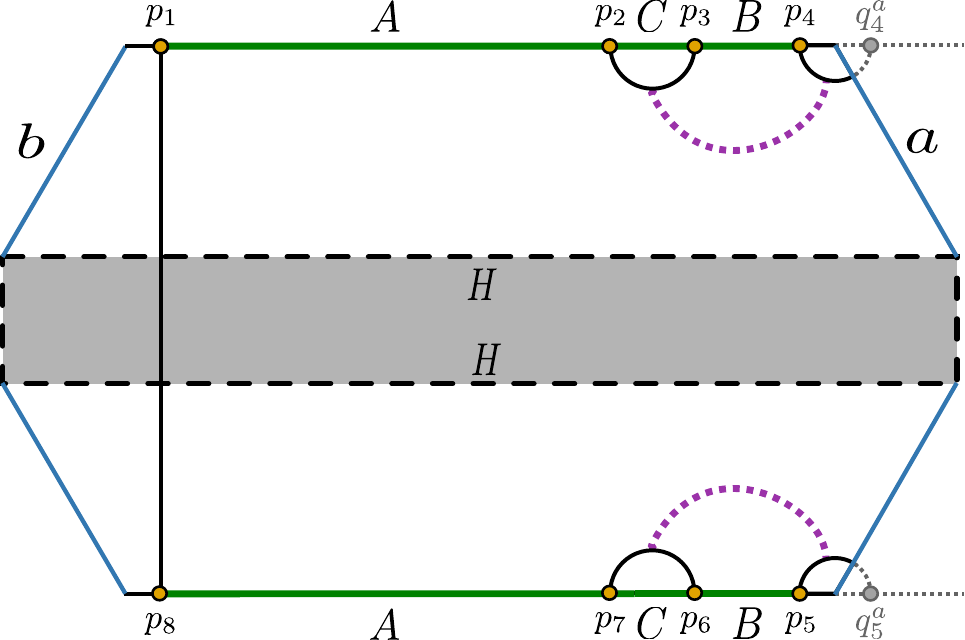}
	\caption{The endpoints of the intervals $A$ and $B$ and the image points in the two $BCFT_2$ copies are identified similarly as in \cref{fig_discase1a}.}\label{fig_discase1o}
\end{figure}
In the bulk dual geometry, the EWCS for this case is described by the violet dotted curve as depicted in \cref{fig_discase1o} which is identical to the one described in \cref{fig_discase1e} of configuration (e). Consequently, the EWCS for this scenario is given by the \cref{EW} with the cross-ratio at a finite temperature in \cref{CrRatio}. However for this configuration, the points in \cref{CrRatio} are identified as $a_1=p_3$, $a_2=p_4$, $b_1=p^a_4$ and $b_2=p_2$ where $q_4^a$ is the image point of $p_4$ with respect to the $a$-boundary in the $BCFT_2$.
As stated in configuration (e), \cref{SRdiscasea,EW} appear to be different with distinct cross-ratios, however one can always transform one expression to another with identical cross-ratios.

\subsubsection*{Configuration (p):}
In this scenario the reflected entropy for the two disjoint intervals $A$ and $B$ is analogous to the one discussed configuration (d). Here, the eight-point function in the corresponding expression for the R\'enyi reflected entropy factorizes in the large central charge limit as shown in \cref{discasep} of appendix \ref{app_disj} and we obtain
\begin{align}
	S_R^{(n,m)}(A:B) &= \frac{1}{1-n}\log \frac{\left< \sigma_{g_A}(p_1)\sigma_{g_A^{-1}}(p_2)\sigma_{g_B}(p_3)\sigma_{g_B^{-1}}(p_4)\sigma_{g_B}(p_5)\sigma_{g_B^{-1}}(p_6)\sigma_{g_A}(p_7)\sigma_{g_A^{-1}}(p_8) \right>_{\mathrm{BCFT}^{\bigotimes mn}}}{\left<\sigma_{g_m}(p_1)\sigma_{g_m^{-1}}(p_2)\sigma_{g_m}(p_3)\sigma_{g_m^{-1}}(p_4)\sigma_{g_m}(p_5)\sigma_{g_m^{-1}}(p_6)\sigma_{g_m}(p_7)\sigma_{g_m^{-1}}(p_8) \right>^n_{\mathrm{BCFT}^{\bigotimes m}}}\notag\\
	&= 2 \frac{1}{1-n}\log \frac{\left<\sigma_{g_A}(q^a_2)\sigma_{g_A^{-1}}(p_2)\sigma_{g_B}(p_3)\sigma_{g_B^{-1}}(q^a_3)\right>_{\mathrm{CFT}^{\bigotimes mn}}}{\left<\sigma_{g_m}(q^a_2)\sigma_{g_m^{-1}}(p_2)\sigma_{g_m}(p_3)\sigma_{g_m^{-1}}(q^a_3)\right>^n_{\mathrm{CFT}^{\bigotimes m}}}, \label{discase1p}
\end{align}
Therefore, the reflected entropy $S_R(A:B)$ is given by the \cref{srcaseb} where the finite temperature cross-ratio $z$ is given by the \cref{CrRatio} with $a_1$, $a_2$, $b_1$, $b_2$ replaced by the points $p_3$, $q_3^a$, $q_2^a$ and $p_2$ respectively.

\begin{figure}[H]
	\centering
	\includegraphics[scale=.9]{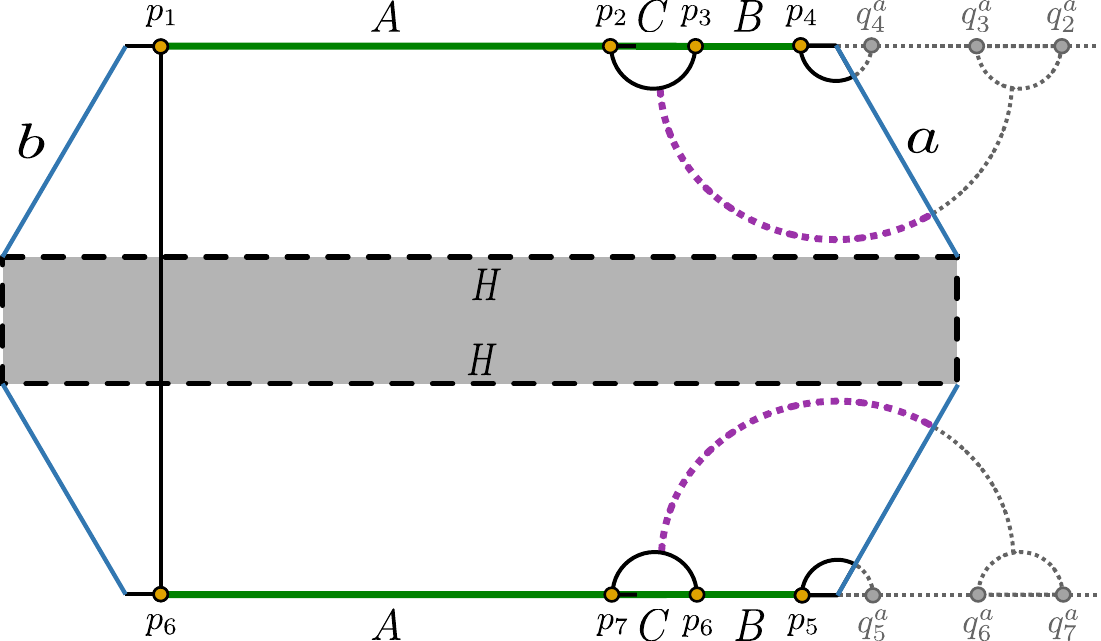}
	\caption{The endpoints of the intervals $A$ and $B$ and the image points in the two $BCFT_2$ copies are identified similarly as in \cref{fig_discase1a,fig_discase1e}.}\label{fig_discase1p}
\end{figure}

We refer to configuration (d) for the computation of the EWCS in this scenario since in both the configurations, the EWCS are identical as seen from the \cref{fig_discase1d,fig_discase1p}. Therefore, the EWCS for this case may expressed as in \cref{Ewcaseb} where once again the finite temperature cross-ratio $z$ is given by the \cref{CrRatio} with $a_1$, $a_2$, $b_1$, $b_2$ replaced by the points $p_3$, $q_3^a$, $q_2^a$ and $p_2$ respectively.

\subsubsection*{Configuration (q):}
Finally the last scenario corresponds to the reflected entropy for the two disjoint intervals $A$ and $B$ which is similar to configuration (d). The factorization of the eight-point function in the corresponding expression for the R\'enyi reflected entropy is shown in \cref{discaseq} of appendix \ref{app_disj} in the large central charge limit. The final expression for the R\'enyi reflected entropy is then obtained as
\begin{align}
	S_R^{(n,m)}(A:B) &= \frac{1}{1-n}\log \frac{\left< \sigma_{g_A}(p_1)\sigma_{g_A^{-1}}(p_2)\sigma_{g_B}(p_3)\sigma_{g_B^{-1}}(p_4)\sigma_{g_B}(p_5)\sigma_{g_B^{-1}}(p_6)\sigma_{g_A}(p_7)\sigma_{g_A^{-1}}(p_8) \right>_{\mathrm{BCFT}^{\bigotimes mn}}}{\left<\sigma_{g_m}(p_1)\sigma_{g_m^{-1}}(p_2)\sigma_{g_m}(p_3)\sigma_{g_m^{-1}}(p_4)\sigma_{g_m}(p_5)\sigma_{g_m^{-1}}(p_6)\sigma_{g_m}(p_7)\sigma_{g_m^{-1}}(p_8) \right>^n_{\mathrm{BCFT}^{\bigotimes m}}}\notag\\
	&= \frac{1}{1-n}\log \frac{\left<\sigma_{g_A^{-1}}(p_2)\sigma_{g_B}(p_3)\sigma_{g_B^{-1}}(p_6)\sigma_{g_A}(p_7)\right>_{\mathrm{CFT}^{\bigotimes mn}}}{\left<\sigma_{g_m^{-1}}(p_2)\sigma_{g_m}(p_3)\sigma_{g_m^{-1}}(p_6)\sigma_{g_m}(p_7)\right>^n_{\mathrm{CFT}^{\bigotimes m}}}, \label{discase1q}
\end{align}
Considering the replica limit, the reflected entropy for the disjoint intervals $A$ and $B$ is given by the \cref{SRdisjcc}.

\begin{figure}[H]
	\centering
	\includegraphics[scale=.9]{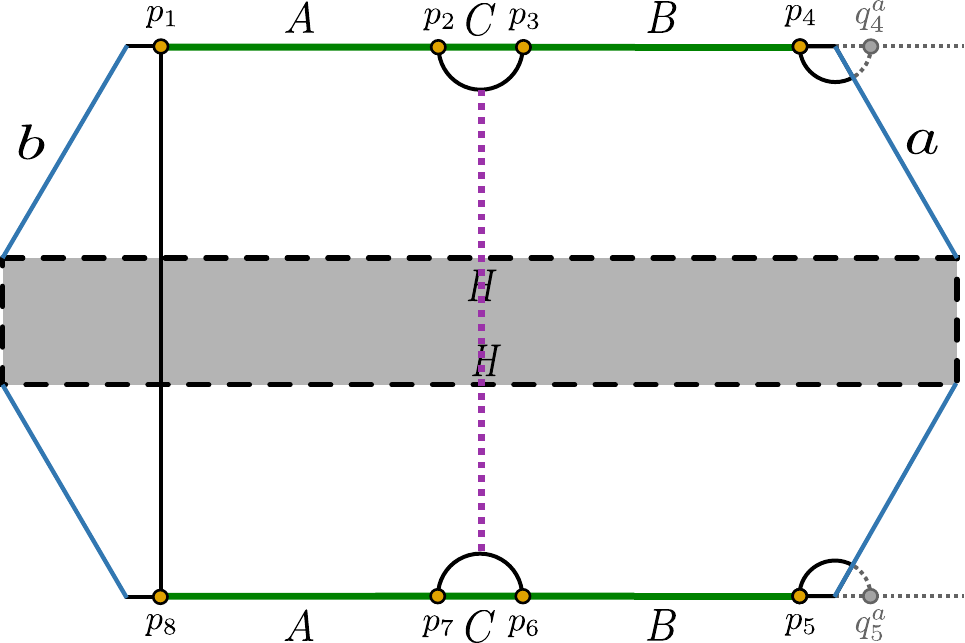}
	\caption{The endpoints of the intervals $A$ and $B$ and the image points in the two $BCFT_2$ copies are identified similarly as in \cref{fig_discase1a}.}\label{fig_discase1q}
\end{figure}

The EWCS for this case is identical to the one described in configuration (c) as seen from the \cref{fig_discase1c,fig_discase1q}. Consequently the corresponding expression for the EWCS may be given by the \cref{SRdiscase1c}.


\section{Markov gap in the braneworld model}\label{MarkovInGeng}
In this section we explore an intriguing issue of the holographic Markov gap described by the difference between the reflected entropy and the mutual information in the context of the braneworld model under consideration. For this purpose we first consider the bipartite mixed state of two adjacent intervals in the two copies of the bath $BCFT_2$s to analyze the behaviour of the Markov gap for various configurations involving the interval sizes and time. Subsequently, the above analysis is repeated for the mixed state of two disjoint intervals in the bath $BCFT_2$s. We observe several phases for the reflected entropy while varying the interval sizes and time with the dominant contributions arising from the different configurations described in \cref{SrEwInGeng} \footnote{Some specific configurations of the reflected entropy (i.e. configurations (j), (k), (m) and (n) for both adjacent and disjoint intervals discussed in \cref{SrEwInGeng}) do not appear in the following plots of the reflected entropy depending upon the size of the bath $BCFT_2$s considered in this paper. The reason behind this is that there exists no extremal solution for the EWCS ending on the HM surface in the dual bulk geometry for such a choices of bath sizes. However, for some special choices of the bath sizes these configurations may appear.}. Similarly the profile for the mutual information also corresponds to various phases due to the various structures of the RT surfaces supported by the intervals with varying sizes and time. For both the scenarios involving the adjacent and disjoint intervals, we observe the holographic Markov gap as expected from \cite{Hayden:2021gno} for multipartite mixed states.

\subsection{Adjacent intervals}
To begin with, we consider various configurations involving two adjacent intervals and comment on the corresponding holographic Markov gap in each case. In this context, the holographic mutual information for two adjacent intervals $A\equiv [p_1,p_2]\cup[p_6,p_5]$ and $B\equiv[p_2+\epsilon,p_3]\cup[p_5+\epsilon,p_4]$ in the bath $BCFT_2$s may be given by the following eq.,
\begin{align}\label{MI_adj}
	I(A:B)&= S(A)+S(B)-S(A\cup B),
\end{align}
where $S(X)$ is the entanglement entropy for a interval $X$ in the $BCFT_2$s described in \cite{Afrasiar:2022ebi}.

\subsubsection*{$\bm{(i)}$ Full system ($\bm{A\cup B}$) fixed, common point varied}\label{adjMG1}
\begin{figure}[H]
	\centering
	\includegraphics[scale=.6]{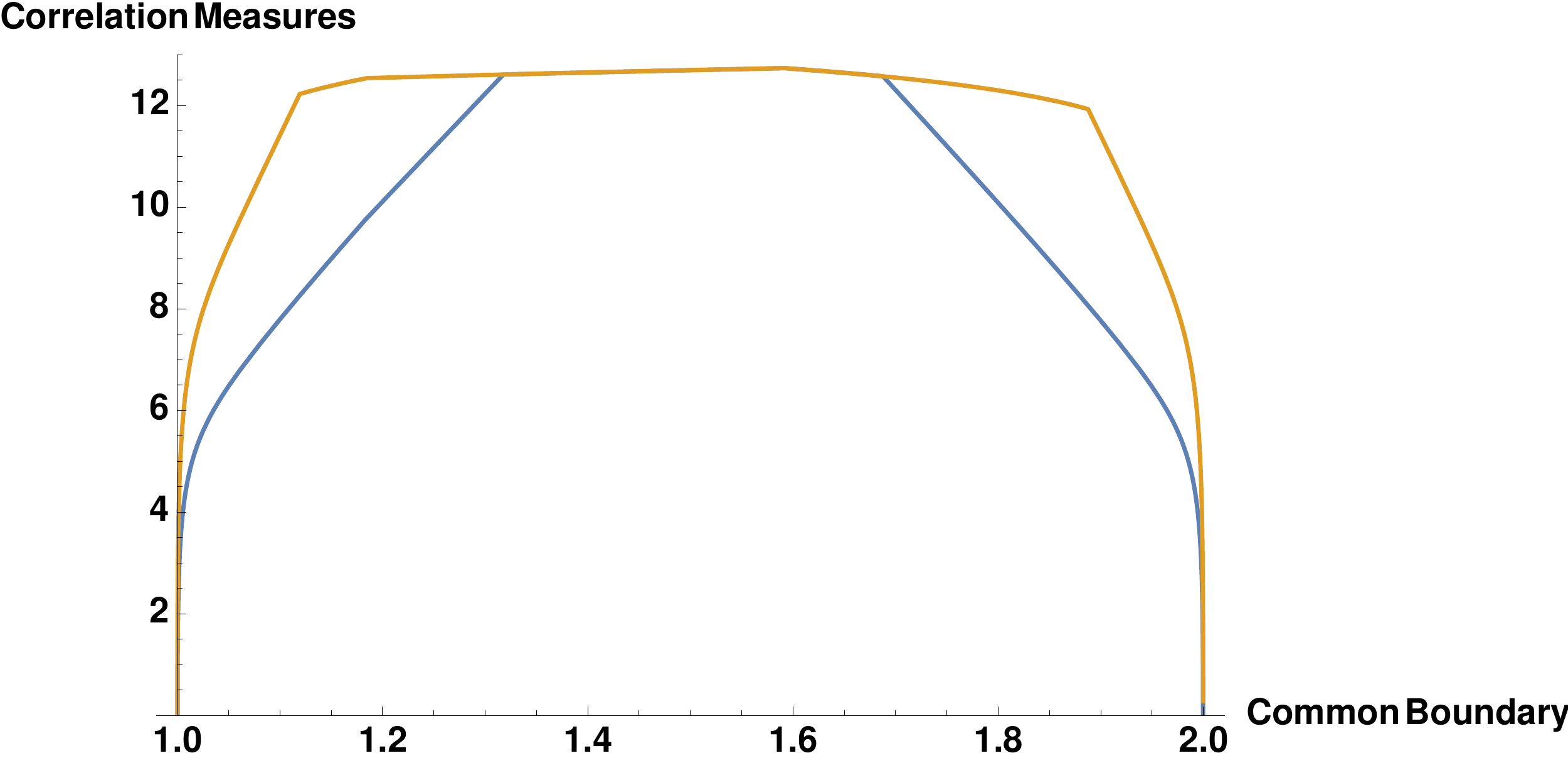}
	\caption{Plot of correlation measures namely reflected entropy (yellow) and mutual information (blue) for two adjacent intervals $A$ and $B$ while the common boundary of the two intervals is shifted from one end point to the other in the bath $BCFT_2$s. In the y-axis, the correlation measures are piloted after scaling over the central charge $c$ of the $BCFT_2$s. Here $r_I=1$, $r_O=2$, $\epsilon=.0001$, $\beta=.1$, $c=500$, $t=.15$, $S_{bdyb}=875$ and $S_{bdya}=850$. }\label{mg_adj_1}
\end{figure}

In the first scenario, we consider the two adjacent intervals $A\equiv[r_I+\epsilon,r]$ and $B\equiv[r,r_O-\epsilon]$ to cover the entire bath $BCFT_2$ regions on a constant time slice and vary the common point $r$ between them. In this case, we compare the corresponding reflected entropy to the holographic mutual information to analyse the characteristics of the Markov gap for various phases
as depicted in the \cref{mg_adj_1}.

In the above figure, the reflected entropy for the adjacent intervals $A$ and $B$ receives dominant contributions from various configurations described in \cref{SrEwInGeng} while shifting the common point between them. Particularly, we observe that configurations (a), (b), (c), (d) and (e) of adjacent intervals in \cref{SrEwInGeng} respectively dominates the consecutive phases of the reflected entropy curve in \cref{mg_adj_1}.

\subsubsection*{$\bm{(ii)}$ interval $\bm{A}$ fixed, $\bm{B}$ varied}\label{adjMG2}
\begin{figure}[H]
	\centering
	\includegraphics[scale=.9]{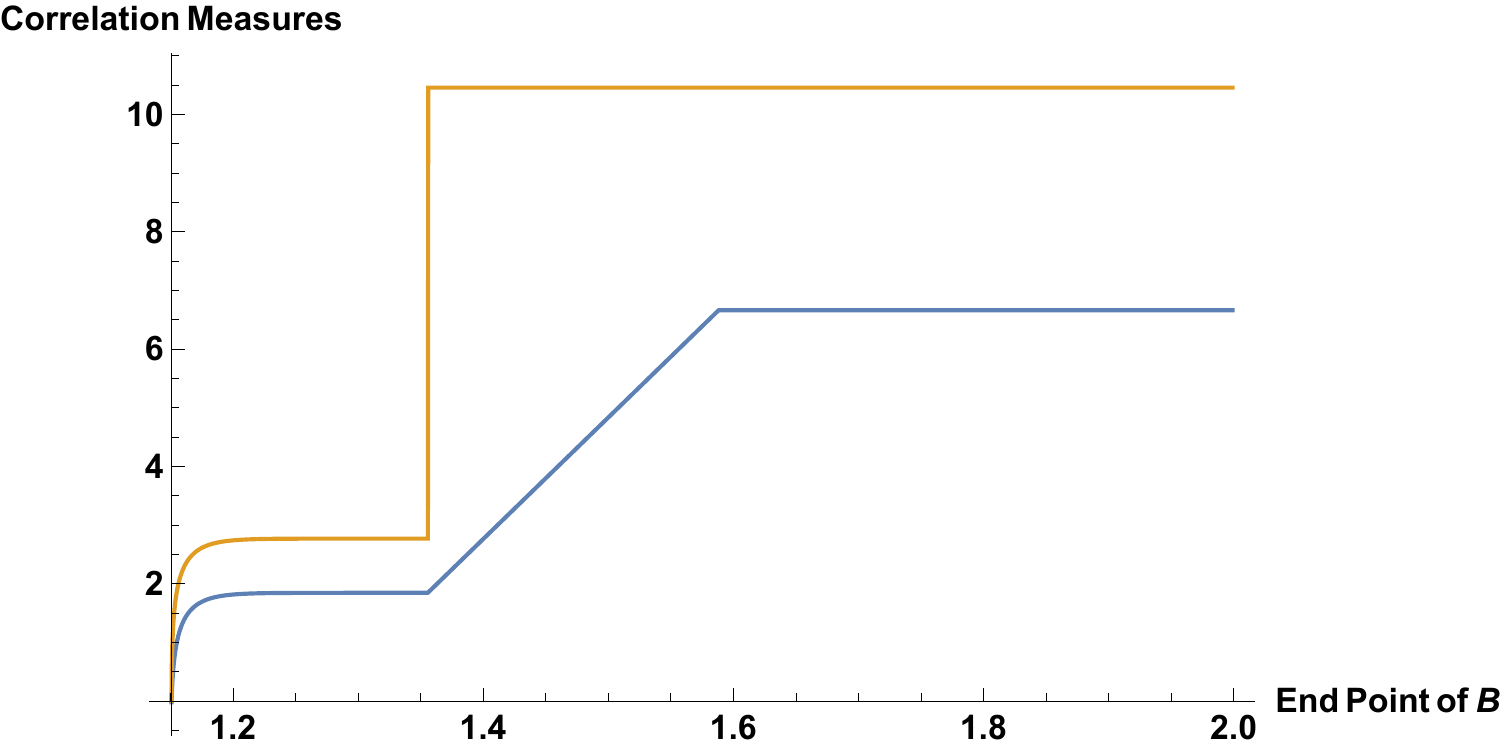}
	\caption{Plot of correlation measures i.e. reflected entropy (yellow) and mutual information (blue) for two adjacent intervals $A$ and $B$ while the endpoint of the interval $B$ is increased in the bath $BCFT_2$s. In the y-axis, the correlation measures are piloted after scaling over the central charge $c$ of the $BCFT_2$s.  Here $r_I=1$, $r_O=2$, $\epsilon=.001$, $\beta=.1$, $c=500$, $t=.15$, $r_1=1.15$, $S_{bdyb}=875$ and $S_{bdya}=850$.}\label{mg_adj_2}
\end{figure}

Next we consider the size of the interval $A=[r_I+\epsilon, r_1]$ to be fixed while varying the size of $B=[r_1, r]$ by shifting the point $r$ on a constant time slice. We then follow the same procedure as in the previous scenario comparing the corresponding holographic reflected entropy with the holographic mutual information. As depicted in \cref{mg_adj_2}, we observe the holographic Markov gap for different phases.

As earlier, the reflected entropy for the adjacent intervals $A$ and $B$ in the scenario receives dominant contributions from various configurations described in \cref{SrEwInGeng} while increasing the size of the interval $B$. In this case, we observe that configurations (f) and (g) of adjacent intervals in \cref{SrEwInGeng} respectively dominates the consecutive phases of the reflected entropy curve in \cref{mg_adj_2}.

\subsubsection*{$\bm{(iii)}$ intervals $\bm{A}$ and $\bm{B}$ fixed, time varied} \label{adjMG3}
\begin{figure}[H]
	\centering
	\includegraphics[scale=.7]{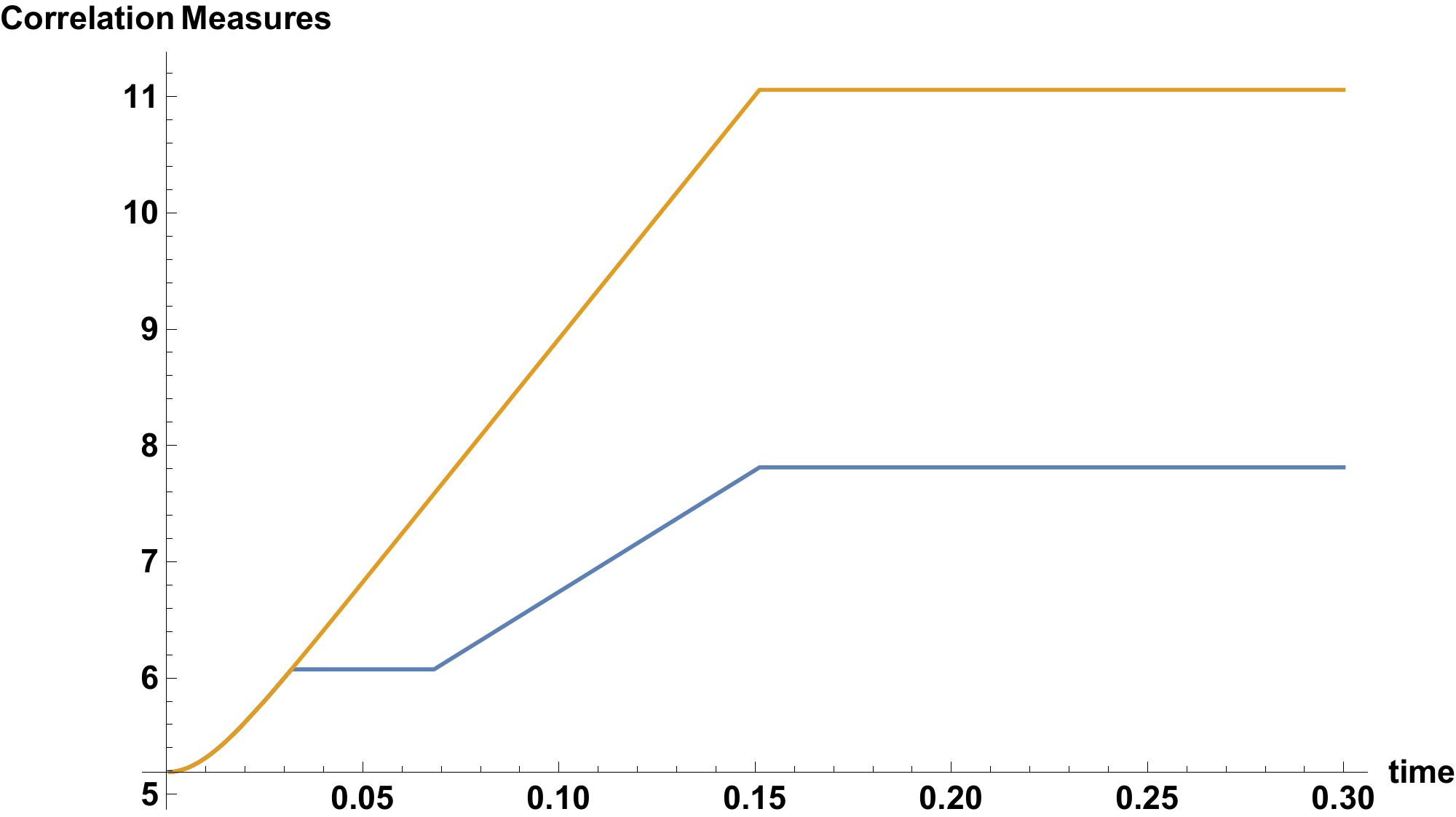}
	\caption{Plot of correlation measures i.e. reflected entropy (yellow) and mutual information (blue) for two adjacent intervals $A$ and $B$ in the dual $BCFT_2$s with increasing time. In the y-axis, the correlation measures are piloted after scaling over the central charge $c$ of the $BCFT_2$s.  Here $r_I=1$, $r_O=2$, $\epsilon=.001$, $\beta=.1$, $c=500$, $S_{bdyb}=875$ and $S_{bdya}=850$.}\label{mg_adj_3}
\end{figure}

Finally we fix the interval sizes of the two adjacent intervals $A=[r_I+\epsilon, 1.15]$, $B=[1.15, r_O-\epsilon]$ and vary the time $t$ to compare the holographic reflected entropy and the mutual information as depicted in \cref{mg_adj_3}. Once again we observe the holographic Markov gap for the various phases with evolving time.

Once again, the reflected entropy for the adjacent intervals $A$ and $B$ in the above figure receives dominant contributions from various configurations described in \cref{SrEwInGeng} with increasing time $t$. In this scenario, we observe the configurations (c) and (b) of adjacent intervals in \cref{SrEwInGeng} respectively to dominate the consecutive phases of the reflected entropy curve in \cref{mg_adj_3}.

\subsection{Disjoint intervals}
We now consider two disjoint intervals in the two copies of the bath $BCFT_2$s and demonstrate the holographic Markov gap for various configurations involving the interval sizes and time. In this connection, the holographic mutual information for the two disjoint intervals $A\equiv[p_1,p_2]\cup[p_8,p_7]$ and $B\equiv[p_3,p_4]\cup[p_6,p_5]$ with $C\equiv[p_2,p_3]\cup[p_7,p_6]$ sandwiched between them may be expressed as
\begin{align}\label{MI_disj}
	I(A:B)&= S(A)+S(B)-S(A\cup B\cup C)-S(C),
\end{align}
where $S(X)$ is again the entanglement entropy for a interval $X$ in the bath $BCFT_2$s described in  \cite{Afrasiar:2022ebi}. Let us now consider the different scenarios for the two disjoint intervals and compare the holographic reflected entropy  with the mutual information.

\subsubsection*{$\bm{(i)}$ interval $\bm{A}$ fixed, $\bm{C}$ varied}\label{disjMG1}
\begin{figure}[H]
	\centering
	\includegraphics[scale=.7]{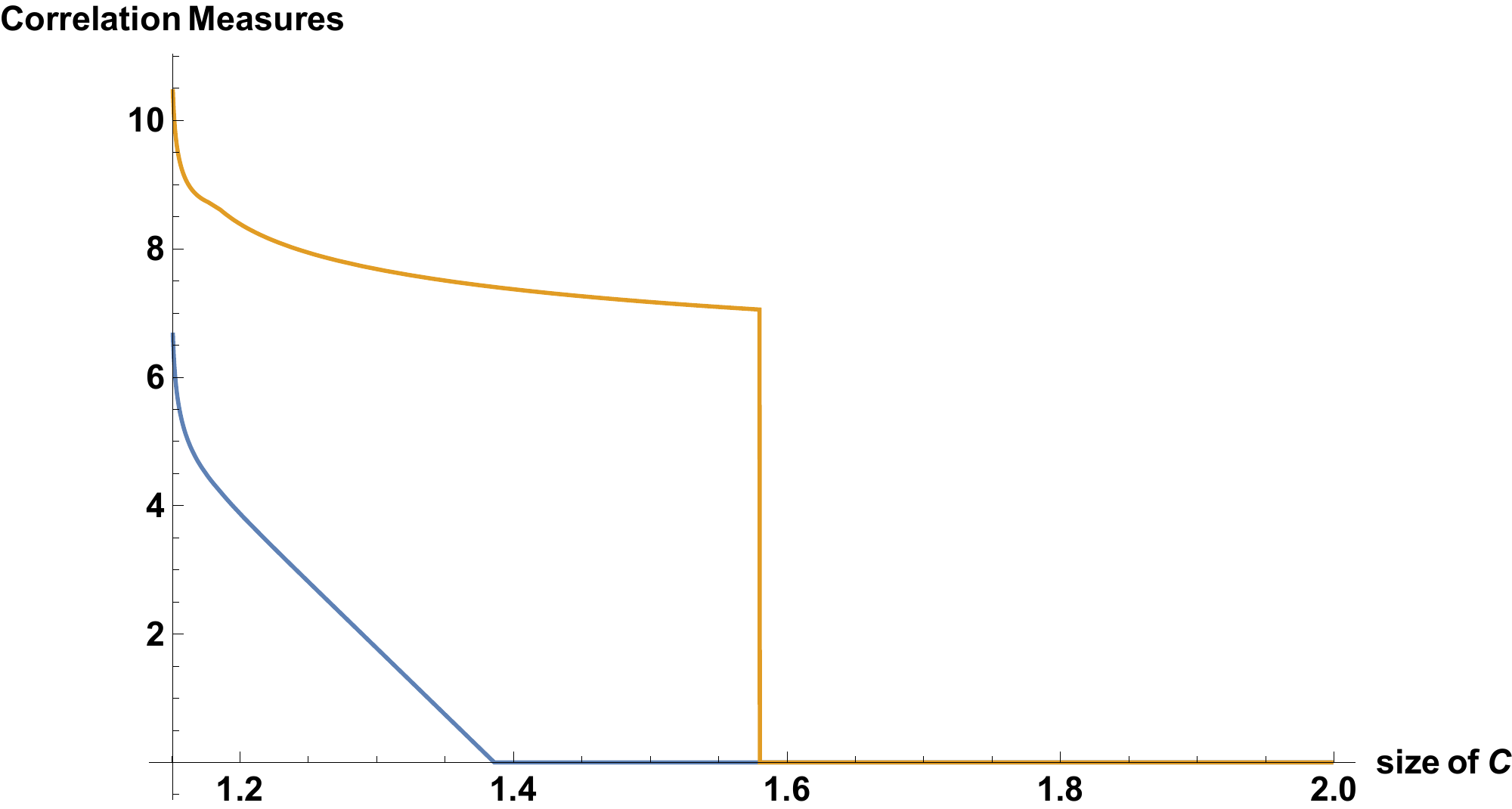}
	\caption{Plot of correlation measures i.e. reflected entropy (yellow) and mutual information (blue) for two disjoint intervals $A$ and $B$ while the endpoint of the interval $C=[r_1,r]$ is increased by shifting the point $r$ in the bath $BCFT_2$s. In the y-axis, the correlation measures are piloted after scaling over the central charge $c$ of the $BCFT_2$s.  Here $r_I=1$, $r_O=2$, $\epsilon=.001$, $\beta=.1$, $c=500$, $t=.15$, $r_1=1.15$, $S_{bdyb}=875$ and $S_{bdya}=850$.}\label{mg_disj_1}
\end{figure}

We begin with the case where the size of the interval $A=[r_I+\epsilon, r_1]$ is fixed and we gradually increase the size of the interval $C=[r_1,r]$ between $A$ and $B$ on a constant time slice. In this scenario we illustrate the holographic Markov gap by comparing the corresponding reflected entropy obtained in \cref{SrEwInGeng} to the mutual information computed through the \cref{MI_disj}.  We observe the holographic Markov gap for different phases as depicted in \cref{mg_disj_1}.

As discussed for the adjacent scenario, the reflected entropy for the disjoint intervals $A$ and $B$ in the above figure receives dominant contributions from various configurations described in \cref{SrEwInGeng} while increasing the size of $C$. We note that the configurations (a) and (c) of disjoint intervals in \cref{SrEwInGeng} respectively dominate the consecutive two phases of the reflected entropy curve in \cref{mg_disj_1} and finally in the last phase the entanglement wedges of $A$ and $B$ become disconnected which results into zero reflected entropy for the two intervals in question.

\subsubsection*{$\bm{(ii)}$ intervals $\bm{A}$ and $\bm{C}$ fixed, $\bm{B}$ varied}\label{disjMG2}
\begin{figure}[H]
	\centering
	\includegraphics[scale=.7]{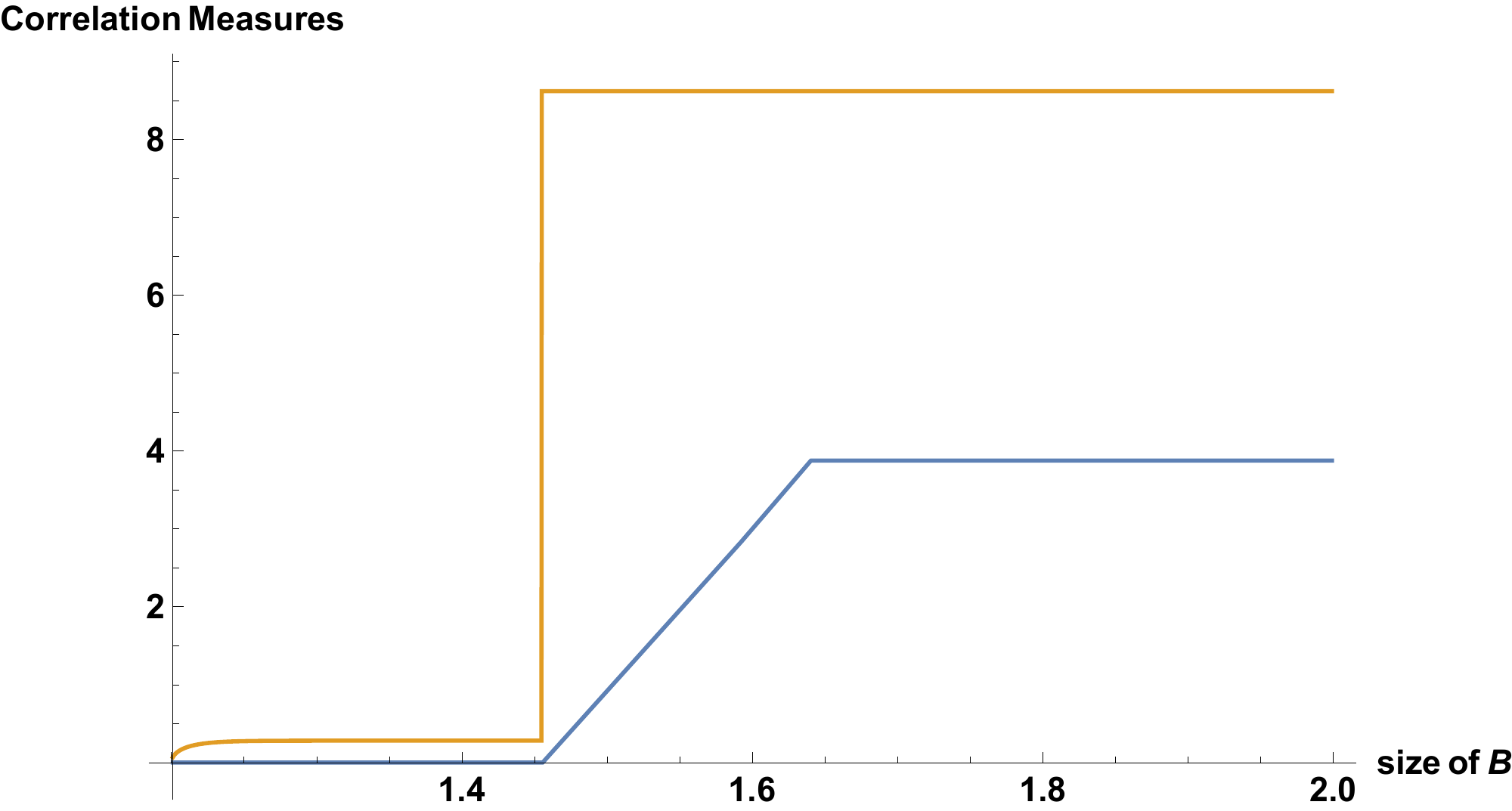}
	\caption{Plot of correlation measures i.e. reflected entropy (yellow) and mutual information (blue) for two disjoint intervals $A$ and $B$ while the endpoint of the interval $B$ is increased in the bath $BCFT_2$s. In the y-axis, the correlation measures are piloted after scaling over the central charge $c$ of the $BCFT_2$s.  Here $r_I=1$, $r_O=2$, $\epsilon=.001$, $\beta=.1$, $c=500$, $r_1=1.15$, $r_2=1.2$, $S_{bdyb}=875$ and $S_{bdya}=850$.}\label{mg_disj_2}
\end{figure}

We proceed to the second scenario where we fix the interval sizes of $A=[r_I+\epsilon,r_1]$ and $C=[r_1,r_2]$ on a constant time slice. We show the holographic Markov gap from \cref{mg_disj_2} where the holographic reflected entropy and the mutual information are plotted together while increasing the size of the interval $B[r_2,r]$ by shifting the point $r$. As earlier, we observe the holographic Markov gap for the various phases in this scenario.

As earlier, different configurations described in \cref{SrEwInGeng} dominate the reflected entropy curve for the disjoint intervals $A$ and $B$ depicted in \cref{mg_disj_2} while increasing the size of $C$. Particularly, the consecutive phases of the corresponding reflected entropy curve receives dominant contributions from configurations (f) and (i) respectively of disjoint intervals in \cref{SrEwInGeng}.

\subsubsection*{$\bm{(iii)}$ intervals $\bm{A}$, $\bm{B}$ and $\bm{C}$ fixed, time varied}\label{disjMG3}
\begin{figure}[H]
	\centering
	\includegraphics[scale=.7]{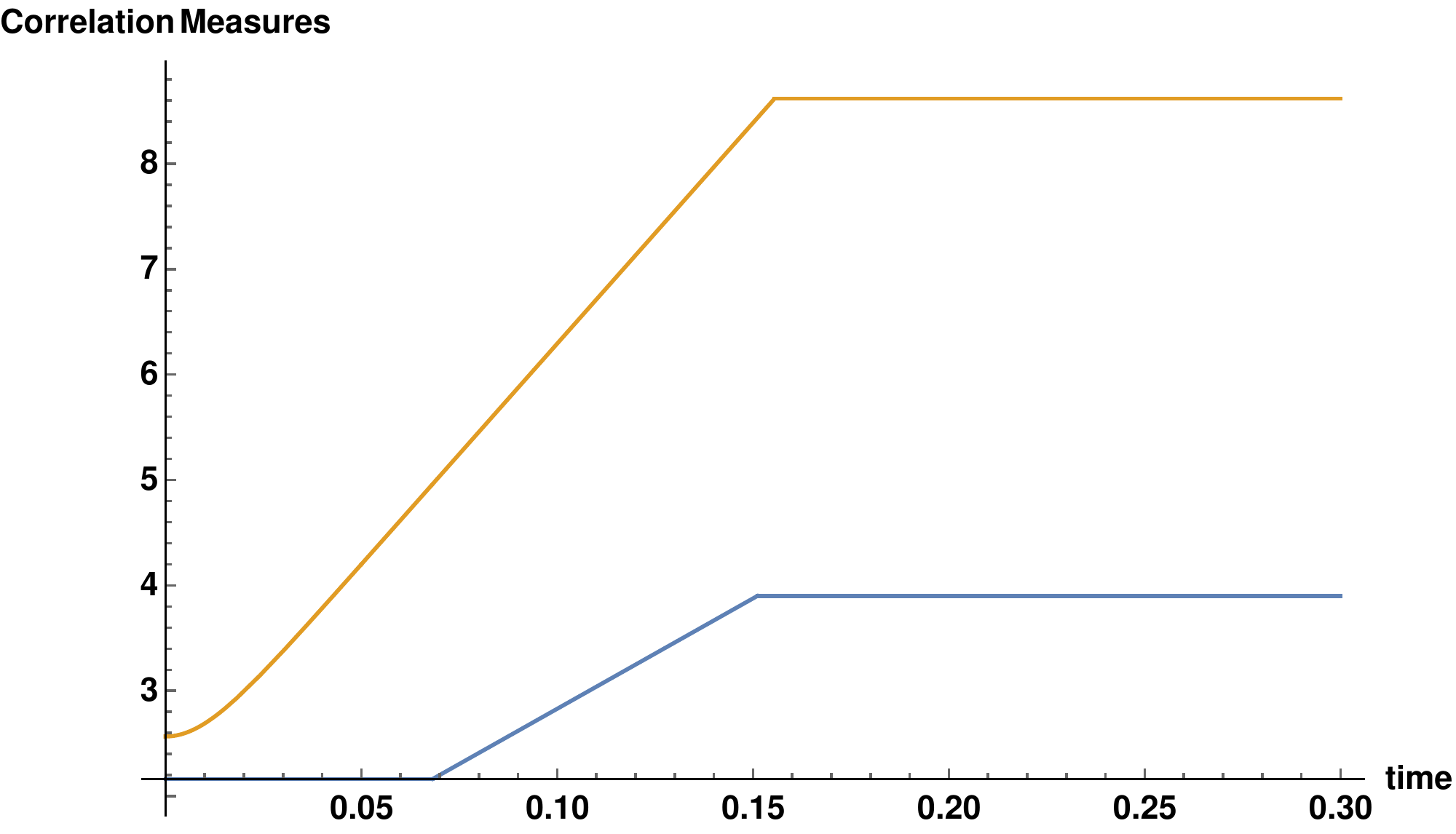}
	\caption{Plot of correlation measures i.e. reflected entropy (yellow) and mutual information (blue) for two disjoint intervals $A$ and $B$ with fixed sizes in the bath $BCFT_2$s while increasing the time $t$. In the y-axis, the correlation measures are piloted after scaling over the central charge $c$ of the $BCFT_2$s.  Here $r_I=1$, $r_O=2$, $\epsilon=.001$, $\beta=.1$, $c=500$, $r_1=1.15$, $r_2=1.2$, $S_{bdyb}=875$ and $S_{bdya}=850$.}\label{mg_disj_3}
\end{figure}

Finally we consider the intervals $A=[r_I+\epsilon,r_1]$, $B=[r_2,r_O-\epsilon]$ and $C=[r_1,r_2]$ with fixed sizes and study the behaviour of the holographic Markov gap from the comparison between the holographic reflected entropy and the mutual information. In \cref{mg_disj_3}, we plot the holographic reflected entropy and the mutual information with increasing time $t$ and observe the holographic Markov gap for different phases as earlier.

Once again, the reflected entropy for the disjoint intervals $A$ and $B$ in the above figure receives dominant contributions from various configurations described in \cref{SrEwInGeng} with increasing time $t$. In this scenario, we observe that the configurations (l) and (q) dominate the phase-1 of the reflected entropy curve, whereas phase-2 receives dominant contribution from the configuration (a) of disjoint intervals described in \cref{SrEwInGeng}.

\newpage
\section{Summary and Discussion}\label{discussion}
To summarize, we have investigated the reflected entropy for various bipartite mixed states at finite temperatures for the communicating black holes configuration in a KR braneworld geometry. For this purpose we considered two copies of $BCFT_2$s at finite temperatures where each $BCFT_2$ was defined on a manifold with two distinct boundaries. The holographic dual of this construction was described by a bulk eternal $BTZ$ black hole truncated by two KR branes. In this scenario, two dimensional black holes were induced on these KR branes from the higher dimensional eternal BTZ black hole. These induced black holes communicate with each other through the shared bath regions described by the $BCFT_2$s. Note that the bath $BCFT_2$s together with one of the KR branes appear to be non-gravitating from the perspective of the other brane. 

We have computed the reflected entropy for various bipartite mixed state configurations of two adjacent and disjoint intervals in the bath $BCFT_2$s for the above scenario. It was observed that for the field theory computations of the reflected entropy involved different possible dominant channels for the multipoint twist correlators in the large central charge limit. The reflected entropy for the mixed state configuration of two adjacent intervals was first computed  in this scenario. This involved the dominant channels and the corresponding factorization of the twist correlators in the large central charge limit. Subsequently the replica technique described in \cite{Dutta:2019gen} were utilized to compute the reflected entropy for the two adjacent intervals under consideration in the above limit. The corresponding EWCS for the different configurations of the two adjacent intervals in the $BCFT_2$s was then computed from the bulk eternal BTZ black hole geometry. Our results clearly verified the holographic duality between the reflected entropy and the EWCS.  Next we followed a similar analysis for the reflected entropy of two disjoint intervals in the bath $BCFT_2$s and  once again obtained the contributions to the reflected entropy arising from the dominant channels for the multi-point twist correlators in the large central charge limit. Subsequently as earlier, we have substantiated these field theory results for the reflected entropy from explicit computations of the corresponding EWCSs in the bulk eternal BTZ black hole geometry verifying the holographic duality mentioned earlier.

As demonstrated in \cite{Hayden:2021gno}, a holographic Markov gap between the reflected entropy and the mutual information appears for multipartite mixed states in the context of the $AdS_3/CFT_2$ correspondence. Keeping this in perspective we have compared the holographic reflected entropies for the bipartite mixed state configurations considered in our work with the corresponding mutual information. Our results clearly demonstrate the holographic Markov gap for all the  mixed state configurations in question and hence provides a strong substantiation of this issue in the island scenario for the bulk braneworld model considered here. Furthermore we have also analyzed the behaviour of the Markov gap for varying interval sizes and time. It is observed that for all the scenarios the holographic Markov gap is non zero. Interestingly we also observe that the Markov gap is non zero even when there are no non-trivial bulk end points of the EWCS which stands in contradiction with its standard geometrical interpretation indicating that there is possibly more to this issue that needs our understanding. 

There are various interesting future directions that can be investigated to provide a clear understanding of the structure of mixed state entanglement in Hawking radiation. One such immediate issue would be the extension of the study of the present article in another brane world geometry described in \cite{Balasubramanian:2021xcm, Afrasiar}. One may also generalize our study to multipartite correlations to elucidate the characteristics of the holographic Markov gap. Following the developments in \cite{Afrasiar:2022ebi, Lu:2022fxb}, it will also be extremely fascinating to explore other mixed state correlation measures in the context of the model utilized in the present article to obtain further insights into their corresponding island constructions and ``Markov gap". We would like to return to these exciting issues in the near future.

\section{Acknowledgement}
The research work of JKB is supported by the grant 110-2636-M-110-008 by the National Science and Technology Council (NSTC) of Taiwan. The work of GS is partially supported by the Dr. Jagmohan Garg Chair Professor position at the Indian Institute of Technology Kanpur, India.

\begin{appendices}

\section{Adjacent intervals}\label{app_adj} 
We consider different configurations of two adjacent intervals at a finite temperature and show how the higher-point twist correlators in the R\'enyi reflected entropy expression factorize into lower-point twist correlator in the large central charge limit for various dominant channels. These factorizations of the twist correlators listed below  are utilized in the reflected entropy computation in \cref{SrEwInGeng}.

\subsubsection*{Configuration (a)}

\begin{align}
	S_R^{(n,m)}(A:B) &= 2\frac{1}{1-n}\log \frac{\left< \sigma_{g_A}(p_1)\sigma_{g_Bg_A^{-1}}(p_2)\sigma_{g_B^{-1}}(p_3) \right>_{\mathrm{BCFT}^{\bigotimes mn}}}{\left<\sigma_{g_m}(p_1)\sigma_{g_m^{-1}}(p_3) \right>^n_{\mathrm{BCFT}^{\bigotimes m}}}\notag\\
	&= 2\frac{1}{1-n}\log \frac{\left< \sigma_{g_A}(p_1)\sigma_{g_Bg_A^{-1}}(p_2)\right>_{\mathrm{BCFT}^{\bigotimes mn}}\left<\sigma_{g_B^{-1}}(p_3) \right>_{\mathrm{BCFT}^{\bigotimes mn}}}{\left(\Big<\sigma_{g_m}(p_1)\Big>_{\mathrm{BCFT}^{\bigotimes m}}\left<\sigma_{g_m^{-1}}(p_3) \right>_{\mathrm{BCFT}^{\bigotimes m}}\right)^n}\notag\\
	&= 2\frac{1}{1-n}\log\frac{\left< \sigma_{g_A}(p_1)\sigma_{g_A^{-1}}(q^b_{1})\sigma_{g_Bg_A^{-1}}(p_2)\right>_{\mathrm{CFT}^{\bigotimes mn}}\left<\sigma_{g_B^{-1}}(p_3) \sigma_{g_B}(q^a_{3})\right>_{\mathrm{CFT}^{\bigotimes mn}}}{\left(\left<\sigma_{g_m}(p_1)\sigma_{g_m^{-1}}(q^b_{1})\right>_{\mathrm{CFT}^{\bigotimes m}}\left<\sigma_{g_m^{-1}}(p_3)\sigma_{g_m^{1}}(q^a_{3}) \right>_{\mathrm{CFT}^{\bigotimes m}}\right)^n}. \label{adjcasea}
\end{align}

\subsubsection*{Configuration (b)}
\begin{align}
	S_R^{(n,m)}(A:B) &= 2\frac{1}{1-n}\log \frac{\left< \sigma_{g_A}(p_1)\sigma_{g_Bg_A^{-1}}(p_2)\sigma_{g_B^{-1}}(p_3) \right>_{\mathrm{BCFT}^{\bigotimes mn}}}{\left<\sigma_{g_m}(p_1)\sigma_{g_m^{-1}}(p_3) \right>^n_{\mathrm{BCFT}^{\bigotimes m}}}\notag\\
	&= 2\frac{1}{1-n}\log \frac{\Big< \sigma_{g_A}(p_1)\Big>_{\mathrm{BCFT}^{\bigotimes mn}}\left<\sigma_{g_Bg_A^{-1}}(p_2)\right>_{\mathrm{BCFT}^{\bigotimes mn}}\left<\sigma_{g_B^{-1}}(p_3) \right>_{\mathrm{BCFT}^{\bigotimes mn}}}{\Big<\sigma_{g_m}(p_1)\Big>^n_{\mathrm{BCFT}^{\bigotimes m}}\left<\sigma_{g_m^{-1}}(p_3) \right>^n_{\mathrm{BCFT}^{\bigotimes m}}}\notag\\
	&= 2\frac{1}{1-n}\log\frac{\left< \sigma_{g_A}(p_1)\sigma_{g_A^{-1}}(q^b_{1})\right>_{\mathrm{CFT}^{\bigotimes mn}}\left<\sigma_{g_B^{-1}}(p_3) \sigma_{g_B}(q^a_{3})\right>_{\mathrm{CFT}^{\bigotimes mn}}}{\left<\sigma_{g_m}(p_1)\sigma_{g_m^{-1}}(q^b_{1})\right>^n_{\mathrm{CFT}^{\bigotimes m}}\left<\sigma_{g_m^{-1}}(p_3)\sigma_{g_m^{1}}(q^a_{3}) \right>^n_{\mathrm{CFT}^{\bigotimes m}}}\notag\\
	&~~~~~~~~~~~~~~~~~~~~~~~~~~~+ 2\frac{1}{1-n}\log\left(\left<\sigma_{g_Bg_A^{-1}}(p_2)\sigma_{g_Ag_B^{-1}}(q^b_{2})\right>_{\mathrm{CFT}^{\bigotimes mn}}\right) \label{adjcaseb}
\end{align}

\subsubsection*{Configuration (c)}
\begin{align}
	S_R^{(n,m)}(A:B) &= \frac{1}{1-n}\log \frac{\left< \sigma_{g_A}(p_1)\sigma_{g_Bg_A^{-1}}(p_2)\sigma_{g_B^{-1}}(p_3)\sigma_{g_B}(p_4)\sigma_{g_Ag_B^{-1}}(p_5)\sigma_{g_A^{-1}}(p_6) \right>_{\mathrm{BCFT}^{\bigotimes mn}}}{\left<\sigma_{g_m}(p_1)\sigma_{g_m^{-1}}(p_3)\sigma_{g_m}(p_4)\sigma_{g_m^{-1}}(p_6) \right>^n_{\mathrm{BCFT}^{\bigotimes m}}}\notag\\
	&= \frac{1}{1-n}\log \frac{\Big< \sigma_{g_A}(p_1)\Big>_{\mathrm{BCFT}^{\bigotimes mn}}\left<\sigma_{g_B^{-1}}(p_3) \right>_{\mathrm{BCFT}^{\bigotimes mn}}}{\left(\Big<\sigma_{g_m}(p_1)\Big>_{\mathrm{BCFT}^{\bigotimes m}}\left<\sigma_{g_m^{-1}}(p_3) \right>_{\mathrm{BCFT}^{\bigotimes m}}\right)^n}\notag\\
	&~~~~~~~~~~~ + \frac{1}{1-n}\log \frac{\Big<\sigma_{g_B}(p_4) \Big>_{\mathrm{BCFT}^{\bigotimes mn}}\left<\sigma_{g_A^{-1}}(p_6) \right>_{\mathrm{BCFT}^{\bigotimes mn}}}{\left(\Big<\sigma_{g_m}(p_4)\Big>_{\mathrm{BCFT}^{\bigotimes m}}\left<\sigma_{g_m^{-1}}(p_6) \right>_{\mathrm{BCFT}^{\bigotimes m}}\right)^n}\notag\\
	&~~~~~~~~~~~~~~~~~~~~~~+ \frac{1}{1-n}\log\left( \left<\sigma_{g_Bg_A^{-1}}(p_2)\sigma_{g_Ag_B^{-1}}(p_5)\right>_{\mathrm{BCFT}^{\bigotimes mn}}\right)\notag\\
	&= \frac{1}{1-n}\log \frac{\Big< \sigma_{g_A}(p_1) \sigma_{g_A^{-1}}(q^b_1)\Big>_{\mathrm{CFT}^{\bigotimes mn}}\left<\sigma_{g_B^{-1}}(p_3)\sigma_{g_B}(q^a_3) \right>_{\mathrm{CFT}^{\bigotimes mn}}}{\left(\Big<\sigma_{g_m}(p_1)\sigma_{g_m^{-1}}(q^b_1)\Big>_{\mathrm{CFT}^{\bigotimes m}}\left<\sigma_{g_m^{-1}}(p_3)\sigma_{g_m}(q^a_3) \right>_{\mathrm{CFT}^{\bigotimes m}}\right)^n}\notag\\
	&~~~~~~~~~~~ + \frac{1}{1-n}\log \frac{\Big<\sigma_{g_B}(p_4)\sigma_{g_B^{-1}}(q^a_4) \Big>_{\mathrm{CFT}^{\bigotimes mn}}\left<\sigma_{g_A^{-1}}(p_6)\sigma_{g_A}(q^b_6) \right>_{\mathrm{CFT}^{\bigotimes mn}}}{\left(\Big<\sigma_{g_m}(p_4)\sigma_{g_m^{-1}}(q^a_4)\Big>_{\mathrm{CFT}^{\bigotimes m}}\left<\sigma_{g_m^{-1}}(p_6)\sigma_{g_m}(q^b_6)  \right>_{\mathrm{CFT}^{\bigotimes m}}\right)^n}\notag\\
	&~~~~~~~~~~~~~~~~~~~~~~+ \frac{1}{1-n}\log\left( \left<\sigma_{g_Bg_A^{-1}}(p_2)\sigma_{g_Ag_B^{-1}}(p_5)\right>_{\mathrm{CFT}^{\bigotimes mn}}\right)\notag\\
	&= \frac{1}{1-n}\log\left( \left<\sigma_{g_Bg_A^{-1}}(p_2)\sigma_{g_Ag_B^{-1}}(p_5)\right>_{\mathrm{CFT}^{\bigotimes mn}}\right), \label{adjcasec}
\end{align}

\subsubsection*{Configuration (d)}
\begin{align}
	S_R^{(n,m)}(A:B) &= 2\frac{1}{1-n}\log \frac{\left< \sigma_{g_A}(p_1)\sigma_{g_Bg_A^{-1}}(p_2)\sigma_{g_B^{-1}}(p_3) \right>_{\mathrm{BCFT}^{\bigotimes mn}}}{\left<\sigma_{g_m}(p_1)\sigma_{g_m^{-1}}(p_3) \right>^n_{\mathrm{BCFT}^{\bigotimes m}}}\notag\\
	&= 2\frac{1}{1-n}\log \frac{\Big< \sigma_{g_A}(p_1)\Big>_{\mathrm{BCFT}^{\bigotimes mn}}\left<\sigma_{g_Bg_A^{-1}}(p_2)\right>_{\mathrm{BCFT}^{\bigotimes mn}}\left<\sigma_{g_B^{-1}}(p_3) \right>_{\mathrm{BCFT}^{\bigotimes mn}}}{\left(\Big<\sigma_{g_m}(p_1)\Big>_{\mathrm{BCFT}^{\bigotimes m}}\left<\sigma_{g_m^{-1}}(p_3) \right>_{\mathrm{BCFT}^{\bigotimes m}}\right)^n}\notag\\
	&= 2\frac{1}{1-n}\log\frac{\left< \sigma_{g_A}(p_1)\sigma_{g_A^{-1}}(q^b_{1})\right>_{\mathrm{CFT}^{\bigotimes mn}}\left<\sigma_{g_B^{-1}}(p_3) \sigma_{g_B}(q^a_{3})\right>_{\mathrm{CFT}^{\bigotimes mn}}}{\left(\left<\sigma_{g_m}(p_1)\sigma_{g_m^{-1}}(q^b_{1})\right>_{\mathrm{CFT}^{\bigotimes m}}\left<\sigma_{g_m^{-1}}(p_3)\sigma_{g_m^{1}}(q^a_{3}) \right>_{\mathrm{CFT}^{\bigotimes m}}\right)^n}\notag\\
	& ~~~~~~~~~~~~~~~~~~~~~~~~~~~~+ 2\frac{1}{1-n}\log\left(\left<\sigma_{g_Bg_A^{-1}}(p_2)\sigma_{g_Ag_B^{-1}}(q^a_{2})\right>_{\mathrm{CFT}^{\bigotimes mn}}\right) \label{adjcased}
\end{align}

\subsubsection*{Configuration (e)}

\begin{align}
	S_R^{(n,m)}(A:B) &= 2\frac{1}{1-n}\log \frac{\left< \sigma_{g_A}(p_1)\sigma_{g_Bg_A^{-1}}(p_2)\sigma_{g_B^{-1}}(p_3) \right>_{\mathrm{BCFT}^{\bigotimes mn}}}{\left<\sigma_{g_m}(p_1)\sigma_{g_m^{-1}}(p_3) \right>^n_{\mathrm{BCFT}^{\bigotimes m}}}\notag\\
	&= 2\frac{1}{1-n}\log \frac{\left< \sigma_{g_A}(p_1)\right>_{\mathrm{BCFT}^{\bigotimes mn}}\left<\sigma_{g_Bg_A^{-1}}(p_2)\sigma_{g_B^{-1}}(p_3) \right>_{\mathrm{BCFT}^{\bigotimes mn}}}{\left(\Big<\sigma_{g_m}(p_1)\Big>_{\mathrm{BCFT}^{\bigotimes m}}\left<\sigma_{g_m^{-1}}(p_3) \right>_{\mathrm{BCFT}^{\bigotimes m}}\right)^n}\notag\\
	&= 2\frac{1}{1-n}\log\frac{\left< \sigma_{g_A}(p_1)\sigma_{g_A^{-1}}(q^b_{1})\right>_{\mathrm{CFT}^{\bigotimes mn}}\left<\sigma_{g_Bg_A^{-1}}(p_2)\sigma_{g_B^{-1}}(p_3) \sigma_{g_B}(q^a_{3})\right>_{\mathrm{CFT}^{\bigotimes mn}}}{\left(\left<\sigma_{g_m}(p_1)\sigma_{g_m^{-1}}(q^b_{1})\right>_{\mathrm{CFT}^{\bigotimes m}}\left<\sigma_{g_m^{-1}}(p_3)\sigma_{g_m^{1}}(q^a_{3}) \right>_{\mathrm{CFT}^{\bigotimes m}}\right)^n} \label{adjcasee}
\end{align}

\subsubsection*{Configuration (g)}

\begin{align}
	S_R^{(n,m)}(A:B) &= \frac{1}{1-n}\log \frac{\left< \sigma_{g_A}(p_1)\sigma_{g_Bg_A^{-1}}(p_2)\sigma_{g_B^{-1}}(p_3)\sigma_{g_B}(p_4)\sigma_{g_Ag_B^{-1}}(p_5)\sigma_{g_A^{-1}}(p_6) \right>_{\mathrm{BCFT}^{\bigotimes mn}}}{\left<\sigma_{g_m}(p_1)\sigma_{g_m^{-1}}(p_3)\sigma_{g_m}(p_4)\sigma_{g_m^{-1}}(p_6) \right>^n_{\mathrm{BCFT}^{\bigotimes m}}}\notag\\
	&= \frac{1}{1-n}\log \frac{\Big< \sigma_{g_A}(p_1)\sigma_{g_Bg_A^{-1}}(p_2)\Big>_{\mathrm{BCFT}^{\bigotimes mn}}}{\Big<\sigma_{g_m}(p_1)\Big>^n_{\mathrm{BCFT}^{\bigotimes m}}} + \frac{1}{1-n}\log \frac{\Big<\sigma_{g_B^{-1}}(p_3)\sigma_{g_B}(p_4) \Big>_{\mathrm{BCFT}^{\bigotimes mn}}}{\Big<\sigma_{g^{-1}_m}(p_3)\sigma_{g_m}(p_4) \Big>^n_{\mathrm{BCFT}^{\bigotimes m}}}\notag\\
	&~~~~~~~~~~~~~~~~~~~~~~~~~~~~~~~~~~~~~~~~~~~~~~~~~~~~~+ \frac{1}{1-n}\log\frac{ \left<\sigma_{g_Ag_B^{-1}}(p_5)\sigma_{g_A^{-1}}(p_6)\right>_{\mathrm{BCFT}^{\bigotimes mn}}}{\left<\sigma_{g_m^{-1}}(p_6) \right>^n_{\mathrm{BCFT}^{\bigotimes m}}}\notag\\
	&= \frac{1}{1-n}\log \frac{\Big< \sigma_{g_A}(p_1)\sigma_{g^{-1}_A}(q^b_1)\sigma_{g_Bg_A^{-1}}(p_2)\Big>_{\mathrm{CFT}^{\bigotimes mn}}}{\Big<\sigma_{g_m}(p_1)\sigma_{g^{-1}_m}(q^b_1)\Big>^n_{\mathrm{CFT}^{\bigotimes m}}}\notag\\
	&~~~~~~~~~~~ + \frac{1}{1-n}\log \frac{\Big<\sigma_{g_B^{-1}}(p_3)\sigma_{g_B}(p_4) \Big>_{\mathrm{CFT}^{\bigotimes mn}}}{\Big<\sigma_{g^{-1}_m}(p_3)\sigma_{g_m}(p_4) \Big>^n_{\mathrm{CFT}^{\bigotimes m}}}\notag\\
	&~~~~~~~~~~~~~~~~~~~~~~+ \frac{1}{1-n}\log\frac{ \left<\sigma_{g_Ag_B^{-1}}(p_5)\sigma_{g_A^{-1}}(p_6)\sigma_{g_A}(q^b_6)\right>_{\mathrm{CFT}^{\bigotimes mn}}}{\left<\sigma_{g_m^{-1}}(p_6)\sigma_{g_m}(q^b_6) \right>^n_{\mathrm{CFT}^{\bigotimes m}}}, \label{adjcaseg}
\end{align}

\subsubsection*{Configuration (h)}

\begin{align}
	S_R^{(n,m)}(A:B) &= \frac{1}{1-n}\log \frac{\left< \sigma_{g_A}(p_1)\sigma_{g_Bg_A^{-1}}(p_2)\sigma_{g_B^{-1}}(p_3)\sigma_{g_B}(p_4)\sigma_{g_Ag_B^{-1}}(p_5)\sigma_{g_A^{-1}}(p_6) \right>_{\mathrm{BCFT}^{\bigotimes mn}}}{\left<\sigma_{g_m}(p_1)\sigma_{g_m^{-1}}(p_3)\sigma_{g_m}(p_4)\sigma_{g_m^{-1}}(p_6) \right>^n_{\mathrm{BCFT}^{\bigotimes m}}}\notag\\
	&= \frac{1}{1-n}\log \frac{\Big< \sigma_{g_A}(p_1)\Big>_{\mathrm{BCFT}^{\bigotimes mn}}\left<\sigma_{g_A^{-1}}(p_6) \right>_{\mathrm{BCFT}^{\bigotimes mn}}\Big<\sigma_{g^{-1}_B}(p_3)\sigma_{g_B}(p_4) \Big>_{\mathrm{BCFT}^{\bigotimes mn}}}{\left(\Big<\sigma_{g_m}(p_1)\Big>_{\mathrm{BCFT}^{\bigotimes m}}\left<\sigma_{g_m^{-1}}(p_6) \right>_{\mathrm{BCFT}^{\bigotimes m}}\left<\sigma_{g^{-1}_m}(p_3)\sigma_{g_m}(p_4) \right>^n_{\mathrm{BCFT}^{\bigotimes m}}\right)^n}\notag\\
	&~~~~~~~~~~~~~~~~~~~~~~+ \frac{1}{1-n}\log\left( \left<\sigma_{g_Bg_A^{-1}}(p_2)\right>_{\mathrm{BCFT}^{\bigotimes mn}}\left<\sigma_{g_Ag_B^{-1}}(p_5)\right>_{\mathrm{BCFT}^{\bigotimes mn}}\right)\notag\\
	&= \frac{1}{1-n}\log \frac{\Big< \sigma_{g_A}(p_1) \sigma_{g_A^{-1}}(q^b_1)\Big>_{\mathrm{CFT}^{\bigotimes mn}}\left<\sigma_{g_A^{-1}}(p_6)\sigma_{g_A}(q^b_6) \right>_{\mathrm{CFT}^{\bigotimes mn}}}{\left(\Big<\sigma_{g_m}(p_1)\sigma_{g_m^{-1}}(q^b_1)\Big>_{\mathrm{CFT}^{\bigotimes m}}\left<\sigma_{g_m^{-1}}(p_6)\sigma_{g_m}(q^b_6) \right>_{\mathrm{CFT}^{\bigotimes m}}\right)^n}\notag\\
	&+ \frac{1}{1-n}\log \frac{\Big<\sigma_{g^{-1}_B}(p_3)\sigma_{g_B}(p_4) \Big>_{\mathrm{CFT}^{\bigotimes mn}}}{\left<\sigma_{g_m^{-1}}(p_3)\sigma_{g_m}(p_4)  \right>^n_{\mathrm{CFT}^{\bigotimes m}}}+ \frac{1}{1-n}\log\left<\sigma_{g_Bg_A^{-1}}(p_2)\sigma_{g_Ag_B^{-1}}(q^b_2)\right>_{\mathrm{CFT}^{\bigotimes mn}}\notag\\
	&+ \frac{1}{1-n}\log \left<\sigma_{g_Ag_B^{-1}}(p_5)\sigma_{g_Bg_A^{-1}}(q^b_5)\right>_{\mathrm{CFT}^{\bigotimes mn}}, \label{adjcaseh}
\end{align}

\subsubsection*{Configuration (i)}

\begin{align}
	S_R^{(n,m)}(A:B) &= \frac{1}{1-n}\log \frac{\left< \sigma_{g_A}(p_1)\sigma_{g_Bg_A^{-1}}(p_2)\sigma_{g_B^{-1}}(p_3)\sigma_{g_B}(p_4)\sigma_{g_Ag_B^{-1}}(p_5)\sigma_{g_A^{-1}}(p_6) \right>_{\mathrm{BCFT}^{\bigotimes mn}}}{\left<\sigma_{g_m}(p_1)\sigma_{g_m^{-1}}(p_3)\sigma_{g_m}(p_4)\sigma_{g_m^{-1}}(p_6) \right>^n_{\mathrm{BCFT}^{\bigotimes m}}}\notag\\
	&= \frac{1}{1-n}\log \frac{\Big< \sigma_{g_A}(p_1)\Big>_{\mathrm{BCFT}^{\bigotimes mn}}\left<\sigma_{g_A^{-1}}(p_6) \right>_{\mathrm{BCFT}^{\bigotimes mn}}}{\left(\Big<\sigma_{g_m}(p_1)\Big>_{\mathrm{BCFT}^{\bigotimes m}}\left<\sigma_{g_m^{-1}}(p_6) \right>_{\mathrm{BCFT}^{\bigotimes m}}\right)^n}\notag\\
	&~~~~~~~~~~~ + \frac{1}{1-n}\log \frac{\Big<\sigma_{g^{-1}_B}(p_3)\sigma_{g_B}(p_4) \Big>_{\mathrm{BCFT}^{\bigotimes mn}}}{\left<\sigma_{g^{-1}_m}(p_3)\sigma_{g_m}(p_4) \right>^n_{\mathrm{BCFT}^{\bigotimes m}}}\notag\\
	&~~~~~~~~~~~~~~~~~~~~~~+ \frac{1}{1-n}\log\left( \left<\sigma_{g_Bg_A^{-1}}(p_2)\sigma_{g_Ag_B^{-1}}(p_5)\right>_{\mathrm{BCFT}^{\bigotimes mn}}\right)\notag\\
	&= \frac{1}{1-n}\log \frac{\Big< \sigma_{g_A}(p_1) \sigma_{g_A^{-1}}(q^b_1)\Big>_{\mathrm{CFT}^{\bigotimes mn}}\left<\sigma_{g_A^{-1}}(p_6)\sigma_{g_A}(q^b_6) \right>_{\mathrm{CFT}^{\bigotimes mn}}}{\left(\Big<\sigma_{g_m}(p_1)\sigma_{g_m^{-1}}(q^b_1)\Big>_{\mathrm{CFT}^{\bigotimes m}}\left<\sigma_{g_m^{-1}}(p_6)\sigma_{g_m}(q^b_6) \right>_{\mathrm{CFT}^{\bigotimes m}}\right)^n}\notag\\
	&~~~~~~~~~~~+ \frac{1}{1-n}\log \frac{\Big<\sigma_{g^{-1}_B}(p_3)\sigma_{g_B}(p_4) \Big>_{\mathrm{CFT}^{\bigotimes mn}}}{\left<\sigma_{g_m^{-1}}(p_3)\sigma_{g_m}(p_4)  \right>^n_{\mathrm{CFT}^{\bigotimes m}}}\notag\\
	&~~~~~~~~~~~~~~~~~~~~~~+\frac{1}{1-n}\log\left( \left<\sigma_{g_Bg_A^{-1}}(p_2)\sigma_{g_Ag_B^{-1}}(p_5)\right>_{\mathrm{CFT}^{\bigotimes mn}}\right), \label{adjcasei}
\end{align}

\subsubsection*{Configuration (j)}

\begin{align}
	S_R^{(n,m)}(A:B) &= \frac{1}{1-n}\log \frac{\left< \sigma_{g_A}(p_1)\sigma_{g_Bg_A^{-1}}(p_2)\sigma_{g_B^{-1}}(p_3)\sigma_{g_B}(p_4)\sigma_{g_Ag_B^{-1}}(p_5)\sigma_{g_A^{-1}}(p_6) \right>_{\mathrm{BCFT}^{\bigotimes mn}}}{\left<\sigma_{g_m}(p_1)\sigma_{g_m^{-1}}(p_3)\sigma_{g_m}(p_4)\sigma_{g_m^{-1}}(p_6) \right>^n_{\mathrm{BCFT}^{\bigotimes m}}}\notag\\
	&= \frac{1}{1-n}\log \frac{\Big< \sigma_{g_A}(p_1)\Big>_{\mathrm{BCFT}^{\bigotimes mn}}\left<\sigma_{g_A^{-1}}(p_6) \right>_{\mathrm{BCFT}^{\bigotimes mn}}}{\left(\Big<\sigma_{g_m}(p_1)\Big>_{\mathrm{BCFT}^{\bigotimes m}}\left<\sigma_{g_m^{-1}}(p_6) \right>_{\mathrm{BCFT}^{\bigotimes m}}\right)^n}\notag\\
	&~~~~~~~~~~~ + \frac{1}{1-n}\log \frac{\Big<\sigma_{g_Bg_A^{-1}}(p_2)\sigma_{g^{-1}_B}(p_3)\sigma_{g_B}(p_4)\sigma_{g_Ag_B^{-1}}(p_5) \Big>_{\mathrm{BCFT}^{\bigotimes mn}}}{\left<\sigma_{g^{-1}_m}(p_3)\sigma_{g_m}(p_4) \right>^n_{\mathrm{BCFT}^{\bigotimes m}}}\notag\\
	&= \frac{1}{1-n}\log \frac{\Big<\sigma_{g_Bg_A^{-1}}(p_2)\sigma_{g^{-1}_B}(p_3)\sigma_{g_B}(p_4)\sigma_{g_Ag_B^{-1}}(p_5) \Big>_{\mathrm{CFT}^{\bigotimes mn}}}{\left<\sigma_{g^{-1}_m}(p_3)\sigma_{g_m}(p_4) \right>^n_{\mathrm{CFT}^{\bigotimes m}}}. \label{adjcasej}
\end{align}

\subsubsection*{Configuration (k)}
\begin{align}
	S_R^{(n,m)}(A:B) &= \frac{1}{1-n}\log \frac{\left< \sigma_{g_A}(p_1)\sigma_{g_Bg_A^{-1}}(p_2)\sigma_{g_B^{-1}}(p_3)\sigma_{g_B}(p_4)\sigma_{g_Ag_B^{-1}}(p_5)\sigma_{g_A^{-1}}(p_6) \right>_{\mathrm{BCFT}^{\bigotimes mn}}}{\left<\sigma_{g_m}(p_1)\sigma_{g_m^{-1}}(p_3)\sigma_{g_m}(p_4)\sigma_{g_m^{-1}}(p_6) \right>^n_{\mathrm{BCFT}^{\bigotimes m}}}\notag\\
	&= \frac{1}{1-n}\log \frac{\Big<\sigma_{g_Bg_A^{-1}}(p_2)\sigma_{g_A}(p_1)\sigma_{g^{-1}_A}(p_6)\sigma_{g_Ag_B^{-1}}(p_5) \Big>_{\mathrm{BCFT}^{\bigotimes mn}}}{\left<\sigma_{g_m}(p_1)\sigma_{g^{-1}_m}(p_6) \right>^n_{\mathrm{BCFT}^{\bigotimes m}}}\notag\\
	&~~~~~~~~~~~ +  \frac{1}{1-n}\log \frac{\Big< \sigma_{g_B^{-1}}(p_3)\sigma_{g_B}(p_4) \Big>_{\mathrm{BCFT}^{\bigotimes mn}}}{\Big<\sigma_{g_m^{-1}}(p_3)\sigma_{g_m}(p_4) \Big>_{\mathrm{BCFT}^{\bigotimes m}}^n}\notag\\
	&= \frac{1}{1-n}\log \frac{\Big<\sigma_{g_Bg_A^{-1}}(p_2)\sigma_{g_A}(p_1)\sigma_{g^{-1}_A}(p_6)\sigma_{g_Ag_B^{-1}}(p_5) \Big>_{\mathrm{CFT}^{\bigotimes mn}}}{\left<\sigma_{g_m}(p_1)\sigma_{g^{-1}_m}(p_6) \right>^n_{\mathrm{CFT}^{\bigotimes m}}}.\label{adjcasek}
\end{align}	

\subsubsection*{Configuration (l)}

\begin{align}
	S_R^{(n,m)}(A:B) &= \frac{1}{1-n}\log \frac{\left< \sigma_{g_A}(p_1)\sigma_{g_Bg_A^{-1}}(p_2)\sigma_{g_B^{-1}}(p_3)\sigma_{g_B}(p_4)\sigma_{g_Ag_B^{-1}}(p_5)\sigma_{g_A^{-1}}(p_6) \right>_{\mathrm{BCFT}^{\bigotimes mn}}}{\left<\sigma_{g_m}(p_1)\sigma_{g_m^{-1}}(p_3)\sigma_{g_m}(p_4)\sigma_{g_m^{-1}}(p_6) \right>^n_{\mathrm{BCFT}^{\bigotimes m}}}\notag\\
	&= \frac{1}{1-n}\log \frac{\Big< \sigma_{g_A}(p_1)\sigma_{g_A^{-1}}(p_6) \Big>_{\mathrm{BCFT}^{\bigotimes mn}}}{\left(\Big<\sigma_{g_m}(p_1)\sigma_{g_m^{-1}}(p_6) \Big>_{\mathrm{BCFT}^{\bigotimes m}}\right)^n}\notag\\
	&~~~~~~~~~~~ + \frac{1}{1-n}\log \frac{\Big<\sigma_{g^{-1}_B}(p_3)\sigma_{g_B}(p_4) \Big>_{\mathrm{BCFT}^{\bigotimes mn}}}{\left<\sigma_{g^{-1}_m}(p_3)\sigma_{g_m}(p_4) \right>^n_{\mathrm{BCFT}^{\bigotimes m}}}\notag\\
	&~~~~~~~~~~~~~~~~~~~~~~+ \frac{1}{1-n}\log\left( \left<\sigma_{g_Bg_A^{-1}}(p_2)\sigma_{g_Ag_B^{-1}}(p_5)\right>_{\mathrm{BCFT}^{\bigotimes mn}}\right)\notag\\
	&= \frac{1}{1-n}\log \frac{\Big< \sigma_{g_A}(p_1)\sigma_{g_A^{-1}}(p_6) \Big>_{\mathrm{CFT}^{\bigotimes mn}}}{\left(\Big<\sigma_{g_m}(p_1)\sigma_{g_m^{-1}}(p_6) \Big>_{\mathrm{CFT}^{\bigotimes m}}\right)^n}\notag\\
	&~~~~~~~~~~~ + \frac{1}{1-n}\log \frac{\Big<\sigma_{g^{-1}_B}(p_3)\sigma_{g_B}(p_4) \Big>_{\mathrm{CFT}^{\bigotimes mn}}}{\left<\sigma_{g^{-1}_m}(p_3)\sigma_{g_m}(p_4) \right>^n_{\mathrm{CFT}^{\bigotimes m}}}\notag\\
	&~~~~~~~~~~~~~~~~~~~~~~+ \frac{1}{1-n}\log\left( \left<\sigma_{g_Bg_A^{-1}}(p_2)\sigma_{g_Ag_B^{-1}}(p_5)\right>_{\mathrm{CFT}^{\bigotimes mn}}\right). \label{adjcasel}
\end{align}

\subsubsection*{Configuration (m)}

\begin{align}
	S_R^{(n,m)}(A:B) &= \frac{1}{1-n}\log \frac{\left< \sigma_{g_A}(p_1)\sigma_{g_Bg_A^{-1}}(p_2)\sigma_{g_B^{-1}}(p_3)\sigma_{g_B}(p_4)\sigma_{g_Ag_B^{-1}}(p_5)\sigma_{g_A^{-1}}(p_6) \right>_{\mathrm{BCFT}^{\bigotimes mn}}}{\left<\sigma_{g_m}(p_1)\sigma_{g_m^{-1}}(p_3)\sigma_{g_m}(p_4)\sigma_{g_m^{-1}}(p_6) \right>^n_{\mathrm{BCFT}^{\bigotimes m}}}\notag\\
	&= \frac{1}{1-n}\log \frac{\left< \sigma_{g_A}(p_1)\sigma_{g_A^{-1}}(p_6) \right>_{\mathrm{BCFT}^{\bigotimes mn}}}{\left<\sigma_{g_m}(p_1)\sigma_{g_m^{-1}}(p_6) \right>_{\mathrm{BCFT}^{\bigotimes m}}^n}\notag\\
	&~~~~~~~~~~~ + \frac{1}{1-n}\log \frac{\Big<\sigma_{g_Bg_A^{-1}}(p_2)\sigma_{g^{-1}_B}(p_3)\sigma_{g_B}(p_4)\sigma_{g_Ag_B^{-1}}(p_5) \Big>_{\mathrm{BCFT}^{\bigotimes mn}}}{\left<\sigma_{g^{-1}_m}(p_3)\sigma_{g_m}(p_4) \right>^n_{\mathrm{BCFT}^{\bigotimes m}}}\notag\\
	&= \frac{1}{1-n}\log \frac{\Big<\sigma_{g_Bg_A^{-1}}(p_2)\sigma_{g^{-1}_B}(p_3)\sigma_{g_B}(p_4)\sigma_{g_Ag_B^{-1}}(p_5) \Big>_{\mathrm{CFT}^{\bigotimes mn}}}{\left<\sigma_{g^{-1}_m}(p_3)\sigma_{g_m}(p_4) \right>^n_{\mathrm{CFT}^{\bigotimes m}}}, \label{adjcasem}
\end{align}

\subsubsection*{Configuration (n)}

\begin{align}
	S_R^{(n,m)}(A:B) &= \frac{1}{1-n}\log \frac{\left< \sigma_{g_A}(p_1)\sigma_{g_Bg_A^{-1}}(p_2)\sigma_{g_B^{-1}}(p_3)\sigma_{g_B}(p_4)\sigma_{g_Ag_B^{-1}}(p_5)\sigma_{g_A^{-1}}(p_6) \right>_{\mathrm{BCFT}^{\bigotimes mn}}}{\left<\sigma_{g_m}(p_1)\sigma_{g_m^{-1}}(p_3)\sigma_{g_m}(p_4)\sigma_{g_m^{-1}}(p_6) \right>^n_{\mathrm{BCFT}^{\bigotimes m}}}\notag\\
	&= \frac{1}{1-n}\log \frac{\Big<\sigma_{g_Bg_A^{-1}}(p_2)\sigma_{g_A}(p_1)\sigma_{g^{-1}_A}(p_6)\sigma_{g_Ag_B^{-1}}(p_5) \Big>_{\mathrm{BCFT}^{\bigotimes mn}}}{\left<\sigma_{g_m}(p_1)\sigma_{g^{-1}_m}(p_6) \right>^n_{\mathrm{BCFT}^{\bigotimes m}}}\notag\\
	&~~~~~~~~~~~ +  \frac{1}{1-n}\log \frac{\Big< \sigma_{g_B^{-1}}(p_3)\Big>_{\mathrm{BCFT}^{\bigotimes mn}}\Big<\sigma_{g_B}(p_4) \Big>_{\mathrm{BCFT}^{\bigotimes mn}}}{\Big<\sigma_{g_m^{-1}}(p_3)\Big>_{\mathrm{BCFT}^{\bigotimes mn}}^n\Big<\sigma_{g_m}(p_4) \Big>_{\mathrm{BCFT}^{\bigotimes m}}^n}\notag\\
	&= \frac{1}{1-n}\log \frac{\Big<\sigma_{g_Bg_A^{-1}}(p_2)\sigma_{g_A}(p_1)\sigma_{g^{-1}_A}(p_6)\sigma_{g_Ag_B^{-1}}(p_5) \Big>_{\mathrm{CFT}^{\bigotimes mn}}}{\left<\sigma_{g_m}(p_1)\sigma_{g^{-1}_m}(p_6) \right>^n_{\mathrm{CFT}^{\bigotimes m}}}, \label{adjcasen}
\end{align}

\subsubsection*{Configuration (o)}

\begin{align}
	S_R^{(n,m)}(A:B) &= \frac{1}{1-n}\log \frac{\left< \sigma_{g_A}(p_1)\sigma_{g_Bg_A^{-1}}(p_2)\sigma_{g_B^{-1}}(p_3)\sigma_{g_B}(p_4)\sigma_{g_Ag_B^{-1}}(p_5)\sigma_{g_A^{-1}}(p_6) \right>_{\mathrm{BCFT}^{\bigotimes mn}}}{\left<\sigma_{g_m}(p_1)\sigma_{g_m^{-1}}(p_3)\sigma_{g_m}(p_4)\sigma_{g_m^{-1}}(p_6) \right>^n_{\mathrm{BCFT}^{\bigotimes m}}}\notag\\
	&= \frac{1}{1-n}\log \frac{\Big<\sigma_{g_A}(p_1)\sigma_{g^{-1}_A}(p_6) \Big>_{\mathrm{BCFT}^{\bigotimes mn}}}{\Big<\sigma_{g_m}(p_1)\sigma_{g_m^{-1}}(p_6) \Big>^n_{\mathrm{BCFT}^{\bigotimes m}}}+\frac{1}{1-n}\log \frac{\Big<\sigma_{g_Bg_A^{-1}}(p_2)\sigma_{g_B^{-1}}(p_3)\Big>_{\mathrm{BCFT}^{\bigotimes mn}}}{\Big<\sigma_{g_m^{-1}}(p_3)\Big>^n_{\mathrm{BCFT}^{\bigotimes m}}}\notag\\
	&~~~~~~~~~~~~~~~~~~~~~~~~~~~~~~~~~~~~~~~~~~~~~~~~~~~~~+ \frac{1}{1-n}\log\frac{ \left<\sigma_{g_B}(p_4)\sigma_{g_Ag_B^{-1}}(p_5)\right>_{\mathrm{BCFT}^{\bigotimes mn}}}{\left<\sigma_{g_m}(p_4) \right>^n_{\mathrm{BCFT}^{\bigotimes m}}}\notag\\
	&= \frac{1}{1-n}\log \frac{\Big<\sigma_{g_A}(p_1)\sigma_{g^{-1}_A}(p_6) \Big>_{\mathrm{CFT}^{\bigotimes mn}}}{\Big<\sigma_{g_m}(p_1)\sigma_{g_m^{-1}}(p_6) \Big>^n_{\mathrm{CFT}^{\bigotimes m}}}\notag\\
	&~~~~~~~~~~~ + \frac{1}{1-n}\log \frac{\Big< \sigma_{g_Bg_A^{-1}}(p_2)\sigma_{g_B^{-1}}(p_3)\sigma_{g_B}(q^a_3)\Big>_{\mathrm{CFT}^{\bigotimes mn}}}{\Big<\sigma_{g_m^{-1}}(p_3)\sigma_{g_m}(q^a_3)\Big>^n_{\mathrm{CFT}^{\bigotimes m}}}\notag\\
	&~~~~~~~~~~~~~~~~~~~~~~+ \frac{1}{1-n}\log\frac{ \left<\sigma_{g_B}(p_4)\sigma_{g_B^{-1}}(q^a_4)\sigma_{g_Ag_B^{-1}}(p_5)\right>_{\mathrm{BCFT}^{\bigotimes mn}}}{\left<\sigma_{g_m}(p_4)\sigma_{g_m^{-1}}(q^a_4) \right>^n_{\mathrm{BCFT}^{\bigotimes m}}}, \label{adjcaseo}
\end{align}

\subsubsection*{Configuration (p)}

\begin{align}
	S_R^{(n,m)}(A:B) &= \frac{1}{1-n}\log \frac{\left< \sigma_{g_A}(p_1)\sigma_{g_Bg_A^{-1}}(p_2)\sigma_{g_B^{-1}}(p_3)\sigma_{g_B}(p_4)\sigma_{g_Ag_B^{-1}}(p_5)\sigma_{g_A^{-1}}(p_6) \right>_{\mathrm{BCFT}^{\bigotimes mn}}}{\left<\sigma_{g_m}(p_1)\sigma_{g_m^{-1}}(p_3)\sigma_{g_m}(p_4)\sigma_{g_m^{-1}}(p_6) \right>^n_{\mathrm{BCFT}^{\bigotimes m}}}\notag\\
	&= \frac{1}{1-n}\log \frac{\Big<\sigma_{g_A}(p_1)\sigma_{g^{-1}_A}(p_6) \Big>_{\mathrm{BCFT}^{\bigotimes mn}}}{\Big<\sigma_{g_m}(p_1)\sigma_{g_m^{-1}}(p_6) \Big>^n_{\mathrm{BCFT}^{\bigotimes m}}}\notag\\
	&~~~~~~~~~~~ +\frac{1}{1-n}\log \left(\Big<\sigma_{g_Bg_A^{-1}}(p_2)\Big>_{\mathrm{BCFT}^{\bigotimes mn}}\Big<\sigma_{g_Ag_B^{-1}}(p_5)\Big>_{\mathrm{BCFT}^{\bigotimes mn}}\right)\notag\\
	&~~~~~~~~~~~~~~~~~~~~~~+ \frac{1}{1-n}\log\frac{ \Big<\sigma_{g_B^{-1}}(p_3)\Big>_{\mathrm{BCFT}^{\bigotimes mn}}\Big<\sigma_{g_B}(p_4)\Big>_{\mathrm{BCFT}^{\bigotimes mn}}}{\Big<\sigma_{g_m^{-1}}(p_3)\Big>^n_{\mathrm{BCFT}^{\bigotimes m}}\left<\sigma_{g_m}(p_4) \right>^n_{\mathrm{BCFT}^{\bigotimes m}}}\notag\\
	&= \frac{1}{1-n}\log \frac{\Big<\sigma_{g_A}(p_1)\sigma_{g^{-1}_A}(p_6) \Big>_{\mathrm{CFT}^{\bigotimes mn}}}{\Big<\sigma_{g_m}(p_1)\sigma_{g_m^{-1}}(p_6) \Big>^n_{\mathrm{CFT}^{\bigotimes m}}}\notag\\
	&+\frac{1}{1-n}\log \left(\Big<\sigma_{g_Bg_A^{-1}}(p_2)\sigma_{g_Ag_B^{-1}}(q^a_2)\Big>_{\mathrm{CFT}^{\bigotimes mn}}\Big<\sigma_{g_Ag_B^{-1}}(p_5)\sigma_{g_Bg_A^{-1}}(q^a_5)\Big>_{\mathrm{CFT}^{\bigotimes mn}}\right)\notag\\
	& +\frac{1}{1-n}\log\frac{ \Big<\sigma_{g_B^{-1}}(p_3)\sigma_{g_B}(q^a_3)\Big>_{\mathrm{CFT}^{\bigotimes mn}}\Big<\sigma_{g_B}(p_4)\sigma_{g_B^{-1}}(q^a_4)\Big>_{\mathrm{CFT}^{\bigotimes mn}}}{\Big<\sigma_{g_m^{-1}}(p_3)\sigma_{g_m}(q^a_3)\Big>^n_{\mathrm{CFT}^{\bigotimes m}}\left<\sigma_{g_m}(p_4)\sigma_{g_m^{-1}}(q^a_4) \right>^n_{\mathrm{CFT}^{\bigotimes m}}}, \label{adjcasep}
\end{align}

\subsubsection*{Configuration (q)}

\begin{align}
	S_R^{(n,m)}(A:B) &= \frac{1}{1-n}\log \frac{\left< \sigma_{g_A}(p_1)\sigma_{g_Bg_A^{-1}}(p_2)\sigma_{g_B^{-1}}(p_3)\sigma_{g_B}(p_4)\sigma_{g_Ag_B^{-1}}(p_5)\sigma_{g_A^{-1}}(p_6) \right>_{\mathrm{BCFT}^{\bigotimes mn}}}{\left<\sigma_{g_m}(p_1)\sigma_{g_m^{-1}}(p_3)\sigma_{g_m}(p_4)\sigma_{g_m^{-1}}(p_6) \right>^n_{\mathrm{BCFT}^{\bigotimes m}}}\notag\\
	&= \frac{1}{1-n}\log \frac{\Big<\sigma_{g_A}(p_1)\sigma_{g^{-1}_A}(p_6) \Big>_{\mathrm{BCFT}^{\bigotimes mn}}}{\Big<\sigma_{g_m}(p_1)\sigma_{g_m^{-1}}(p_6) \Big>^n_{\mathrm{BCFT}^{\bigotimes m}}}\notag\\
	&~~~~~~~~~~~ +\frac{1}{1-n}\log \left(\Big<\sigma_{g_Bg_A^{-1}}(p_2)\sigma_{g_Ag_B^{-1}}(p_5)\Big>_{\mathrm{BCFT}^{\bigotimes mn}}\right)\notag\\
	&~~~~~~~~~~~~~~~~~~~~~~+ \frac{1}{1-n}\log\frac{ \Big<\sigma_{g_B^{-1}}(p_3)\Big>_{\mathrm{BCFT}^{\bigotimes mn}}\Big<\sigma_{g_B}(p_4)\Big>_{\mathrm{BCFT}^{\bigotimes mn}}}{\Big<\sigma_{g_m^{-1}}(p_3)\Big>^n_{\mathrm{BCFT}^{\bigotimes m}}\left<\sigma_{g_m}(p_4) \right>^n_{\mathrm{BCFT}^{\bigotimes m}}}\notag\\
	&= \frac{1}{1-n}\log \frac{\Big<\sigma_{g_A}(p_1)\sigma_{g^{-1}_A}(p_6) \Big>_{\mathrm{CFT}^{\bigotimes mn}}}{\Big<\sigma_{g_m}(p_1)\sigma_{g_m^{-1}}(p_6) \Big>^n_{\mathrm{CFT}^{\bigotimes m}}}\notag\\
	&+\frac{1}{1-n}\log \left(\Big<\sigma_{g_Bg_A^{-1}}(p_2)\sigma_{g_Ag_B^{-1}}(p_5)\Big>_{\mathrm{CFT}^{\bigotimes mn}}\right)\notag\\
	&+ \frac{1}{1-n}\log\frac{ \Big<\sigma_{g_B^{-1}}(p_3)\sigma_{g_B}(q^a_3)\Big>_{\mathrm{CFT}^{\bigotimes mn}}\Big<\sigma_{g_B}(p_4)\sigma_{g_B^{-1}}(q^a_4)\Big>_{\mathrm{CFT}^{\bigotimes mn}}}{\Big<\sigma_{g_m^{-1}}(p_3)\sigma_{g_m}(q^a_3)\Big>^n_{\mathrm{CFT}^{\bigotimes m}}\left<\sigma_{g_m}(p_4)\sigma_{g_m^{-1}}(q^a_4) \right>^n_{\mathrm{CFT}^{\bigotimes m}}}, \label{adjcaseq}
\end{align}

\section{Disjoint intervals}\label{app_disj}
Next we consider different configurations of two disjoint intervals $A$ and $B$ at a finite temperature and show how the higher-point twist correlators in the R\'enyi reflected entropy expression factorize into lower-point twist correlator in the large central charge limit for various dominant channels. These factorizations of the twist correlators listed below are utilized in the reflected entropy computation in \cref{SrEwInGeng}.

\subsubsection*{Configuration (a)}
\begin{align}
	S_R^{(n,m)}(A:B) &= 2\frac{1}{1-n}\log \frac{\left< \sigma_{g_A}(p_1)\sigma_{g_A^{-1}}(p_2)\sigma_{g_B}(p_3)\sigma_{g_B^{-1}}(p_4) \right>_{\mathrm{BCFT}^{\bigotimes mn}}}{\left<\sigma_{g_m}(p_1)\sigma_{g_m^{-1}}(p_2)\sigma_{g_m}(p_3)\sigma_{g_m^{-1}}(p_4) \right>^n_{\mathrm{BCFT}^{\bigotimes m}}}\notag\\
	&= 2\frac{1}{1-n}\log \frac{\left< \sigma_{g_A}(p_1)\sigma_{g_A^{-1}}(p_2)\sigma_{g_B}(p_3)\right>_{\mathrm{BCFT}^{\bigotimes mn}}\left<\sigma_{g_B^{-1}}(p_4) \right>_{\mathrm{BCFT}^{\bigotimes mn}}}{\left<\sigma_{g_m}(p_1)\sigma_{g_m^{-1}}(p_2)\sigma_{g_m}(p_3)\right>^n_{\mathrm{BCFT}^{\bigotimes m}}\left<\sigma_{g_m^{-1}}(p_4) \right>^n_{\mathrm{BCFT}^{\bigotimes m}}}\notag\\
	&= 2\frac{1}{1-n}\log \frac{\left< \sigma_{g_A^{-1}}(q^b_1)\sigma_{g_A}(p_1)\sigma_{g_A^{-1}}(p_2)\sigma_{g_B}(p_3)\right>_{\mathrm{CFT}^{\bigotimes mn}}\left<\sigma_{g_B^{-1}}(p_4)\sigma_{g_B}(q^a_4) \right>_{\mathrm{CFT}^{\bigotimes mn}}}{\left<\sigma_{g_m^{-1}}(q^b_1)\sigma_{g_m}(p_1)\sigma_{g_m^{-1}}(p_2)\sigma_{g_m}(p_3)\right>^n_{\mathrm{CFT}^{\bigotimes m}}\left<\sigma_{g_m^{-1}}(p_4)\sigma_{g_m}(q^a_4) \right>^n_{\mathrm{CFT}^{\bigotimes m}}}\notag\\
	&= 2\frac{1}{1-n}\log \frac{\left< \sigma_{g_A^{-1}}(q^b_1)\sigma_{g_A}(p_1)\sigma_{g_A^{-1}}(p_2)\sigma_{g_B}(p_3)\right>_{\mathrm{CFT}^{\bigotimes mn}}}{\left<\sigma_{g_m^{-1}}(q^b_1)\sigma_{g_m}(p_1)\sigma_{g_m^{-1}}(p_2)\sigma_{g_m}(p_3)\right>^n_{\mathrm{CFT}^{\bigotimes m}}},\label{discasea}
\end{align}

\subsubsection*{Configuration (b)}
\begin{align}
	S_R^{(n,m)}(A:B) &= 2\frac{1}{1-n}\log \frac{\left< \sigma_{g_A}(p_1)\sigma_{g_A^{-1}}(p_2)\sigma_{g_B}(p_3)\sigma_{g_B^{-1}}(p_4) \right>_{\mathrm{BCFT}^{\bigotimes mn}}}{\left<\sigma_{g_m}(p_1)\sigma_{g_m^{-1}}(p_2)\sigma_{g_m}(p_3)\sigma_{g_m^{-1}}(p_4) \right>^n_{\mathrm{BCFT}^{\bigotimes m}}}\notag\\
	&= 2\frac{1}{1-n}\log \frac{\Big< \sigma_{g_A}(p_1)\Big>_{\mathrm{BCFT}^{\bigotimes mn}}\left<\sigma_{g_A^{-1}}(p_2)\sigma_{g_B}(p_3)\right>_{\mathrm{BCFT}^{\bigotimes mn}}\left<\sigma_{g_B^{-1}}(p_4) \right>_{\mathrm{BCFT}^{\bigotimes mn}}}{\Big<\sigma_{g_m}(p_1)\Big>^n_{\mathrm{BCFT}^{\bigotimes m}}\left<\sigma_{g_m^{-1}}(p_2)\sigma_{g_m}(p_3)\right>^n_{\mathrm{BCFT}^{\bigotimes m}}\left<\sigma_{g_m^{-1}}(p_4) \right>^n_{\mathrm{BCFT}^{\bigotimes m}}}\notag\\
	&= 2\frac{1}{1-n}\log \frac{\left<\sigma_{g_A^{-1}}(q^b_1) \sigma_{g_A}(p_1)\right>_{\mathrm{CFT}^{\bigotimes mn}}}{\left<\sigma_{g_m^{-1}}(q^b_1)\sigma_{g_m}(p_1)\right>^n_{\mathrm{CFT}^{\bigotimes m}}} + 2 \frac{1}{1-n}\log \frac{\left<\sigma_{g_B^{-1}}(p_4)\sigma_{g_B}(q^a_4) \right>_{\mathrm{CFT}^{\bigotimes mn}}}{\left<\sigma_{g_m^{-1}}(p_4)\sigma_{g_m}(q^a_4) \right>^n_{\mathrm{CFT}^{\bigotimes m}}}\notag\\
	&~~~~~~~~~~~~~~~~~~~~~~~~~~~~~~~~+2 \frac{1}{1-n}\log \frac{\left<\sigma_{g_A}(q^b_2)\sigma_{g_A^{-1}}(p_2)\sigma_{g_B}(p_3)\sigma_{g_B^{-1}}(q^b_3)\right>_{\mathrm{CFT}^{\bigotimes mn}}}{\left<\sigma_{g_m}(q^b_2)\sigma_{g_m^{-1}}(p_2)\sigma_{g_m}(p_3)\sigma_{g_m^{-1}}(q^b_3)\right>^n_{\mathrm{CFT}^{\bigotimes m}}}\notag\\
	&=2 \frac{1}{1-n}\log \frac{\left<\sigma_{g_A}(q^b_2)\sigma_{g_A^{-1}}(p_2)\sigma_{g_B}(p_3)\sigma_{g_B^{-1}}(q^b_3)\right>_{\mathrm{CFT}^{\bigotimes mn}}}{\left<\sigma_{g_m}(q^b_2)\sigma_{g_m^{-1}}(p_2)\sigma_{g_m}(p_3)\sigma_{g_m^{-1}}(q^b_3)\right>^n_{\mathrm{CFT}^{\bigotimes m}}},\label{discaseb}
\end{align}

\subsubsection*{Configuration (c)}
\begin{align}
	S_R^{(n,m)}(A:B) &= \frac{1}{1-n}\log \frac{\left< \sigma_{g_A}(p_1)\sigma_{g_A^{-1}}(p_2)\sigma_{g_B}(p_3)\sigma_{g_B^{-1}}(p_4)\sigma_{g_B}(p_5)\sigma_{g_B^{-1}}(p_6)\sigma_{g_A}(p_7)\sigma_{g_A^{-1}}(p_8) \right>_{\mathrm{BCFT}^{\bigotimes mn}}}{\left<\sigma_{g_m}(p_1)\sigma_{g_m^{-1}}(p_2)\sigma_{g_m}(p_3)\sigma_{g_m^{-1}}(p_4)\sigma_{g_m}(p_5)\sigma_{g_m^{-1}}(p_6)\sigma_{g_m}(p_7)\sigma_{g_m^{-1}}(p_8) \right>^n_{\mathrm{BCFT}^{\bigotimes m}}}\notag\\
	&= \frac{1}{1-n}\log \frac{\Big< \sigma_{g_A}(p_1)\Big>_{\mathrm{BCFT}^{\bigotimes mn}}\left<\sigma_{g_B^{-1}}(p_4) \right>_{\mathrm{BCFT}^{\bigotimes mn}}}{\Big<\sigma_{g_m}(p_1)\Big>^n_{\mathrm{BCFT}^{\bigotimes m}}\left<\sigma_{g_m^{-1}}(p_4) \right>^n_{\mathrm{BCFT}^{\bigotimes m}}}\notag\\
	&~~~~~~~~~~~~+ \frac{1}{1-n}\log \frac{\Big< \sigma_{g_B}(p_5)\Big>_{\mathrm{BCFT}^{\bigotimes mn}}\left<\sigma_{g_A^{-1}}(p_8) \right>_{\mathrm{BCFT}^{\bigotimes mn}}}{\Big<\sigma_{g_m}(p_5)\Big>^n_{\mathrm{BCFT}^{\bigotimes m}}\left<\sigma_{g_m^{-1}}(p_8) \right>^n_{\mathrm{BCFT}^{\bigotimes m}}}\notag\\
	&~~~~~~~~~~~~~~~~~~~~+ \frac{1}{1-n}\log \frac{\left< \sigma_{g_A^{-1}}(p_2)\sigma_{g_B}(p_3)\sigma_{g_B^{-1}}(p_6)\sigma_{g_A}(p_7)\right>_{\mathrm{BCFT}^{\bigotimes mn}}}{\left<\sigma_{g_m^{-1}}(p_2)\sigma_{g_m}(p_3)\sigma_{g_m^{-1}}(p_6)\sigma_{g_m}(p_7) \right>^n_{\mathrm{BCFT}^{\bigotimes m}}}\notag\\
	&= \frac{1}{1-n}\log \frac{\Big<\sigma_{g_A^{-1}}(q^b_1) \sigma_{g_A}(p_1)\Big>_{\mathrm{CFT}^{\bigotimes mn}}\left<\sigma_{g_B^{-1}}(p_4)\sigma_{g_B}(q^a_4) \right>_{\mathrm{CFT}^{\bigotimes mn}}}{\Big<\sigma_{g_m^{-1}}(q^b_1)\sigma_{g_m}(p_1)\Big>^n_{\mathrm{CFT}^{\bigotimes m}}\left<\sigma_{g_m^{-1}}(p_4)\sigma_{g_m}(q^a_4) \right>^n_{\mathrm{CFT}^{\bigotimes m}}}\notag\\
	&~~~~~~~~~~~~+ \frac{1}{1-n}\log \frac{\Big< \sigma_{g_B}(p_5)\sigma_{g_B^{-1}}(q^a_5)\Big>_{\mathrm{CFT}^{\bigotimes mn}}\left<\sigma_{g_A^{-1}}(p_8)\sigma_{g_A}(q^b_8) \right>_{\mathrm{CFT}^{\bigotimes mn}}}{\Big<\sigma_{g_m}(p_5)\sigma_{g_m^{-1}}(q^a_5)\Big>^n_{\mathrm{CFT}^{\bigotimes m}}\left<\sigma_{g_m^{-1}}(p_8)\sigma_{g_m}(q^b_8) \right>^n_{\mathrm{CFT}^{\bigotimes m}}}\notag\\
	&~~~~~~~~~~~~~~~~~~~~+ \frac{1}{1-n}\log \frac{\left< \sigma_{g_A^{-1}}(p_2)\sigma_{g_B}(p_3)\sigma_{g_B^{-1}}(p_6)\sigma_{g_A}(p_7)\right>_{\mathrm{CFT}^{\bigotimes mn}}}{\left<\sigma_{g_m^{-1}}(p_2)\sigma_{g_m}(p_3)\sigma_{g_m^{-1}}(p_6)\sigma_{g_m}(p_7) \right>^n_{\mathrm{CFT}^{\bigotimes m}}}\notag\\
	&= \frac{1}{1-n}\log \frac{\left< \sigma_{g_A^{-1}}(p_2)\sigma_{g_B}(p_3)\sigma_{g_B^{-1}}(p_6)\sigma_{g_A}(p_7)\right>_{\mathrm{CFT}^{\bigotimes mn}}}{\left<\sigma_{g_m^{-1}}(p_2)\sigma_{g_m}(p_3)\sigma_{g_m^{-1}}(p_6)\sigma_{g_m}(p_7) \right>^n_{\mathrm{CFT}^{\bigotimes m}}}, \label{discasec}
\end{align}

\subsubsection*{Configuration (d):}
\begin{align}
	S_R^{(n,m)}(A:B) &= 2\frac{1}{1-n}\log \frac{\left< \sigma_{g_A}(p_1)\sigma_{g_A^{-1}}(p_2)\sigma_{g_B}(p_3)\sigma_{g_B^{-1}}(p_4)\right>_{\mathrm{BCFT}^{\bigotimes mn}}}{\left<\sigma_{g_m}(p_1)\sigma_{g_m^{-1}}(p_2)\sigma_{g_m}(p_3)\sigma_{g_m^{-1}}(p_4) \right>^n_{\mathrm{BCFT}^{\bigotimes m}}}\notag\\
	&= 2\frac{1}{1-n}\log \frac{\Big< \sigma_{g_A}(p_1)\Big>_{\mathrm{BCFT}^{\bigotimes mn}}\left<\sigma_{g_A^{-1}}(p_2)\sigma_{g_B}(p_3)\right>_{\mathrm{BCFT}^{\bigotimes mn}}\left<\sigma_{g_B^{-1}}(p_4) \right>_{\mathrm{BCFT}^{\bigotimes mn}}}{\Big<\sigma_{g_m}(p_1)\Big>^n_{\mathrm{BCFT}^{\bigotimes m}}\left<\sigma_{g_m^{-1}}(p_2)\sigma_{g_m}(p_3)\right>^n_{\mathrm{BCFT}^{\bigotimes m}}\left<\sigma_{g_m^{-1}}(p_4) \right>^n_{\mathrm{BCFT}^{\bigotimes m}}}\notag\\
	&= 2\frac{1}{1-n}\log \frac{\left<\sigma_{g_A^{-1}}(q^b_1) \sigma_{g_A}(p_1)\right>_{\mathrm{CFT}^{\bigotimes mn}}}{\left<\sigma_{g_m^{-1}}(q^b_1)\sigma_{g_m}(p_1)\right>^n_{\mathrm{CFT}^{\bigotimes m}}} + 2 \frac{1}{1-n}\log \frac{\left<\sigma_{g_B^{-1}}(p_4)\sigma_{g_B}(q^a_4) \right>_{\mathrm{CFT}^{\bigotimes mn}}}{\left<\sigma_{g_m^{-1}}(p_4)\sigma_{g_m}(q^a_4) \right>^n_{\mathrm{CFT}^{\bigotimes m}}}\notag\\
	&~~~~~~~~~~~~~~~~~~~~~~~~~~~~~~~~+2 \frac{1}{1-n}\log \frac{\left<\sigma_{g_A}(q^a_2)\sigma_{g_A^{-1}}(p_2)\sigma_{g_B}(p_3)\sigma_{g_B^{-1}}(q^a_3)\right>_{\mathrm{CFT}^{\bigotimes mn}}}{\left<\sigma_{g_m}(q^a_2)\sigma_{g_m^{-1}}(p_2)\sigma_{g_m}(p_3)\sigma_{g_m^{-1}}(q^a_3)\right>^n_{\mathrm{CFT}^{\bigotimes m}}}\notag\\
	&= 2 \frac{1}{1-n}\log \frac{\left<\sigma_{g_A}(q^a_2)\sigma_{g_A^{-1}}(p_2)\sigma_{g_B}(p_3)\sigma_{g_B^{-1}}(q^a_3)\right>_{\mathrm{CFT}^{\bigotimes mn}}}{\left<\sigma_{g_m}(q^a_2)\sigma_{g_m^{-1}}(p_2)\sigma_{g_m}(p_3)\sigma_{g_m^{-1}}(q^a_3)\right>^n_{\mathrm{CFT}^{\bigotimes m}}}, \label{discased}
\end{align}

\subsubsection*{Configuration (e):}
\begin{align}
	S_R^{(n,m)}(A:B) &= 2\frac{1}{1-n}\log \frac{\left< \sigma_{g_A}(p_1)\sigma_{g_A^{-1}}(p_2)\sigma_{g_B}(p_3)\sigma_{g_B^{-1}}(p_4)\right>_{\mathrm{BCFT}^{\bigotimes mn}}}{\left<\sigma_{g_m}(p_1)\sigma_{g_m^{-1}}(p_2)\sigma_{g_m}(p_3)\sigma_{g_m^{-1}}(p_4) \right>^n_{\mathrm{BCFT}^{\bigotimes m}}}\notag\\
	&= 2\frac{1}{1-n}\log \frac{\Big<\sigma_{g_A}(p_1) \Big>_{\mathrm{BCFT}^{\bigotimes mn}}\left<\sigma_{g_A^{-1}}(p_2)\sigma_{g_B}(p_3)\sigma_{g_B^{-1}}(p_4)\right>_{\mathrm{BCFT}^{\bigotimes mn}}}{\Big<\sigma_{g_m}(p_1) \Big>^n_{\mathrm{BCFT}^{\bigotimes m}}\left<\sigma_{g_m^{-1}}(p_2)\sigma_{g_m}(p_3)\sigma_{g_m^{-1}}(p_4)\right>^n_{\mathrm{BCFT}^{\bigotimes m}}}\notag\\
	&= 2\frac{1}{1-n}\log \frac{\Big<\sigma_{g_A^{-1}}(q^b_1)\sigma_{g_A}(p_1) \Big>_{\mathrm{CFT}^{\bigotimes mn}}}{\Big<\sigma_{g_m^{-1}}(q^b_1)\sigma_{g_m}(p_1) \Big>^n_{\mathrm{CFT}^{\bigotimes m}}}\notag\\
	&~~~~~~~~+ 2\frac{1}{1-n}\log \frac{\left< \sigma_{g_A^{-1}}(p_2)\sigma_{g_B}(p_3)\sigma_{g_B^{-1}}(p_4)\sigma_{g_B}(q^a_4)\right>_{\mathrm{CFT}^{\bigotimes mn}}}{\left<\sigma_{g_m^{-1}}(p_2)\sigma_{g_m}(p_3)\sigma_{g_m^{-1}}(p_4)\sigma_{g_m}(q^a_4) \right>^n_{\mathrm{CFT}^{\bigotimes m}}}\notag\\
	&= 2\frac{1}{1-n}\log \frac{\left< \sigma_{g_A^{-1}}(p_2)\sigma_{g_B}(p_3)\sigma_{g_B^{-1}}(p_4)\sigma_{g_B}(q^a_4)\right>_{\mathrm{CFT}^{\bigotimes mn}}}{\left<\sigma_{g_m^{-1}}(p_2)\sigma_{g_m}(p_3)\sigma_{g_m^{-1}}(p_4)\sigma_{g_m}(q^a_4) \right>^n_{\mathrm{CFT}^{\bigotimes m}}}, \label{discasee}
\end{align}

\subsubsection*{Configuration (g):}
\begin{align}
	S_R^{(n,m)}(A:B) &= \frac{1}{1-n}\log \frac{\left< \sigma_{g_A}(p_1)\sigma_{g_A^{-1}}(p_2)\sigma_{g_B}(p_3)\sigma_{g_B^{-1}}(p_4)\sigma_{g_B}(p_5)\sigma_{g_B^{-1}}(p_6)\sigma_{g_A}(p_7)\sigma_{g_A^{-1}}(p_8) \right>_{\mathrm{BCFT}^{\bigotimes mn}}}{\left<\sigma_{g_m}(p_1)\sigma_{g_m^{-1}}(p_2)\sigma_{g_m}(p_3)\sigma_{g_m^{-1}}(p_4)\sigma_{g_m}(p_5)\sigma_{g_m^{-1}}(p_6)\sigma_{g_m}(p_7)\sigma_{g_m^{-1}}(p_8) \right>^n_{\mathrm{BCFT}^{\bigotimes m}}}\notag\\
	&= 2\frac{1}{1-n}\log \frac{\left< \sigma_{g_A}(p_1)\sigma_{g_A^{-1}}(p_2)\sigma_{g_B}(p_3)\right>_{\mathrm{BCFT}^{\bigotimes mn}}\left< \sigma_{g_B^{-1}}(p_4)\sigma_{g_B}(p_5)\right>_{\mathrm{BCFT}^{\bigotimes mn}}}{\left<\sigma_{g_m}(p_1)\sigma_{g_m^{-1}}(p_2)\sigma_{g_m}(p_3)\right>^n_{\mathrm{BCFT}^{\bigotimes m}}\left<\sigma_{g_m^{-1}}(p_4)\sigma_{g_m}(p_5)\right>^n_{\mathrm{BCFT}^{\bigotimes m}}}\notag\\
	&= 2\frac{1}{1-n}\log \frac{\left< \sigma_{g_A^{-1}}(q^b_1)\sigma_{g_A}(p_1)\sigma_{g_A^{-1}}(p_2)\sigma_{g_B}(p_3)\right>_{\mathrm{CFT}^{\bigotimes mn}}}{\left<\sigma_{g_m^{-1}}(q^b_1)\sigma_{g_m}(p_1)\sigma_{g_m^{-1}}(p_2)\sigma_{g_m}(p_3)\right>^n_{\mathrm{CFT}^{\bigotimes m}}}, \label{discaseg}
\end{align}

\subsubsection*{Configuration (h):}
\begin{align}
	S_R^{(n,m)}(A:B) &= \frac{1}{1-n}\log \frac{\left< \sigma_{g_A}(p_1)\sigma_{g_A^{-1}}(p_2)\sigma_{g_B}(p_3)\sigma_{g_B^{-1}}(p_4)\sigma_{g_B}(p_5)\sigma_{g_B^{-1}}(p_6)\sigma_{g_A}(p_7)\sigma_{g_A^{-1}}(p_8) \right>_{\mathrm{BCFT}^{\bigotimes mn}}}{\left<\sigma_{g_m}(p_1)\sigma_{g_m^{-1}}(p_2)\sigma_{g_m}(p_3)\sigma_{g_m^{-1}}(p_4)\sigma_{g_m}(p_5)\sigma_{g_m^{-1}}(p_6)\sigma_{g_m}(p_7)\sigma_{g_m^{-1}}(p_8) \right>^n_{\mathrm{BCFT}^{\bigotimes m}}}\notag\\
	&= 2\frac{1}{1-n}\log \frac{\Big< \sigma_{g_A}(p_1)\Big>_{\mathrm{BCFT}^{\bigotimes mn}}\left<\sigma_{g_A^{-1}}(p_2)\sigma_{g_B}(p_3)\right>_{\mathrm{BCFT}^{\bigotimes mn}}}{\Big<\sigma_{g_m}(p_1)\Big>^n_{\mathrm{BCFT}^{\bigotimes m}}\left<\sigma_{g_m^{-1}}(p_2)\sigma_{g_m}(p_3)\right>^n_{\mathrm{BCFT}^{\bigotimes m}}}\notag\\
	&~~~~~~~~~~~~~~~~~~~~~~~~~~~~~+ \frac{1}{1-n}\log \frac{\Big< \sigma_{g_B^{-1}}(p_4)\sigma_{g_B}(p_5)\Big>_{\mathrm{BCFT}^{\bigotimes mn}}}{\Big<\sigma_{g_m^{-1}}(p_4)\sigma_{g_m}(p_5)\Big>^n_{\mathrm{BCFT}^{\bigotimes m}}}\notag\\
	&= 2 \frac{1}{1-n}\log \frac{\left<\sigma_{g_A}(q^b_2)\sigma_{g_A^{-1}}(p_2)\sigma_{g_B}(p_3)\sigma_{g_B^{-1}}(q^b_3)\right>_{\mathrm{CFT}^{\bigotimes mn}}}{\left<\sigma_{g_m}(q^b_2)\sigma_{g_m^{-1}}(p_2)\sigma_{g_m}(p_3)\sigma_{g_m^{-1}}(q^b_3)\right>^n_{\mathrm{CFT}^{\bigotimes m}}}, \label{discaseh}
\end{align}

\subsubsection*{Configuration (i):}
\begin{align}
	S_R^{(n,m)}(A:B) &= \frac{1}{1-n}\log \frac{\left< \sigma_{g_A}(p_1)\sigma_{g_A^{-1}}(p_2)\sigma_{g_B}(p_3)\sigma_{g_B^{-1}}(p_4)\sigma_{g_B}(p_5)\sigma_{g_B^{-1}}(p_6)\sigma_{g_A}(p_7)\sigma_{g_A^{-1}}(p_8) \right>_{\mathrm{BCFT}^{\bigotimes mn}}}{\left<\sigma_{g_m}(p_1)\sigma_{g_m^{-1}}(p_2)\sigma_{g_m}(p_3)\sigma_{g_m^{-1}}(p_4)\sigma_{g_m}(p_5)\sigma_{g_m^{-1}}(p_6)\sigma_{g_m}(p_7)\sigma_{g_m^{-1}}(p_8) \right>^n_{\mathrm{BCFT}^{\bigotimes m}}}\notag\\
	&= \frac{1}{1-n}\log \frac{\Big< \sigma_{g_A}(p_1)\Big>_{\mathrm{BCFT}^{\bigotimes mn}}\left<\sigma_{g_A^{-1}}(p_8)\right>_{\mathrm{BCFT}^{\bigotimes mn}}}{\Big<\sigma_{g_m}(p_1)\Big>^n_{\mathrm{BCFT}^{\bigotimes m}}\left<\sigma_{g_m^{-1}}(p_8)\right>^n_{\mathrm{BCFT}^{\bigotimes m}}}\notag\\
	&~~~~~~~~+ \frac{1}{1-n}\log \frac{\left< \sigma_{g_A^{-1}}(p_2)\sigma_{g_B}(p_3)\sigma_{g_B^{-1}}(p_6)\sigma_{g_A}(p_7)\right>_{\mathrm{BCFT}^{\bigotimes mn}}}{\left<\sigma_{g_m^{-1}}(p_2)\sigma_{g_m}(p_3)\sigma_{g_m^{-1}}(p_6)\sigma_{g_m}(p_7) \right>^n_{\mathrm{BCFT}^{\bigotimes m}}}\notag\\
	&~~~~~~~~~~~~~~~~~~~~~~~~~~~~~+ \frac{1}{1-n}\log \frac{\Big< \sigma_{g_B^{-1}}(p_4)\sigma_{g_B}(p_5)\Big>_{\mathrm{BCFT}^{\bigotimes mn}}}{\Big<\sigma_{g_m^{-1}}(p_4)\sigma_{g_m}(p_5)\Big>^n_{\mathrm{BCFT}^{\bigotimes m}}}\notag\\
	&= \frac{1}{1-n}\log \frac{\left< \sigma_{g_A^{-1}}(p_2)\sigma_{g_B}(p_3)\sigma_{g_B^{-1}}(p_6)\sigma_{g_A}(p_7)\right>_{\mathrm{CFT}^{\bigotimes mn}}}{\left<\sigma_{g_m^{-1}}(p_2)\sigma_{g_m}(p_3)\sigma_{g_m^{-1}}(p_6)\sigma_{g_m}(p_7) \right>^n_{\mathrm{CFT}^{\bigotimes m}}}. \label{discasei}
\end{align}

\subsubsection*{Configuration (j):}
\begin{align}
	S_R^{(n,m)}(A:B) &= \frac{1}{1-n}\log \frac{\left< \sigma_{g_A}(p_1)\sigma_{g_A^{-1}}(p_2)\sigma_{g_B}(p_3)\sigma_{g_B^{-1}}(p_4)\sigma_{g_B}(p_5)\sigma_{g_B^{-1}}(p_6)\sigma_{g_A}(p_7)\sigma_{g_A^{-1}}(p_8) \right>_{\mathrm{BCFT}^{\bigotimes mn}}}{\left<\sigma_{g_m}(p_1)\sigma_{g_m^{-1}}(p_2)\sigma_{g_m}(p_3)\sigma_{g_m^{-1}}(p_4)\sigma_{g_m}(p_5)\sigma_{g_m^{-1}}(p_6)\sigma_{g_m}(p_7)\sigma_{g_m^{-1}}(p_8) \right>^n_{\mathrm{BCFT}^{\bigotimes m}}}\notag\\
	&= \frac{1}{1-n}\log \frac{\Big< \sigma_{g_A}(p_1)\Big>_{\mathrm{BCFT}^{\bigotimes mn}}}{\Big< \sigma_{g_m}(p_1)\Big>^n_{\mathrm{BCFT}^{\bigotimes mn}}} \notag\\
	&~~~~~~~+ \frac{1}{1-n}\log \frac{\left< \sigma_{g_A^{-1}}(p_2)\sigma_{g_B}(p_3)\sigma_{g_B^{-1}}(p_4)\sigma_{g_B}(p_5)\sigma_{g_B^{-1}}(p_6)\sigma_{g_A}(p_7)\right>_{\mathrm{BCFT}^{\bigotimes mn}}}{\left<\sigma_{g_m^{-1}}(p_2)\sigma_{g_m}(p_3)\sigma_{g_m^{-1}}(p_4)\sigma_{g_m}(p_5)\sigma_{g_m^{-1}}(p_6)\sigma_{g_m}(p_7)\right>^n_{\mathrm{BCFT}^{\bigotimes m}}}\notag\\
	&= \frac{1}{1-n}\log \frac{\left< \sigma_{g_A^{-1}}(p_2)\sigma_{g_B}(p_3)\sigma_{g_B^{-1}}(p_4)\sigma_{g_B}(p_5)\sigma_{g_B^{-1}}(p_6)\sigma_{g_A}(p_7)\right>_{\mathrm{CFT}^{\bigotimes mn}}}{\left<\sigma_{g_m^{-1}}(p_2)\sigma_{g_m}(p_3)\sigma_{g_m^{-1}}(p_4)\sigma_{g_m}(p_5)\sigma_{g_m^{-1}}(p_6)\sigma_{g_m}(p_7)\right>^n_{\mathrm{CFT}^{\bigotimes m}}}, \label{discasej}
\end{align}

\subsubsection*{Configuration (k):}
\begin{align}
	S_R^{(n,m)}(A:B) &= \frac{1}{1-n}\log \frac{\left< \sigma_{g_A}(p_1)\sigma_{g_A^{-1}}(p_2)\sigma_{g_B}(p_3)\sigma_{g_B^{-1}}(p_4)\sigma_{g_B}(p_5)\sigma_{g_B^{-1}}(p_6)\sigma_{g_A}(p_7)\sigma_{g_A^{-1}}(p_8) \right>_{\mathrm{BCFT}^{\bigotimes mn}}}{\left<\sigma_{g_m}(p_1)\sigma_{g_m^{-1}}(p_2)\sigma_{g_m}(p_3)\sigma_{g_m^{-1}}(p_4)\sigma_{g_m}(p_5)\sigma_{g_m^{-1}}(p_6)\sigma_{g_m}(p_7)\sigma_{g_m^{-1}}(p_8) \right>^n_{\mathrm{BCFT}^{\bigotimes m}}}\notag\\
	&= \frac{1}{1-n}\log \frac{\left< \sigma_{g_A}(p_1)\sigma_{g_A^{-1}}(p_2)\sigma_{g_B}(p_3)\sigma_{g_B^{-1}}(p_6)\sigma_{g_A}(p_7)\sigma_{g_A^{-1}}(p_8) \right>_{\mathrm{BCFT}^{\bigotimes mn}}}{\left<\sigma_{g_m}(p_1)\sigma_{g_m^{-1}}(p_2)\sigma_{g_m}(p_3)\sigma_{g_m^{-1}}(p_6)\sigma_{g_m}(p_7)\sigma_{g_m^{-1}}(p_8) \right>^n_{\mathrm{BCFT}^{\bigotimes m}}} \notag\\
	&~~~~~~~~~~~~~~~~~~~~~~~~~~~~~~~~~~~~~~~~~~+ \frac{1}{1-n}\log \frac{\Big< \sigma_{g_B^{-1}}(p_4)\sigma_{g_B}(p_5)\Big>_{\mathrm{BCFT}^{\bigotimes mn}}}{\Big<\sigma_{g_m^{-1}}(p_4)\sigma_{g_m}(p_5)\Big>^n_{\mathrm{BCFT}^{\bigotimes m}}}\notag\\
	&= \frac{1}{1-n}\log \frac{\left< \sigma_{g_A}(p_1)\sigma_{g_A^{-1}}(p_2)\sigma_{g_B}(p_3)\sigma_{g_B^{-1}}(p_6)\sigma_{g_A}(p_7)\sigma_{g_A^{-1}}(p_8) \right>_{\mathrm{CFT}^{\bigotimes mn}}}{\left<\sigma_{g_m}(p_1)\sigma_{g_m^{-1}}(p_2)\sigma_{g_m}(p_3)\sigma_{g_m^{-1}}(p_6)\sigma_{g_m}(p_7)\sigma_{g_m^{-1}}(p_8) \right>^n_{\mathrm{CFT}^{\bigotimes m}}}, \label{discasek}
\end{align}

\subsubsection*{Configuration (l):}
\begin{align}
	S_R^{(n,m)}(A:B) &= \frac{1}{1-n}\log \frac{\left< \sigma_{g_A}(p_1)\sigma_{g_A^{-1}}(p_2)\sigma_{g_B}(p_3)\sigma_{g_B^{-1}}(p_4)\sigma_{g_B}(p_5)\sigma_{g_B^{-1}}(p_6)\sigma_{g_A}(p_7)\sigma_{g_A^{-1}}(p_8) \right>_{\mathrm{BCFT}^{\bigotimes mn}}}{\left<\sigma_{g_m}(p_1)\sigma_{g_m^{-1}}(p_2)\sigma_{g_m}(p_3)\sigma_{g_m^{-1}}(p_4)\sigma_{g_m}(p_5)\sigma_{g_m^{-1}}(p_6)\sigma_{g_m}(p_7)\sigma_{g_m^{-1}}(p_8) \right>^n_{\mathrm{BCFT}^{\bigotimes m}}}\notag\\
	&= \frac{1}{1-n}\log \frac{\Big< \sigma_{g_A}(p_1)\sigma_{g_A^{-1}}(p_8)\Big>_{\mathrm{BCFT}^{\bigotimes mn}}}{\Big<\sigma_{g_m^{-1}}(p_4)\sigma_{g_m}(p_5)\Big>^n_{\mathrm{BCFT}^{\bigotimes m}}}\notag\\
	&~~~~~~~~+ \frac{1}{1-n}\log \frac{\left< \sigma_{g_A^{-1}}(p_2)\sigma_{g_B}(p_3)\sigma_{g_B^{-1}}(p_6)\sigma_{g_A}(p_7)\right>_{\mathrm{BCFT}^{\bigotimes mn}}}{\left<\sigma_{g_m^{-1}}(p_2)\sigma_{g_m}(p_3)\sigma_{g_m^{-1}}(p_6)\sigma_{g_m}(p_7) \right>^n_{\mathrm{BCFT}^{\bigotimes m}}}\notag\\
	&~~~~~~~~~~~~~~~~~~~~~~~~~~~~~+ \frac{1}{1-n}\log \frac{\Big< \sigma_{g_B^{-1}}(p_4)\sigma_{g_B}(p_5)\Big>_{\mathrm{BCFT}^{\bigotimes mn}}}{\Big<\sigma_{g_m^{-1}}(p_4)\sigma_{g_m}(p_5)\Big>^n_{\mathrm{BCFT}^{\bigotimes m}}}\notag\\
	&= \frac{1}{1-n}\log \frac{\left< \sigma_{g_A^{-1}}(p_2)\sigma_{g_B}(p_3)\sigma_{g_B^{-1}}(p_6)\sigma_{g_A}(p_7)\right>_{\mathrm{CFT}^{\bigotimes mn}}}{\left<\sigma_{g_m^{-1}}(p_2)\sigma_{g_m}(p_3)\sigma_{g_m^{-1}}(p_6)\sigma_{g_m}(p_7) \right>^n_{\mathrm{CFT}^{\bigotimes m}}}, \label{discasel}
\end{align}

\subsubsection*{Configuration (m):}
\begin{align}
	S_R^{(n,m)}(A:B) &= \frac{1}{1-n}\log \frac{\left< \sigma_{g_A}(p_1)\sigma_{g_A^{-1}}(p_2)\sigma_{g_B}(p_3)\sigma_{g_B^{-1}}(p_4)\sigma_{g_B}(p_5)\sigma_{g_B^{-1}}(p_6)\sigma_{g_A}(p_7)\sigma_{g_A^{-1}}(p_8) \right>_{\mathrm{BCFT}^{\bigotimes mn}}}{\left<\sigma_{g_m}(p_1)\sigma_{g_m^{-1}}(p_2)\sigma_{g_m}(p_3)\sigma_{g_m^{-1}}(p_4)\sigma_{g_m}(p_5)\sigma_{g_m^{-1}}(p_6)\sigma_{g_m}(p_7)\sigma_{g_m^{-1}}(p_8) \right>^n_{\mathrm{BCFT}^{\bigotimes m}}}\notag\\
	&= \frac{1}{1-n}\log \frac{\Big< \sigma_{g_A}(p_1)\sigma_{g_A^{-1}}(p_8)\Big>_{\mathrm{BCFT}^{\bigotimes mn}}}{\Big<\sigma_{g_m^{-1}}(p_4)\sigma_{g_m}(p_5)\Big>^n_{\mathrm{BCFT}^{\bigotimes m}}}\notag\\
	&~~~~~~~~~+ \frac{1}{1-n}\log \frac{\left< \sigma_{g_A^{-1}}(p_2)\sigma_{g_B}(p_3)\sigma_{g_B^{-1}}(p_4)\sigma_{g_B}(p_5)\sigma_{g_B^{-1}}(p_6)\sigma_{g_A}(p_7)\right>_{\mathrm{BCFT}^{\bigotimes mn}}}{\left<\sigma_{g_m^{-1}}(p_2)\sigma_{g_m}(p_3)\sigma_{g_m^{-1}}(p_4)\sigma_{g_m}(p_5)\sigma_{g_m^{-1}}(p_6)\sigma_{g_m}(p_7)\right>^n_{\mathrm{BCFT}^{\bigotimes m}}}\notag\\
	&= \frac{1}{1-n}\log \frac{\left< \sigma_{g_A^{-1}}(p_2)\sigma_{g_B}(p_3)\sigma_{g_B^{-1}}(p_4)\sigma_{g_B}(p_5)\sigma_{g_B^{-1}}(p_6)\sigma_{g_A}(p_7)\right>_{\mathrm{CFT}^{\bigotimes mn}}}{\left<\sigma_{g_m^{-1}}(p_2)\sigma_{g_m}(p_3)\sigma_{g_m^{-1}}(p_4)\sigma_{g_m}(p_5)\sigma_{g_m^{-1}}(p_6)\sigma_{g_m}(p_7)\right>^n_{\mathrm{CFT}^{\bigotimes m}}}. \label{discasem}
\end{align}

\subsubsection*{Configuration (n):}
\begin{align}
	S_R^{(n,m)}(A:B) &= \frac{1}{1-n}\log \frac{\left< \sigma_{g_A}(p_1)\sigma_{g_A^{-1}}(p_2)\sigma_{g_B}(p_3)\sigma_{g_B^{-1}}(p_4)\sigma_{g_B}(p_5)\sigma_{g_B^{-1}}(p_6)\sigma_{g_A}(p_7)\sigma_{g_A^{-1}}(p_8) \right>_{\mathrm{BCFT}^{\bigotimes mn}}}{\left<\sigma_{g_m}(p_1)\sigma_{g_m^{-1}}(p_2)\sigma_{g_m}(p_3)\sigma_{g_m^{-1}}(p_4)\sigma_{g_m}(p_5)\sigma_{g_m^{-1}}(p_6)\sigma_{g_m}(p_7)\sigma_{g_m^{-1}}(p_8) \right>^n_{\mathrm{BCFT}^{\bigotimes m}}}\notag\\
	&= \frac{1}{1-n}\log \frac{\left< \sigma_{g_A}(p_1)\sigma_{g_A^{-1}}(p_2)\sigma_{g_B}(p_3)\sigma_{g_B^{-1}}(p_6)\sigma_{g_A}(p_7)\sigma_{g_A^{-1}}(p_8) \right>_{\mathrm{BCFT}^{\bigotimes mn}}}{\left<\sigma_{g_m}(p_1)\sigma_{g_m^{-1}}(p_2)\sigma_{g_m}(p_3)\sigma_{g_m^{-1}}(p_6)\sigma_{g_m}(p_7)\sigma_{g_m^{-1}}(p_8) \right>^n_{\mathrm{BCFT}^{\bigotimes m}}}\notag\\
	&~~~~~~~~~~~~~~~~~~~+\frac{1}{1-n}\log \frac{\left<\sigma_{g_B^{-1}}(p_4) \right>_{\mathrm{BCFT}^{\bigotimes mn}}\Big<\sigma_{g_B}(p_5) \Big>_{\mathrm{BCFT}^{\bigotimes mn}}}{\left<\sigma_{g_m^{-1}}(p_4) \right>^n_{\mathrm{BCFT}^{\bigotimes m}}\Big<\sigma_{g_m}(p_5) \Big>^n_{\mathrm{BCFT}^{\bigotimes m}}}\notag\\
	&= \frac{1}{1-n}\log \frac{\left< \sigma_{g_A}(p_1)\sigma_{g_A^{-1}}(p_2)\sigma_{g_B}(p_3)\sigma_{g_B^{-1}}(p_6)\sigma_{g_A}(p_7)\sigma_{g_A^{-1}}(p_8) \right>_{\mathrm{CFT}^{\bigotimes mn}}}{\left<\sigma_{g_m}(p_1)\sigma_{g_m^{-1}}(p_2)\sigma_{g_m}(p_3)\sigma_{g_m^{-1}}(p_6)\sigma_{g_m}(p_7)\sigma_{g_m^{-1}}(p_8) \right>^n_{\mathrm{CFT}^{\bigotimes m}}}, \label{discasen}
\end{align}

\subsubsection*{Configuration (o):}
\begin{align}
	S_R^{(n,m)}(A:B) &= \frac{1}{1-n}\log \frac{\left< \sigma_{g_A}(p_1)\sigma_{g_A^{-1}}(p_2)\sigma_{g_B}(p_3)\sigma_{g_B^{-1}}(p_4)\sigma_{g_B}(p_5)\sigma_{g_B^{-1}}(p_6)\sigma_{g_A}(p_7)\sigma_{g_A^{-1}}(p_8) \right>_{\mathrm{BCFT}^{\bigotimes mn}}}{\left<\sigma_{g_m}(p_1)\sigma_{g_m^{-1}}(p_2)\sigma_{g_m}(p_3)\sigma_{g_m^{-1}}(p_4)\sigma_{g_m}(p_5)\sigma_{g_m^{-1}}(p_6)\sigma_{g_m}(p_7)\sigma_{g_m^{-1}}(p_8) \right>^n_{\mathrm{BCFT}^{\bigotimes m}}}\notag\\
	&= \frac{1}{1-n}\log \frac{\Big< \sigma_{g_A}(p_1)\sigma_{g_A^{-1}}(p_8)\Big>_{\mathrm{BCFT}^{\bigotimes mn}}}{\Big<\sigma_{g_m^{-1}}(p_4)\sigma_{g_m}(p_5)\Big>^n_{\mathrm{BCFT}^{\bigotimes m}}}\notag\\
	&~~~~~~~~~+ 2\frac{1}{1-n}\log \frac{\left< \sigma_{g_A^{-1}}(p_2)\sigma_{g_B}(p_3)\sigma_{g_B^{-1}}(p_4)\right>_{\mathrm{BCFT}^{\bigotimes mn}}}{\left<\sigma_{g_m^{-1}}(p_2)\sigma_{g_m}(p_3)\sigma_{g_m^{-1}}(p_4)\right>^n_{\mathrm{BCFT}^{\bigotimes m}}}\notag\\
	&= 2\frac{1}{1-n}\log \frac{\left< \sigma_{g_A^{-1}}(p_2)\sigma_{g_B}(p_3)\sigma_{g_B^{-1}}(p_4)\sigma_{g_B}(q^a_4)\right>_{\mathrm{CFT}^{\bigotimes mn}}}{\left<\sigma_{g_m^{-1}}(p_2)\sigma_{g_m}(p_3)\sigma_{g_m^{-1}}(p_4)\sigma_{g_m}(q^a_4) \right>^n_{\mathrm{CFT}^{\bigotimes m}}}, \label{discaseo}
\end{align}

\subsubsection*{Configuration (p):}
\begin{align}
	S_R^{(n,m)}(A:B) &= \frac{1}{1-n}\log \frac{\left< \sigma_{g_A}(p_1)\sigma_{g_A^{-1}}(p_2)\sigma_{g_B}(p_3)\sigma_{g_B^{-1}}(p_4)\sigma_{g_B}(p_5)\sigma_{g_B^{-1}}(p_6)\sigma_{g_A}(p_7)\sigma_{g_A^{-1}}(p_8) \right>_{\mathrm{BCFT}^{\bigotimes mn}}}{\left<\sigma_{g_m}(p_1)\sigma_{g_m^{-1}}(p_2)\sigma_{g_m}(p_3)\sigma_{g_m^{-1}}(p_4)\sigma_{g_m}(p_5)\sigma_{g_m^{-1}}(p_6)\sigma_{g_m}(p_7)\sigma_{g_m^{-1}}(p_8) \right>^n_{\mathrm{BCFT}^{\bigotimes m}}}\notag\\
	&= \frac{1}{1-n}\log \frac{\Big< \sigma_{g_A}(p_1)\sigma_{g_A^{-1}}(p_8)\Big>_{\mathrm{BCFT}^{\bigotimes mn}}}{\Big<\sigma_{g_m}(p_1)\sigma_{g_m^{-1}}(p_8)\Big>^n_{\mathrm{BCFT}^{\bigotimes m}}} + 2 \frac{1}{1-n}\log \frac{\left<\sigma_{g_A^{-1}}(p_2)\sigma_{g_B}(p_3)\right>_{\mathrm{BCFT}^{\bigotimes mn}}}{\left<\sigma_{g_m^{-1}}(p_2)\sigma_{g_m}(p_3)\right>^n_{\mathrm{BCFT}^{\bigotimes m}}}\notag\\
	&~~~~~~~~~~~~~~~~~~~~~~~~~~~~~~~~~~~~~~+ \frac{1}{1-n}\log \frac{\left<\sigma_{g_B^{-1}}(p_4) \right>_{\mathrm{BCFT}^{\bigotimes mn}}\Big<\sigma_{g_B}(p_5) \Big>_{\mathrm{BCFT}^{\bigotimes mn}}}{\left<\sigma_{g_m^{-1}}(p_4) \right>^n_{\mathrm{BCFT}^{\bigotimes m}}\Big<\sigma_{g_m}(p_5) \Big>^n_{\mathrm{BCFT}^{\bigotimes m}}}\notag\\
	&= 2 \frac{1}{1-n}\log \frac{\left<\sigma_{g_A}(q^a_2)\sigma_{g_A^{-1}}(p_2)\sigma_{g_B}(p_3)\sigma_{g_B^{-1}}(q^a_3)\right>_{\mathrm{CFT}^{\bigotimes mn}}}{\left<\sigma_{g_m}(q^a_2)\sigma_{g_m^{-1}}(p_2)\sigma_{g_m}(p_3)\sigma_{g_m^{-1}}(q^a_3)\right>^n_{\mathrm{CFT}^{\bigotimes m}}}, \label{discasep}
\end{align}

\subsubsection*{Configuration (q):}
\begin{align}
	S_R^{(n,m)}(A:B) &= \frac{1}{1-n}\log \frac{\left< \sigma_{g_A}(p_1)\sigma_{g_A^{-1}}(p_2)\sigma_{g_B}(p_3)\sigma_{g_B^{-1}}(p_4)\sigma_{g_B}(p_5)\sigma_{g_B^{-1}}(p_6)\sigma_{g_A}(p_7)\sigma_{g_A^{-1}}(p_8) \right>_{\mathrm{BCFT}^{\bigotimes mn}}}{\left<\sigma_{g_m}(p_1)\sigma_{g_m^{-1}}(p_2)\sigma_{g_m}(p_3)\sigma_{g_m^{-1}}(p_4)\sigma_{g_m}(p_5)\sigma_{g_m^{-1}}(p_6)\sigma_{g_m}(p_7)\sigma_{g_m^{-1}}(p_8) \right>^n_{\mathrm{BCFT}^{\bigotimes m}}}\notag\\
	&= \frac{1}{1-n}\log \frac{\Big< \sigma_{g_A}(p_1)\sigma_{g_A^{-1}}(p_8)\Big>_{\mathrm{BCFT}^{\bigotimes mn}}}{\Big<\sigma_{g_m}(p_1)\sigma_{g_m^{-1}}(p_8)\Big>^n_{\mathrm{BCFT}^{\bigotimes m}}}\notag\\
	&~~~~~~~~~~~~~~ + \frac{1}{1-n}\log \frac{\left<\sigma_{g_A^{-1}}(p_2)\sigma_{g_B}(p_3)\sigma_{g_B^{-1}}(p_6)\sigma_{g_A}(p_7)\right>_{\mathrm{BCFT}^{\bigotimes mn}}}{\left<\sigma_{g_m^{-1}}(p_2)\sigma_{g_m}(p_3)\sigma_{g_m^{-1}}(p_6)\sigma_{g_m}(p_7)\right>^n_{\mathrm{BCFT}^{\bigotimes m}}}\notag\\
	&~~~~~~~~~~~~~~~~~~~~~~~~~~~~~~~~~~~~~~+ \frac{1}{1-n}\log \frac{\left<\sigma_{g_B^{-1}}(p_4) \right>_{\mathrm{BCFT}^{\bigotimes mn}}\Big<\sigma_{g_B}(p_5) \Big>_{\mathrm{BCFT}^{\bigotimes mn}}}{\left<\sigma_{g_m^{-1}}(p_4) \right>^n_{\mathrm{BCFT}^{\bigotimes m}}\Big<\sigma_{g_m}(p_5) \Big>^n_{\mathrm{BCFT}^{\bigotimes m}}}\notag\\
	&= \frac{1}{1-n}\log \frac{\left<\sigma_{g_A^{-1}}(p_2)\sigma_{g_B}(p_3)\sigma_{g_B^{-1}}(p_6)\sigma_{g_A}(p_7)\right>_{\mathrm{CFT}^{\bigotimes mn}}}{\left<\sigma_{g_m^{-1}}(p_2)\sigma_{g_m}(p_3)\sigma_{g_m^{-1}}(p_6)\sigma_{g_m}(p_7)\right>^n_{\mathrm{CFT}^{\bigotimes m}}}, \label{discaseq}
\end{align}

\end{appendices}

\bibliographystyle{jhep}
\bibliography{SR_Brane}

\end{document}